\pdfoutput=1

\documentclass[12pt,a4paper,hyperref]{JHEP3}
\let\ifpdf\relax
\usepackage{ifpdf}
\usepackage{color}
\let\normalcolor\relax
\usepackage{mathtools}
\usepackage{cite}

\definecolor{paper_blue}{rgb}{0.3,0.2,0.75}
\definecolor{cut2}{rgb}{0.18824,0.18824,0.48}
\definecolor{cut1}{rgb}{0.48,0.02824,0.18824}
\definecolor{varcolor}{rgb}{0.08,0.44,0.2}
\definecolor{functioncolor}{rgb}{0.08,0.28,0.6}
\newcommand{\mathematica}[3]{\vspace{0.35cm}\noindent\boxed{\begin{minipage}{#1\textwidth}\begin{tabular}{lp{11cm}}{\color{paper_blue}{\scriptsize{\tt In[1]:}}\raisebox{-0.65pt}{{\scriptsize{\tt=}}}}&{\tt #2}\\{\color{paper_blue}{\scriptsize {\tt Out[1]:}}\raisebox{-0.65pt}{{\scriptsize{\tt=}}}}&{\tt #3}\end{tabular}\end{minipage}}\vspace{0.35cm}}
\newcommand{\packageName}{{\tt two\uscore loop\uscore amplitudes}}
\newcommand{\vardef}[1]{{\color{varcolor}{\sl #1}\rule[-1.05pt]{7.5pt}{.75pt}}}
\newcommand{\vardefms}[1]{{\color{varcolor}{\sl #1}\rule[-1.05pt]{15pt}{.75pt}}}
\newcommand{\vardefo}[1]{{\color{varcolor}{\sl #1}\rule[-1.05pt]{7.5pt}{.75pt}{\bf{\sl :}}}}

\newcommand{\defn}[3]{~\\[-35pt]\begin{itemize}\item[]\indent\hspace{-21pt}$\bullet$\hspace{-.75pt} {\tt {\color{functioncolor}#1}\![}#2{\tt\,]\!:\hspace{2pt}}#3\end{itemize}\vspace{-10pt}}
\newcommand{\defnNA}[3]{~\\[-35pt]\begin{itemize}\item[]\indent\hspace{-21pt}$\bullet$\hspace{-.75pt} {\tt {\color{functioncolor}#1}\!}#2{\tt\,\!:\hspace{2pt}}#3\end{itemize}\vspace{-10pt}}
\newcommand{\defntb}[4]{~\\[-35pt]\begin{itemize}\item[]\indent\hspace{-21pt}$\bullet$\hspace{-.75pt} {\tt {\color{functioncolor}#1}\![}#2{\tt\,]\![}#3{\tt\,]\!:\hspace{2pt}}#4\end{itemize}\vspace{-10pt}}

\newcommand{\var}[1]{{\tt{\color{varcolor}{\sl#1}}}}
\newcommand{\ind}{\hspace{4ex}}
\newcommand{\fun}[1]{{\color{functioncolor}#1}}
\newcommand{\uscore}{\rule[-1.05pt]{7.5pt}{.75pt}}
\newcommand{\ns}{\hspace{-0.75pt}}
\newcommand{\paren}[1]{(\ns #1\ns)}
\newcommand{\fwbox}[2]{\text{\makebox[#1][c]{$\hspace{-150pt}\displaystyle#2\hspace{-150pt}$}}}
\newcommand{\fwboxL}[2]{\text{\makebox[#1][l]{$#2$}}}
\newcommand{\fwboxR}[2]{\text{\makebox[#1][r]{$#2$}}}
\renewcommand{\bar}{\overline}
\renewcommand{\hat}{\widehat}

\newcommand{\eq}[1]{\vspace{-0pt}\begin{equation}\hspace{-500pt}#1\hspace{-500pt}\vspace{-2.5pt}\end{equation}}
\newcommand{\eqs}[1]{\vspace{-0pt}\begin{equation}\begin{split}#1\end{split}\vspace{-2.5pt}\end{equation}}
\newcommand{\ab}[1]{\langle #1\rangle}
\newcommand{\x}[2]{(#1,#2)}
\newcommand{\newcap}{\mathrm{\raisebox{0.75pt}{{$\,\bigcap\,$}}}}
\newcommand{\tcap}{\scalebox{1}{$\!\newcap\!$}}
\newcommand{\tncap}{\scalebox{0.8}{$\!\newcap\!$}}
\newcommand{\ahat}{\hat{a}}
\newcommand{\mi}{{\rm\rule[2.4pt]{6pt}{0.65pt}}}
\newcommand{\pl}{\hspace{0.5pt}\text{{\small+}}\hspace{-0.5pt}}
\DeclareMathOperator*{\Res}{\mathrm{Res}}
\newcommand{\merge}{\raisebox{-1.5pt}{\scalebox{1.5}{$\otimes$}}}

\title{\mbox{\hspace{-0cm}{\LARGE Local Integrand Representations of All}}\\
\mbox{\hspace{-0cm}{\LARGE Two-Loop Amplitudes in Planar SYM}}}
\author{Jacob L. Bourjaily$^{a}$ and Jaroslav Trnka$^{b}$\\
\mbox{{\it $^{a}$ Niels Bohr International Academy and Discovery Center, Copenhagen, Denmark}}\\
\mbox{$^{b}$ {\it Walter Burke Institute for Theoretical Physics, California Institute of Technology,}}\\{{\it \,\,\,\,Pasadena, CA 91125, USA}}\vspace{-8pt}}
\preprint{2015}
\abstract{
We use generalized unitarity at the integrand-level to directly construct local, manifestly dual-conformally invariant formulae for all two-loop scattering amplitudes in planar, maximally supersymmetric Yang-Mills theory (SYM). This representation separates contributions into manifestly finite and divergent terms---in a way that makes manifest the exponentiation of infrared divergences at the integrand-level. These results perfectly match the all-loop BCFW recursion relations, to which we provide a closed-form solution valid through two-loop-order. Finally, we describe and document a {\sc Mathematica} package which implements these results, available as part of this work's source files on the {\tt arXiv}.}
\preprint{CALT-TH-2015-026}

\begin{document}

\newpage
\section{Introduction and Overview}\label{introduction_section}\vspace{-10pt}
Generalized unitarity has proven to be an extremely powerful tool for studying scattering amplitudes in quantum field theory beyond the leading-order of perturbation theory. One of its earliest triumphs was to show that any one-loop amplitude could be represented in terms of a basis of pre-chosen integrals, with coefficients computed in terms of tree-amplitudes (glued together into `on-shell functions'), \cite{Bern:1994zx,Bern:1994cg,Britto:2004nj,Britto:2004nc,Cachazo:2008vp,Bourjaily:2013mma}. Despite the enormous success of generalized unitarity at one-loop order, its extension to two or more loops---while straight-forward in principle---proved surprisingly difficult in practice until quite recently, when renewed interest from collider experiments was met with more powerful theoretical techniques (and more powerful computers), \cite{Kosower:2011ty,Johansson:2012sf,Johansson:2012zv,Zhang:2012ce,Henn:2013woa,Sogaard:2013fpa,Johansson:2013sda,Badger:2013sta,Badger:2014cva,Henn:2014lfa}. 

In addition to its practical applications, generalized unitarity has led to many important insights regarding scattering amplitudes, including the discovery of tree-level recursion relations for amplitudes \cite{Britto:2004ap,Britto:2005fq} and their all-loop generalization (at least for amplitudes in certain theories), \cite{ArkaniHamed:2010kv}. Moreover, it was learned through generalized unitarity that loop-amplitude integrands in planar, maximally supersymmetric ($\mathcal{N}\!=\!4$) Yang-Mills theory (`SYM') are conformally-invariant in dual-momentum space (`dual-conformally invariant') \cite{Drummond:2006rz,Drummond:2007aua}, a symmetry that was later recognized as a new superconformal symmetry of all scattering amplitudes in planar SYM, \cite{Drummond:2008vq}. When combined with the ordinary superconformal symmetry defining the theory, the two generate an infinite-dimensional symmetry algebra of scattering amplitudes known as the Yangian, \cite{Drummond:2009fd}. The desire to make dual-superconformal invariance manifest was partly responsible for the development of powerful new tools for analyzing scattering amplitudes, including new, compact representations of tree-amplitudes, \cite{Drummond:2008cr}, and the momentum-twistor variables introduced in \mbox{ref.\ \cite{Hodges:2009hk}}.

More recently, considerations of the general aspects of the (maximal) cuts (leading singularities) of scattering amplitudes (see e.g.\ \cite{ArkaniHamed:2009dn}) led to a new proposal for perturbative quantum field theory described in \mbox{ref.\ \cite{ArkaniHamed:2012nw}}. And for the particular case of planar SYM, there now exists a completely geometric, dual formulation of the $S$-matrix to all orders of perturbation theory, defined as the volume(-form) on a natural geometric space called the {\it amplituhedron}, \cite{Arkani-Hamed:2013jha,Arkani-Hamed:2013kca}. In this picture, the (all-loop) recursion relations (are believed to) provide a Yangian-invariant triangulation of the amplituhedron which can be understood in terms of on-shell diagrams. 

Although the recursion relations provide extremely efficient representations of scattering amplitudes (that someday may become easy to evaluate numerically), it is natural to seek representations that involve only local poles in the loop-momenta. This amounts to revisiting generalized unitarity, but now using the improved knowledge about scattering amplitudes and their symmetries at the integrand-level; and having access to the all-loop recursion relations greatly facilitates our ability to check any conjectures that we may have. Indeed, soon after the recursion relations became available, compact local expressions were guessed for all NMHV amplitudes through two-loops, and all MHV amplitudes through three-loops, \cite{ArkaniHamed:2010gh}. 

More systematically, an integrand-level enhancement of generalized unitarity was described in \mbox{ref.\ \cite{Bourjaily:2013mma}}, where one-loop integrands were fixed by listing a minimal (but complete) set of on-shell data, and tailoring specific integrands to match each cut individually. In this work, we describe the generalization of this approach to two-loop amplitudes in planar SYM. The representation we find follows from a similarly minimal set of independent on-shell data, sufficient to fix any two-loop scattering amplitude---specifically, the following six classes of on-shell functions:\\[-12pt]
\vspace{-10pt}\eq{\hspace{-520pt}\begin{array}{@{}l@{}c@{}c@{}c@{}c@{}c@{}c@{}c@{}r@{}}\fwboxL{0pt}{\left\{\rule[-20pt]{0pt}{65pt}\right.}\hspace{5pt}&\fwbox{100pt}{\raisebox{-47.5pt}{\includegraphics[scale=1]{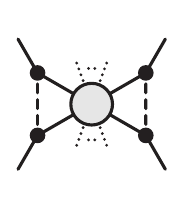}}}&,&\fwbox{120pt}{\raisebox{-47.5pt}{\includegraphics[scale=1]{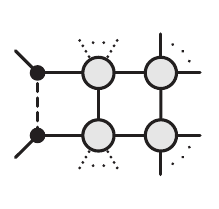}}}&,&\fwbox{130pt}{\hspace{-10pt}\raisebox{-47.5pt}{\includegraphics[scale=1]{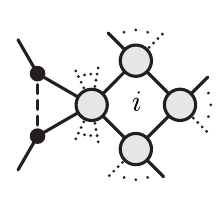}}}&\hspace{-10pt},&\\[-17pt]&\fwbox{100pt}{\raisebox{-47.5pt}{\includegraphics[scale=1]{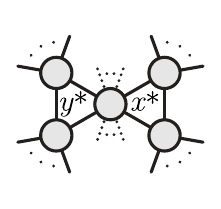}}}&,&\fwbox{120pt}{\raisebox{-47.5pt}{\includegraphics[scale=1]{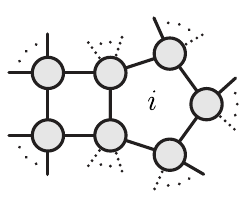}}}&,&\fwbox{135pt}{\raisebox{-47.5pt}{\includegraphics[scale=1]{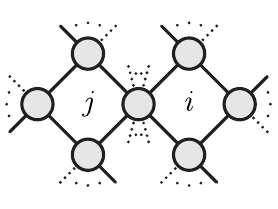}}}&&\fwboxL{0pt}{\left.\rule[-20pt]{0pt}{65pt}\right\}.}\\[-10pt]\end{array}\hspace{-500pt}\vspace{-2.5pt}\label{two_loop_on_shell_data_list_intro}}
To each of these, we simply attach a corresponding integrand, 
\vspace{-10pt}\eq{\hspace{-520pt}\begin{array}{@{}l@{}c@{}c@{}c@{}c@{}c@{}c@{}c@{}r@{}}\fwboxL{0pt}{\left\{\rule[-20pt]{0pt}{65pt}\right.}\hspace{5pt}&\fwbox{100pt}{\hspace{10pt}\raisebox{-47.5pt}{\includegraphics[scale=1]{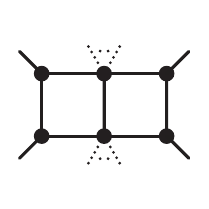}}\hspace{10pt}}&,&\fwbox{100pt}{\hspace{-20pt}\raisebox{-47.5pt}{\includegraphics[scale=1]{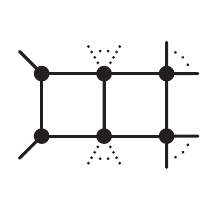}}\hspace{-10pt}}&,&\fwbox{100pt}{\hspace{-10pt}\raisebox{-47.5pt}{\includegraphics[scale=1]{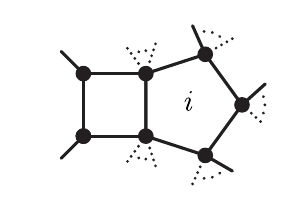}}}&\hspace{-10pt},&\\[-17pt]&\fwbox{100pt}{\hspace{10pt}\raisebox{-47.5pt}{\includegraphics[scale=1]{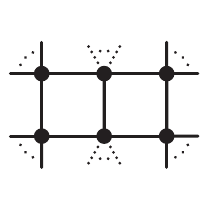}}\hspace{10pt}}&,&\fwbox{120pt}{\hspace{-20pt}\raisebox{-47.5pt}{\includegraphics[scale=1]{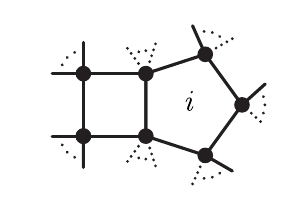}}\hspace{-10pt}}&,&\fwbox{135pt}{\raisebox{-47.5pt}{\includegraphics[scale=1]{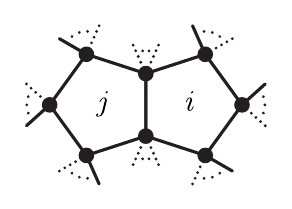}}}&&\fwboxL{0pt}{\left.\rule[-20pt]{0pt}{65pt}\right\},}\\[-10pt]\end{array}\hspace{-500pt}\vspace{-2.5pt}\label{two_loop_on_shell_int_list_intro}}
uniquely tailored to match the corresponding cut of the amplitude.

Notice that not all of the on-shell data listed in (\ref{two_loop_on_shell_data_list_intro}) are {\it leading} singularities. In particular, the following on-shell function may appear somewhat unusual:\\[-12pt]
\vspace{0pt}\eq{\fwbox{100pt}{\raisebox{-31.2pt}{\includegraphics[scale=1]{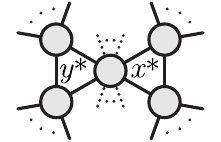}}}.\vspace{-15pt}\label{double_triangle_picture_from_intro}}
This represents a co-dimension six residue of the amplitude. As such, it depends on two further loop-integration variables, denoted $(x,y)$, which we evaluate at some conventional (but arbitrary) point $(x^*\hspace{-1pt},y^*\hspace{-1pt})$. Although the choice of this point is arbitrary, it will affect the forms of the integrands in (\ref{two_loop_on_shell_int_list_intro}): the double-box integral,\\[-12pt]
\vspace{-15pt}\eq{\fwbox{100pt}{\raisebox{-47.5pt}{\includegraphics[scale=1]{double_triangle_int}}},\vspace{-15pt}}
is tailored so that it evaluates to the identity at this point along the hexa-cut (\ref{double_triangle_picture_from_intro}); and each of the integrals, 
\vspace{-10pt}\eq{\fwbox{130pt}{\raisebox{-47.5pt}{\includegraphics[scale=1]{pentabox_int}}},\qquad\fwbox{130pt}{\raisebox{-47.5pt}{\includegraphics[scale=1]{kissing_boxes_int}}},\vspace{-10pt}}
are tailored so that each of their double-triangle hexa-cuts vanish at these points.

The use of such unfamiliar on-shell diagrams is required by the fact that scattering amplitudes at two-loops (even in planar SYM) cannot be fixed by their leading singularities alone. This is for a very simple reason: there are co-dimension seven cuts of amplitudes that support no further residues. The easiest way to see this is to observe, as noticed in \mbox{ref.\ \cite{CaronHuot:2012ab}}, that there exists a particular helicity component of the 10-particle N$^3$MHV two-loop amplitude which is entirely represented by a single Feynman diagram (in $\varphi^4$ theory):
\vspace{-2.5pt}\eq{\hspace{-40pt}\phantom{.}\raisebox{-47.5pt}{\includegraphics[scale=1]{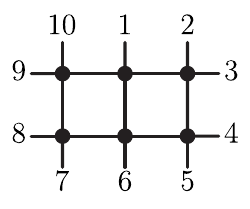}}\hspace{-5pt}\equiv\!\!\int\!\!d^4\ell_1d^4\ell_2\frac{1}{\x{\ell_1}{2}\x{\ell_1}{4}\x{\ell_1}{6}\x{\ell_1}{\ell_2}\x{\ell_2}{7}\x{\ell_2}{9}\x{\ell_2}{1}}.\vspace{-7.5pt}\label{elliptic_double_box_eg}}
It is not hard to see that this two-loop integral has no co-dimension eight residues. And so, this example proves that any representation of the amplitude must include data about less-than-maximal (co-dimension eight) cuts. In the representation we describe here, this is achieved by matching the hexa-cut, (\ref{double_triangle_picture_from_intro}).

This work is organized as follows. In \mbox{section \ref{amplitude_cuts_section}}, we provide a broad introduction to how generalized unitarity can be used to reconstruct loop amplitude {\it integrands}, using one-loop amplitudes as the primary example, outlined in \mbox{section \ref{one_loop_integrand_review}}. We review how relations among on-shell diagrams can be understood as residue theorems in \mbox{section \ref{boundaries_of_diagrams_and_residue_theorems_section}}, and we discuss how infrared singularities of amplitudes are encoded in the local amplitude expressions at the integrand-level in \mbox{section \ref{one_loop_IR_divergences_subsection}}. 

Our main results are described in \mbox{section \ref{two_loop_amplitude_section}}, where we provide a local, closed-form representation of all two-loop amplitude integrands in planar, maximally supersymmetric ($\mathcal{N}\!=\!4$) SYM. In \mbox{section \ref{on_shell_data_at_two_loops_subsection}}, we describe the construction of integrand contributions designed to explicitly match a minimal but complete set of on-shell data (cuts) term-by-term; this construction is summarized in \mbox{section \ref{local_two_loop_amplitude_formula_section}}, where a closed-form expression for all two-loop amplitudes is given. And finally, we show in \mbox{section \ref{two_loop_ratio_function_finiteness_section}} that the representation of two-loop amplitudes we have described makes manifest the exponentiation of infrared divergences at the integrand-level.

Because we expect the broad outlines of our approach to be accessible to most researchers familiar with the methods of generalized unitarity, we have taken care to avoid any unnecessary notational or conventional complications in the main body of this work. But for the sake of concreteness and reference, we provide explicit expressions for all the necessary ingredients using momentum-twistor variables in the appendices. Specifically, we review dual-momentum coordinates and momentum-twistor variables (and related notations and conventions) in \mbox{appendix \ref{overview_of_momentum_twistors_section}}; we provide explicit formulae for all the on-shell functions needed to represent amplitudes in \mbox{appendix \ref{explicit_formulae_for_on_shell_diagrams_appendix}}; and we give expressions for the necessary integrands in \mbox{appendix \ref{momentum_twistor_reps_of_integrands}}.

Our confidence in the correctness of our local representations of two-loop amplitudes follows in part from direct (numerical) comparison with the all-loop recursion relations, \cite{ArkaniHamed:2010kv}, to which we provide a closed-form solution which is valid through two-loop-order in \mbox{appendix \ref{two_loop_bcfw_appendix}}. Finally, we have made available a {\sc Mathematica} package `\packageName' which makes available our results. This package is documented and described in \mbox{appendix \ref{mathematica_appendix}}, and can be used to generate explicit expressions for all two-loop amplitudes in planar SYM using either the all-loop recursion relations or the local integrand representations we describe here. The package is available to download as part of this work's submission files on the {\tt arXiv}.

We have explicitly checked the equivalence of the local and recursed representations of all two-loop N$^k$MHV amplitude integrands involving as many as fourteen particles---well beyond the last appearance of any novel functional structures (the last of which---elliptic contributions---first are needed for 10-particle N$^3$MHV amplitudes). In addition to verifying the correctness of our local form of loop-amplitudes, the fact that these two representations agree provides strong evidence that both have been correctly implemented in the {\sc Mathematica} package (free of bugs or typos).

\newpage
\section{(Re-)Constructing Amplitudes by Their Cuts (Residues)}\label{amplitude_cuts_section}\vspace{-10pt}

In the traditional, path integral approach to quantum field theory, loop-corrections to scattering amplitudes are determined by summing over all the Feynman diagrams and integrating over all the loop-momenta $\{\ell_i\}$. At least for planar quantum field theories, an unambiguous definition of {\it the} loop integrand---the sum of Feynman diagrams prior to integration---can be provided using dual-momentum coordinates (reviewed in \mbox{appendix \ref{overview_of_momentum_twistors_section}}) and symmetrizing over all the loop variables $\{\ell_i\}$. Thus, we may unambiguously refer to {\it the} integrand, denoted $\mathcal{A}_{n}^{(k),l}$ for the $n$-particle, $l$-loop, N$^{k}$MHV amplitude in planar, maximally supersymmetric ($\mathcal{N}\!=\!4$) SYM.

Broadly speaking, the method of generalized unitarity is based on the observation that a loop amplitude integrand, viewed as a rational function of loop-momenta (analytically continued to the complex plane), can be reconstructed (up to terms without poles) using knowledge of its residues, also called its `cuts'. And these residues can always be computed in terms of strictly lower-loop amplitudes. Recall for example the familiar unitarity-cut:
\vspace{-0pt}\eq{\hspace{-200pt}\phantom{.}\raisebox{-24.5pt}{\includegraphics[scale=1]{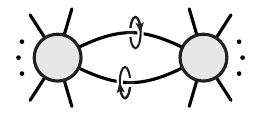}}.\hspace{-200pt}\vspace{-3pt}\label{unitarity_cut_figure}}
This picture represents a co-dimension two residue of a loop integrand where two internal propagators are put on shell. The unitarity-cut shown above is perhaps the most familiar and historically most important example of a much broader class of physically meaningful functions called {\it on-shell diagrams}.

More generally, locality and unitarity can provide a precise physical meaning to any network (graph) of scattering amplitudes connected by internal, on-shell particles---which we will call an on-shell diagram. For any graph $\Gamma$ involving amplitudes $\mathcal{A}_{v\in V}$ connected by edges indexed by $i\!\in\!I$, we may associate with it an {\it on-shell function} $f_\Gamma$ defined according to, \cite{ArkaniHamed:2012nw}:
\vspace{5pt}\eq{f_\Gamma\equiv \int\!\prod_{i\in I}\!d^{3|4}\mathrm{LIPS}_i\left(\prod_{v\in V}\mathcal{A}_v\right)\!,\label{general_on_shell_function_definition}\vspace{2.5pt}}
where $d^{3|4}\mathrm{LIPS}_i$ represents the measure on the Lorentz-invariant super-phase-space for the $i^{\mathrm{th}}$ internal particle.\footnote{In a more general quantum field theory, the super-phase-space integrals of (\ref{general_on_shell_function_definition}) would be replaced with ordinary phase-space integrals, together with a summation over the possible quantum numbers (helicity, colour, etc.) for the internal particle $i\!\in\!I$.} Notice that this definition is completely general, and provides an invariant meaning to any graph built out of amplitudes---not just those of planar maximally supersymmetric Yang-Mills theory. 

(While on-shell functions can be defined for any quantum field theory, those of planar SYM are especially simple: all such functions and all their relations can be classified by permutations and computed combinatorially (see \mbox{refs.\ \cite{ArkaniHamed:2012nw,Bourjaily:2012gy}}). Because of this, all the functions needed for our present work could be systematically described by permutations. Nevertheless, such sophisticated machinery will not be necessary for us here, as it is relatively straight-forward to write closed-form expressions for every on-shell function we will need---which we provide in \mbox{appendix \ref{explicit_formulae_for_on_shell_diagrams_appendix}}.)

An important characteristic of any on-shell function is number of non-trivial phase-space integrations that remain after trivializing as many of them as possible using the momentum-conserving $\delta$-functions of the amplitudes at each vertex. Because each internal line represents a three-dimensional phase-space integral, and there are four momentum-conserving $\delta$-functions at each vertex (four of which always impose overall momentum conservation), a diagram with $n_I$ internal lines and $n_V$  vertices represents an integral over $(3n_I\,\mi\, 4n_V\pl\,4)$ internal degrees of freedom. Thus, the unitarity cut, (\ref{unitarity_cut_figure}), represents a two-dimensional phase-space integral, matching its interpretation as a co-dimension two residue of a four-dimensional loop integrand.

Because all the residues of loop amplitudes correspond to on-shell diagrams that can be computed according to (\ref{general_on_shell_function_definition}), it is in principle possible to reconstruct any scattering amplitude given any complete basis of $l$-loop integrals: it becomes a straight-forward (if computationally onerous) problem of linear algebra to find the coefficients of integrals in the bais which ensure that every cut matches field theory. This is the traditional way in which generalized unitarity is used to represent loop amplitudes. 

However, there are many difficulties with this approach in practice. Beyond one-loop, for example, it is surprisingly difficult to even find a complete basis of integrals (let alone choose a good basis). And even if a complete basis were known, the linear algebra required to determine the right coefficients from the on-shell data grows in complexity quite rapidly with multiplicity. And finally, the isolated on-shell diagrams (those which put $4l$ propagators on-shell, greatly simplifying the linear algebra involved) prove to be insufficient as data to determine amplitudes beyond one-loop in general---a fact that greatly complicates matters.

The strategy we describe here is very different. Rather than starting with a basis of integrands and solving for coefficients, we will {\it directly construct} amplitudes from data about their cuts, encoded as on-shell diagrams. Indeed, the representation of two-loop amplitudes we describe in \mbox{section \ref{two_loop_amplitude_section}} will involve far fewer integrands than needed for a complete basis (nor will the all integrands used be independent). Each term is constructed to match a specific cut of the amplitude; and matching these cuts will ensure that we match field theory everywhere. This approach to generalized unitarity seems applicable to all loop-orders (for general quantum field theories).

It is natural to wonder what advantage (if any) there is to matching field theory at the level of the integrand. After all, representations which differ by terms that are parity-odd or total derivatives will agree after integration; and it may seem more economical, for example, to discard parity-odd contributions as in the standard approach to one-loop generalized unitarity. Our motivations for matching the full Feynman integrand are two-fold. First, matching field theory at the level of the integrand will help make manifest all the symmetries of the theory (and allow us to compare our results with BCFW in the case of planar SYM). Second, and much more importantly, the requirement that we match every cut individually is a {\it stronger} one---strong enough to lead to a unique representation. Without this stronger constraint, it is likely that no closed-form representation of two-loop amplitudes in planar SYM would have been found; and we expect this to be similarly important for finding representations of amplitudes in more general theories.

This new approach to generalized unitarity was first described in \mbox{ref.\ \cite{Bourjaily:2013mma}}, where it was used to construct local integrand-level representations of all one-loop amplitudes in planar SYM, and we will extend it here to match all two-loop amplitudes in planar SYM. But this strategy has obvious applications to any quantum field theory at arbitrary loop-order. Both in order to put the present work in the context of this more general philosophy and to introduce some essential ingredients for our construction of two-loop amplitudes, let us begin with a brief review of the construction at one-loop.

\subsection{Review of One-Loop, Integrand-Level Generalized Unitarity}\label{one_loop_integrand_review}
Prior to the integrand-level construction described in \mbox{ref.\ \cite{Bourjaily:2013mma}}, generalized unitarity would typically be used to represent (the cut-constructible part of) any one-loop amplitude in terms of scalar box, triangle, and bubble integrals. The coefficients of the scalar boxes would be determined using co-dimension four residues (leading singularities) corresponding to on-shell diagrams with the topology of a box. Although this approach does correctly reproduce integrated expressions for amplitudes, it is incapable of matching field theory at the integrand-level for a very simple reason: no collection of scalar integrals (boxes, triangles, bubbles, etc.) can form a complete basis of one-loop {\it integrands}. To illustrate the obstruction and how to ameliorate it, let us briefly review the possible co-dimension four residues of one-loop amplitudes.

In any four-dimensional quantum field theory, there are two types of isolated points in (complexified) loop-momentum space where co-dimension four residues have support. These either involve cutting three or four distinct propagators, the latter of which should be quite familiar. When four propagators are put on-shell, the resulting on-shell diagram has the topology of a box:
\eq{\hspace{00pt}\phantom{\;\;\mathrm{with}\;\;\ell\!\mapsto\!\ell^i\!\in\!\{{\color{cut1}Q^1},{\color{cut2}Q^2}\},}f^i_{a,b,c,d}\equiv\!\!\!\raisebox{-47.25pt}{\includegraphics[scale=1]{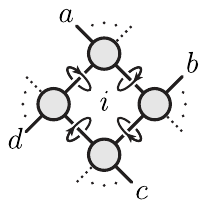}}\;\;\mathrm{with}\;\;\ell\!\mapsto\!\ell^i\!\!\in\!\{{\color{cut1}Q^1},{\color{cut2}Q^2}\},\phantom{f^i_{a,b,c,d}\equiv\!\!\!}\label{one_loop_quad_cut_figure}\vspace{-5pt}}
where $\{{\color{cut1}Q^1},{\color{cut2}Q^2}\}$ are the two possible `quad-cuts'---the places in loop-momentum space that simultaneously solve the four quadratic constraints,
\vspace{5pt}\eq{\x{\ell}{a}=\x{\ell}{b}=\x{\ell}{c}=\x{\ell}{d}=0,\label{quad_cut_equations}}
which make the internal propagators on-shell. Here, we have used dual-momentum coordinates (reviewed in \mbox{appendix \ref{overview_of_momentum_twistors_section}}) to denote the ordinary Feynman propagators of (\ref{one_loop_quad_cut_figure}). Specifically, we write $p_a\!\equiv\!x_{a+1}\mi\,x_a$, and make use of the Lorentz-invariants,\\[-12pt]
\vspace{0pt}\eq{\hspace{-30pt}\x{a}{b}=\x{b}{a}\equiv(x_b\,\mi\,x_a)^2=(p_a+p_{a+1}+\ldots+p_{b-1})^2,\;\;\;\x{\ell}{a}\equiv(\ell\,\mi\,x_{a})^2.\vspace{0.pt}}

It is important to note that the on-shell function involving internal momentum $\ell\!=\!{\color{cut1}Q^1}$ is almost always unequal to the function involving $\ell\!=\!{\color{cut2}Q^2}$:
\vspace{-0pt}\eq{\;{\color{cut1}f^{1}_{a,b,c,d}}\equiv\!\!\!\raisebox{-47.25pt}{\includegraphics[scale=1]{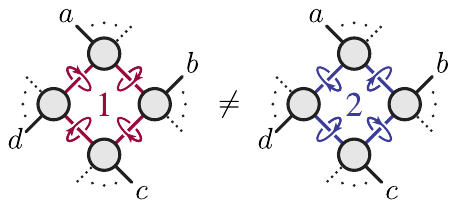}}\!\!\!\equiv {\color{cut2}f^{2}_{a,b,c,d}}\;.\vspace{-5pt}}
The inequality of the amplitude's residues about $\{{\color{cut1}Q^1},{\color{cut2}Q^2}\}$ in loop-momentum space reflects the general {\it chirality} of loop integrands in field theory as ${\color{cut1}Q^1}$ and ${\color{cut2}Q^2}$ are exchanged under parity. This makes it impossible to match field theory at the integrand-level using {\it scalar} integrals for a very simple reason: the {\it Global Residue Theorem} (see e.g.\ \mbox{ref.\ \cite{GriffithsHarris}}) tells us that the sum of all the non-vanishing residues of a multidimensional integral must vanish; for a scalar box integral (one involving four propagators) this implies that the two residues are related by (for generic $a,b,c,d$):\\[-10pt]
\vspace{5pt}\eq{\Res_{\,\ell={\color{cut1}Q^1}}\!\!\left(\frac{d^4\ell}{\x{\ell}{a}\x{\ell}{b}\x{\ell}{c}\x{\ell}{d}}\!\right)+\Res_{\,\ell={\color{cut2}Q^2}}\!\!\left(\frac{d^4\ell}{\x{\ell}{a}\x{\ell}{b}\x{\ell}{c}\x{\ell}{d}}\!\right)=0.}
(This also provides a simple way of seeing that there must be (at least) two solutions to the quad-cut equations, (\ref{quad_cut_equations}).)

Because any integral with support on one of the two quad-cuts must involve at least the four relevant propagators, and because any integrand involving only four propagators will {\it necessarily} have support on both quad-cuts (with residues equal in magnitude), the simplest possible integrand with support on {\it only one} of the two physical quad-cuts must involve at least five propagators; and in order to ensure that the additional propagator introduces no other physical cuts, we are encouraged to choose the fifth propagator to be entirely spurious---of the form $\x{\ell}{X}$ for some {\it arbitrary} point $X$ in dual-momentum space, not among those points associated with the external momenta. In this way, we can construct integrals $\mathcal{I}^i_{a,b,c,d}$ with support on exactly one of the physical quad-cuts $Q^i$ for each topology,\footnote{Of course, each integrand will have support on unphysical cuts involving the propagator $\x{\ell}{X}$; and we will need to make sure that that the full amplitude is free of support on such spurious cuts.}
\eq{\mathcal{I}^i_{a,b,c,d}\equiv\!\!\!\raisebox{-47.5pt}{\includegraphics[scale=1]{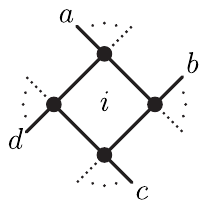}}\!\!\!\equiv \frac{\x{X}{Y_i(\ell)}}{\x{\ell}{a}\x{\ell}{b}\x{\ell}{c}\x{\ell}{d}\x{\ell}{X}},\vspace{-5pt}\label{chiral_one_loop_integrands_as_decorations}}
where the numerator $\x{X}{Y_i(\ell)}$ is chosen so that the integrand's residue on $Q^i$ is of unit magnitude, and that it vanishes on the other cut (when $\ell\!\mapsto\!Q^{j\neq i}$).\footnote{There is actually one additional criterion needed to fully fix the form of the numerators $Y_i(\ell)$ for every topology (which also plays a role in ensuring that the amplitude is independent of $X$): we should also require that the integrand vanishes on any parity-even contour enclosing $\x{\ell}{X}\!=\!0$.} Explicit expressions for the numerators $Y_i(\ell)$ which satisfy these criteria for every topology are given in \mbox{Table \ref{chiral_box_integrands_table}} of \mbox{appendix \ref{momentum_twistor_reps_of_integrands}}.

Thus, by decorating each on-shell diagram $f^{i}_{a,b,c,d}$ by its corresponding chiral integrand $\mathcal{I}^i_{a,b,c,d}$, we will have an integrand which precisely matches field theory on all co-dimension four residues {\it involving four distinct propagators}. But not all co-dimension four residues of the field theory integrand involve four distinct propagators.

The other co-dimension four residues are supported on points involving only three distinct propagators---corresponding to the so-called {\it composite} leading singularities. In order for this to be possible, at least one of the three propagators must factorize on the support of the other two. This in fact happens whenever any two propagators are {consecutive}---of the form $\x{\ell}{a\mi1},\x{\ell}{a}$ for some leg $a$: on the support of $\x{\ell}{a\mi1}\!=\!0$, $\x{\ell}{a}$ factorizes, and vice versa. Thus, for example, any three propagators for which two are consecutive can define support a co-dimension four residue where one of the consecutive pair is cut twice, as in the following example:
\eq{\phantom{\;\;\mathrm{with}\;\;\ell\!\mapsto\!x_a\phantom{\,.}}\raisebox{-37.2pt}{\includegraphics[scale=1]{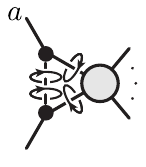}}\;\;\mathrm{with}\;\;\ell\!\mapsto\!x_a\,.\vspace{-7.5pt}\label{composite_triangle_on_shell_diagram}}
It is not hard to see that when both factors of a propagator vanish, the loop-momentum through that line must vanish; and so, the physical residue is simple:\\[-12pt] 
\vspace{-5pt}\eq{\hspace{-200pt}\phantom{\hspace{-5pt}=\mathcal{A}_{n}^{\text{tree}}.}\raisebox{-37.25pt}{\includegraphics[scale=1]{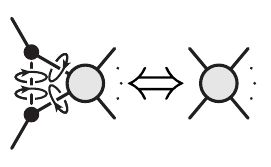}}\hspace{-5pt}=\mathcal{A}_{n}^{\text{tree}}.\hspace{-200pt}\vspace{-5pt}}
And because no momentum flows through one internal leg on such a cut, it is easy to see that these are the {\it only} composite quad-cuts for which the amplitude is non-vanishing: all other cases would correspond to bubble-corrections to trees (which vanish in SYM). 

In order to reconstruct the field theory loop-integrand, we must also match all the composite leading singularities of the form (\ref{composite_triangle_on_shell_diagram}). Because these quad-cuts are parity-even, all of the integrands $\mathcal{I}^i_{a,b,c,d}$ vanish at these points in loop-momentum space; therefore, we must match them separately by attaching to each on-shell diagram of the form (\ref{composite_triangle_on_shell_diagram}), an integrand engineered precisely to match this cut (and no others):\\[-10pt]
\vspace{-10pt}\eq{\mathcal{I}_{\mathrm{}}^a\equiv\hspace{-10pt}\raisebox{-42.25pt}{\includegraphics[scale=1]{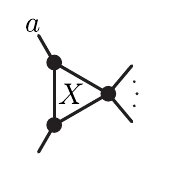}}\hspace{-20pt}\equiv \frac{\x{X}{Y^a}}{\x{\ell}{a\mi1}\x{\ell}{a}\x{\ell}{a\pl1}\x{\ell}{X}}.\vspace{-17.5pt}\label{divergent_triangle_integrand_definition}}
The numerator of this integrand is uniquely fixed by the criterion that it have unit residue on the corresponding (composite) quad-cut. An explicit expression for $Y^a$ is given in \mbox{Table \ref{chiral_box_integrands_table}} of \mbox{appendix \ref{momentum_twistor_reps_of_integrands}}.

We have now exhausted the list of points in loop-momentum-space where one-loop amplitudes can support co-dimension four residues:
\vspace{-0pt}\eq{\hspace{-200pt}\phantom{.}\left\{\rule[5pt]{0pt}{35pt}\right.\hspace{-5pt}\raisebox{-37.25pt}{\includegraphics[scale=1]{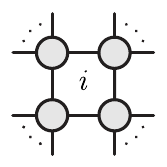}},\hspace{-0pt}\raisebox{-37.25pt}{\includegraphics[scale=1]{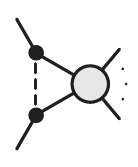}}\hspace{-0pt}\left.\rule[5pt]{0pt}{35pt}\right\}.\hspace{-200pt}\vspace{-0pt}\label{independent_one_loop_data}}
Decorating each of these diagrams by its individually-tailored integrand, 
\vspace{-0pt}\eq{\hspace{-200pt}\begin{array}{@{}c@{$\,\,$}c@{}c@{}c@{}c@{}@{}c@{$\,\,$}l@{}}f^i_{a,b,c,d}&\equiv&\raisebox{-47.5pt}{\includegraphics[scale=1]{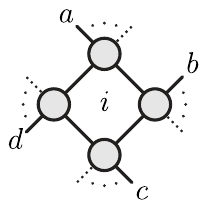}}&\hspace{-0pt}\raisebox{-3pt}{\scalebox{2}{$\Rightarrow$}}\hspace{-0pt}&\raisebox{-47.5pt}{\includegraphics[scale=1]{one_loop_integrand_decoration}}&\equiv&\mathcal{I}_{a,b,c,d}^{i},\\[-20pt]
\mathcal{A}_{n}^{\text{tree}}&\equiv&\hspace{-5pt}\raisebox{-47.5pt}{\includegraphics[scale=1]{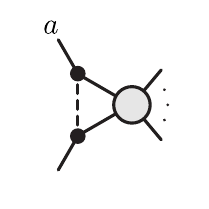}}\hspace{5pt}&\hspace{-0pt}\raisebox{-3pt}{\scalebox{2}{$\Rightarrow$}}\hspace{-0pt}&\hspace{5pt}\raisebox{-42.25pt}{\includegraphics[scale=1]{one_loop_triangle}}\hspace{-5pt}&\equiv&\mathcal{I}_{\mathrm{}}^{a},\end{array}\hspace{-200pt}\vspace{-10pt}\label{decorating_one_loop_diagrams}}
we will have constructed an integrand which perfectly matches field theory on all its co-dimension four cuts:
\vspace{5pt}\eq{\sum_{a,b,c,d}\sum_{i}f_{a,b,c,d}^i\mathcal{I}_{a,b,c,d}^i+\mathcal{A}_{n}^{\mathrm{tree}}\sum_a\mathcal{I}^a\;.\label{quad_cut_matching_one_loop_expression}}
(For a general quantum field theory, we would need to continue this discussion to match residues of lower co-dimension (not already matched); but in SYM, which is free of any poles at infinity, the above analysis is complete.)

In addition to showing that (\ref{quad_cut_matching_one_loop_expression}) matches all the co-dimension four residues of field theory, however, we must show that it is free of any unphysical singularities---those involving the propagator $\x{\ell}{X}$. The independence of $X$ in (\ref{quad_cut_matching_one_loop_expression}) is easiest to demonstrate using residue theorems, which we review in the following subsection.

\subsection{Boundaries of On-Shell Diagrams: Residue Theorems and Identities}\label{boundaries_of_diagrams_and_residue_theorems_section}
Not all on-shell diagrams are independent as functions; they satisfy many identities referred to as {\it residue theorems}. These identities can be understood homologically in terms of the geometry of the positroid stratification described in \mbox{ref.\ \cite{ArkaniHamed:2012nw}}, or as applications of Cauchy's residue theorem when they are considered as residues of loop-amplitude integrands. Consider any residue of the loop amplitude with next-to-maximal co-dimension $(4l\,\mi\,1)$; such a cut represents a one-dimensional integral. (From our previous discussion, this corresponds to any on-shell diagram satisfying $3n_I\mi\,4n_V\pl\,4\,\!=\!1$.) Cauchy's theorem tells us that the sum of all the co-dimension one residues of this cut will be zero.

The co-dimension one residues of any cut are very easy to classify as they correspond to the poles of the corresponding on-shell function. These arise from the boundaries from each vertex tree-amplitude (which are factorization channels), or from deleting any edge connecting two three-point amplitudes:\footnote{Here, blue/filled (white/empty) trivalent vertices indicate MHV ($\bar{\text{MHV}}$) $3$-particle amplitudes.}
\vspace{7.5pt}\eq{\raisebox{-19.625pt}{\includegraphics[scale=1]{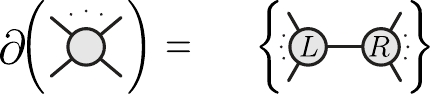}}\fwboxL{0pt}{\hspace{-110pt}\raisebox{0.75pt}{\scalebox{1.5}{$\displaystyle\sum_{L,R}$}}},\quad\mathrm{}\quad\raisebox{-19.625pt}{\includegraphics[scale=1]{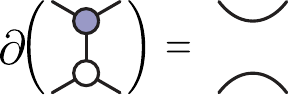}}\;.\phantom{(2.14)}\vspace{2.5pt}\label{general_diagrammatic_boundary_rule}}
Notice that both operations reduce the number of integrals, ($3n_I\mi\,4n_V\pl\,4$), by one. 

Starting with any on-shell diagram corresponding to a $(4l\,\mi\,1)$-cut, we therefore have an identity among $4l$-cuts by taking boundaries to all sub-diagrams according to (\ref{general_diagrammatic_boundary_rule}). For one-loop leading singularities, these identities are generated from triple-cut diagrams. When all corners are massive (that is, when each corner involves at least two legs), these identities take the form:
\vspace{-5pt}\eq{\hspace{-240pt}\raisebox{-40.75pt}{\includegraphics[scale=1]{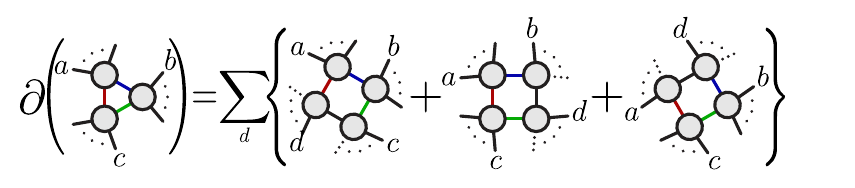}}\hspace{-32pt}\raisebox{-1.75pt}{\scalebox{1.5}{$=\!0.$}}\hspace{-200pt}\vspace{-7.5pt}\label{one_loop_grt_1}}
And when two of the corners are massless, we get two types of contributions:
\vspace{-15pt}\eq{\hspace{-200pt}\raisebox{-2.75pt}{\scalebox{1.75}{$\displaystyle\partial$}}\hspace{-9.5pt}\raisebox{-52.5pt}{\includegraphics[scale=1]{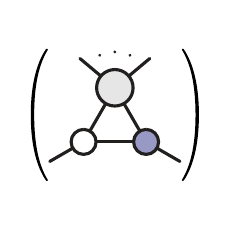}}\hspace{-7.5pt}\raisebox{-1.75pt}{\scalebox{1.5}{$\displaystyle=$}}\hspace{-32.5pt}\raisebox{-52.5pt}{\includegraphics[scale=1]{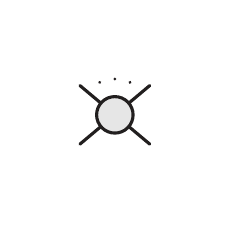}}\hspace{-30pt}\raisebox{-1.75pt}{\scalebox{1.5}{$\displaystyle-$}}\hspace{5pt}\raisebox{0.75pt}{\scalebox{1.5}{$\displaystyle\sum_{L,R}$}}\hspace{-10pt}\raisebox{-52.5pt}{\includegraphics[scale=1]{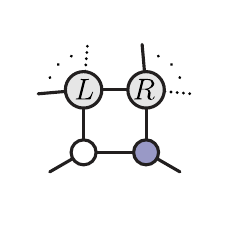}}\hspace{-10pt}\raisebox{-1.75pt}{\scalebox{1.5}{$\displaystyle=0.$}}\hspace{-200pt}\vspace{-22pt}\label{one_loop_grt_2}}
(We should point out that this residue theorem played a very important role in the development of our understanding of scattering amplitudes: it can be understood as representing a tree-amplitude in terms of diagrams which are themselves built out of amplitudes with strictly fewer legs. This was the way in which recursion relations for (tree-)amplitudes were first discovered in \mbox{ref.\ \cite{Britto:2004ap}}.)

We now have all the ingredients needed to show that the combination of $X$-dependent integrands in (\ref{quad_cut_matching_one_loop_expression}) is in fact independent of $X$, and thereby show that (\ref{quad_cut_matching_one_loop_expression}) must match the full one-loop amplitude integrand. To prove this, we need only show that all the `spurious' co-dimension four residues of (\ref{quad_cut_matching_one_loop_expression}) involving the propagator $\x{\ell}{X}$ vanish.

Consider the spurious quad-cut supported at the point where
\vspace{5pt}\eq{\x{\ell}{X}\!=\!\x{\ell}{a}\!=\!\x{\ell}{b}\!=\!\x{\ell}{c}\!=\!0.\label{generic_spurious_quad_cut}}
In the expression (\ref{quad_cut_matching_one_loop_expression}), there are many integrands which have support on this spurious quad-cut---in particular, those involving integrands $\{\mathcal{I}^i_{a,b,c,d},\mathcal{I}^i_{a,b,d,c},\mathcal{I}^i_{a,d,b,c}\}$. Conveniently, the coefficients of these integrands in (\ref{quad_cut_matching_one_loop_expression}) correspond exactly to the collection of on-shell diagrams appearing in the right-hand side of the identity (\ref{one_loop_grt_1}). 

A separate case to consider is when the three physical propagators in (\ref{generic_spurious_quad_cut}) are consecutive---say, $\x{\ell}{a\mi1},\x{\ell}{a},\x{\ell}{a\pl1}$. In this case, the triangle integral $\mathcal{I}^a$ also has support on this cut.\footnote{There are actually two spurious cuts of this topology which are parity-conjugates. The argument in both cases is the same.} The other terms which contribute to this cut are those involving boxes with these propagators. Thus, the coefficients of all the integrands in (\ref{quad_cut_matching_one_loop_expression}) which have support on such a spurious quad-cut are precisely those appearing in the right-hand side of the identity (\ref{one_loop_grt_2}). 

Therefore, we have shown that dressing each on-shell diagram in (\ref{independent_one_loop_data}) by its corresponding integrand according to (\ref{decorating_one_loop_diagrams}) must result in an integrand which is free of any dependence on $X$ and therefore must match field theory everywhere:
\vspace{5pt}\eq{\boxed{\mathcal{A}_n^{(k),1}=\sum_{a,b,c,d}\sum_{i}f_{a,b,c,d}^i\mathcal{I}_{a,b,c,d}^i+\mathcal{A}_{n}^{\mathrm{tree}}\sum_a\mathcal{I}^a\;.}\label{local_one_loop_amplitude_expression}}
This was the form of one-loop amplitude integrands derived in \mbox{ref.\ \cite{Bourjaily:2013mma}}. Of course, this representation matches the more familiar scalar box expansion {\it after integration}; at the integrand-level, however, (\ref{local_one_loop_amplitude_expression}) differs from the scalar box expansion by parity-odd contributions (which vanish when integrated over the parity-invariant contour).

Before moving on, let us briefly note that because the representation (\ref{local_one_loop_amplitude_expression}) is independent of $X$, while each term involves an $X$-dependent factor of the form $\x{X}{Y_q}/\x{\ell}{X}$, it must be the case that sum of terms in the numerator (put over a common denominator) factorizes to become proportional to $\x{\ell}{X}$ directly. This fact will be useful in \mbox{section \ref{two_loop_ratio_function_finiteness_section}}.

\newpage
\subsection{Infrared Divergences of Amplitudes and Finiteness of Observables}\label{one_loop_IR_divergences_subsection}

For any theory involving massless particles, loop amplitudes have physically meaningful infrared singularities. These singularities arise due to very specific regions of the loop integration, which can be fully understood at the integrand-level. In particular, logarithmic divergences arise from the so-called {\it collinear} regions (which are co-dimension three), where $\ell\!\mapsto\!\alpha\, p_a$ for some (massless) external momentum $p_a$.

In dual-momentum coordinates (see \mbox{appendix \ref{overview_of_momentum_twistors_section}}), collinear regions correspond to regions where $\ell$ approaches the line between $x_a$ and $x_{a+1}$. This is achieved by the triple-cut involving any two consecutive propagators:
\vspace{5pt}\eq{\raisebox{-23.5pt}{\includegraphics[scale=1]{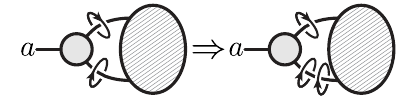}}.\vspace{-0pt}\label{collinear_region_figure}}
Recall that every propagator $\x{\ell}{a}$ factorizes on the support of either $\x{\ell}{a\mi1}$ or $\x{\ell}{a\pl1}$, and that when both factors of a propagator are cut, the momentum flowing through the corresponding line vanishes. 

Collinear regions are responsible for simple logarithmic divergences---terms proportional to $1/\epsilon$, $\log(m^2)$, or $\log(\epsilon)$ in dimensional-regularization, mass-regularization, or the scale-invariant regularization scheme described in \mbox{ref.\ \cite{Bourjaily:2013mma}}, respectively. Further divergences arise when the integrand has support on mutually overlapping collinear regions, which give rise to terms proportional to $1/\epsilon^2$, $\log^2(m^2)$, or $\log^2(\epsilon)$. These co-dimension four regions correspond to the places where $\ell\!\mapsto\!x_a$, giving rise to so-called `soft-collinear' divergences. 

Importantly, regions of collinear divergence are parity-even: a double-cut involving two consecutive propagators must be either one of the two parity conjugates,\\[-12pt]
\vspace{5pt}\eq{\raisebox{-23.5pt}{\includegraphics[scale=1]{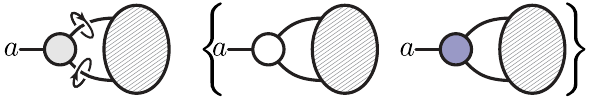}}\fwboxL{0pt}{\hspace{-200pt}\raisebox{-1.2pt}{\scalebox{1.5}{$\in$}}\hspace{90pt}\raisebox{-1.2pt}{\scalebox{1.5}{$,$}}};}
and the co-dimension three collinear region of (\ref{collinear_region_figure}) corresponds to an intersection of these two conditions. Because the integrands $\mathcal{I}^i_{a,b,c,d}$ defined in (\ref{chiral_one_loop_integrands_as_decorations}) were constructed explicitly to exclude one of these regions or the other, each of these integrands {\it manifestly} vanishes in all such regions of collinear divergence and is therefore `manifestly finite'---can be evaluated without any regularization. 

In contrast, the triangle integrands, $\mathcal{I}^a$, defined in (\ref{divergent_triangle_integrand_definition}) were {specifically} constructed to have support on the regions supporting (soft-)collinear divergences. Thus, each term in the representation (\ref{local_one_loop_amplitude_expression}) involving integrands $\mathcal{I}^i_{a,b,c,d}$ is manifestly finite, and each term involving the triangle integrands $\mathcal{I}^a$ is manifestly divergent. Therefore, letting $\mathcal{I}_{\mathrm{div}}\!\equiv\!\sum_{a}\mathcal{I}^a$, we see that our representation of one-loop amplitudes admits the very convenient separation,
\vspace{5pt}\eq{\hspace{-500pt}\mathcal{A}_{n}^{(k),1}\!=\,\underbrace{\fwbox{0pt}{\phantom{\sum}}\mathcal{A}_{n}^{(k),0}\mathcal{I}_{\mathrm{div}}}_{\displaystyle\mathcal{A}_{n,\mathrm{div}}^{(k),1}}+\underbrace{\displaystyle\sum f_{a,b,c,d}^i\mathcal{I}_{a,b,c,d}^i}_{\displaystyle\mathcal{A}_{n,\mathrm{fin}}^{(k),1}}\equiv\mathcal{A}_{n,\text{div}}^{(k),1}+\mathcal{A}_{n,\mathrm{fin}}^{(k),1}.\hspace{-500pt}\vspace{-0pt}\label{one_loop_divergent_finite_split}}

Because in this representation, all of the infrared singularities are universally proportional to the same $\mathcal{I}_{\mathrm{div}}$, the ratio of any two helicity amplitudes with the same multiplicity will always be finite. (This continues to be true to all orders of perturbation theory, in fact.) The {\it ratio function} is the canonical observable of this type, defined as the ratio of the N$^k$MHV amplitude to the N$^{(k=0)}$MHV amplitude:
\vspace{7.5pt}\eq{\hspace{-35pt}\mathcal{R}_{n}^{(k)}\equiv\frac{\mathcal{A}_n^{(k)}}{\mathcal{A}_{n}^{(0)}}\equiv\frac{\mathcal{A}_{n}^{(k),0}\!\!+\!\hbar\mathcal{A}_{n}^{(k),1}\!\!+\!\hbar^2\mathcal{A}_{n}^{(k),2}\!\!+\!\ldots}{\mathcal{A}_{n}^{(0),0}\!\!+\!\hbar\mathcal{A}_{n}^{(0),1}\!\!+\!\hbar^2\mathcal{A}_{n}^{(0),2}\!\!+\!\ldots}\equiv\mathcal{R}_{n}^{(k),0}\!\!+\hbar\mathcal{R}_{n}^{(k),1}\!\!+\hbar^2\mathcal{R}_{n}^{(k),2}\!\!+\!\ldots\,.\label{general_ratio_definition}\vspace{5pt}}
Notice that by expanding this as a formal power series in $\hbar$ allows us to define an \mbox{$l$-loop} ratio function $\mathcal{R}_{n}^{(k),l}$ for each order of perturbation theory. Expressing all tree-amplitudes using momentum-twistor variables (for which $\mathcal{A}_{n}^{(0),0}\!=\!1$), we find that to one-loop order,\\[-12pt]
\vspace{5pt}\eq{\mathcal{R}_{n}^{(k),1}=\mathcal{A}_n^{(k),1}-\mathcal{A}_n^{(k),0}\mathcal{A}_{n}^{(0),1}.\vspace{0pt}}
And using the explicit representation of one-loop amplitudes, (\ref{one_loop_divergent_finite_split}), we find that,
\vspace{7.5pt}\eq{\boxed{\mathcal{R}_{n}^{(0),1}=\mathcal{A}_{n,\mathrm{fin}}^{(k),1}-\mathcal{A}_{n}^{(k),0}\mathcal{A}_{n,\mathrm{fin}}^{(0),1}.}\label{manifestly_finite_one_loop_ratio}\vspace{5pt}}
Importantly, the representation (\ref{manifestly_finite_one_loop_ratio}) for the ratio function is given in terms of {manifestly} finite integrals---requiring no regularization whatsoever to be evaluated. We will see in \mbox{section \ref{two_loop_ratio_function_finiteness_section}} that the same magic {\it nearly} happens for two-loop ratio functions. (And we strongly expect something like this to persist to all loop-orders.) There is a good reason why we expect this to be possible: the integrand for the ratio function vanishes in all kinematical regions responsible for IR divergences; and therefore, we expect we can use only manifestly IR finite integrals to match it. This goal is not achieved here---as the merger of finite one-loop integrands (defined below) is not guaranteed to be finite; but what we describe below is a step in the right direction. 

\newpage

\section{\mbox{Local Integrand Representations of Two-Loop Amplitudes}}\label{two_loop_amplitude_section}
Although our basic strategy for determining two-loop amplitude integrands is the same as that for one-loop amplitudes, there are several important ways in which the details will differ. Recall from \mbox{section \ref{one_loop_integrand_review}} that one-loop amplitude integrands could be constructed by simply listing {\it all} of the maximal co-dimension (four) residues of an amplitude, and decorating each of these with an integrand engineered to match field theory on that cut (and none others); and recall that in order to construct integrands matching precisely one physical cut at a time required the introduction of an artificial propagator $\x{\ell}{X}$, supporting unphysical cuts term-by-term. One difference at two loops will be that we need not introduce any artificial propagators. 

Another important distinction is that for two-loop amplitudes, listing (and matching) every co-dimension eight residue of the amplitude individually turns out to be both {\it unnecessary} {and} {\it insufficient}. Let us first describe why it is not necessary to match every `octa-cut' individually. Consider the only leading singularities involving only six propagators:\\[-12pt]
\vspace{-2pt}\eq{\hspace{-15pt}\phantom{.}\raisebox{-34.65pt}{\includegraphics[scale=1]{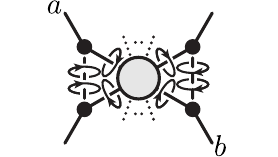}}\hspace{-15pt}.\vspace{-2.5pt}\label{double_composite_cut_figure}}
(From \mbox{section \ref{amplitude_cuts_section}}, we know that the corresponding residue is simply the tree-amplitude.) Now, there is a unique, dual-conformally invariant two-loop integrand (involving no other external propagators) with unit-residue on this cut:
\vspace{-5pt}\eq{\hspace{-30pt}\raisebox{-34.65pt}{\includegraphics[scale=1]{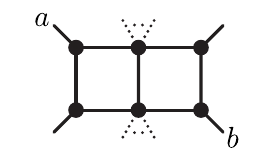}}\hspace{-20pt}\equiv\frac{\x{a\mi1}{a\pl1}\x{a}{b}\x{b\mi1}{b\pl1}}{\x{\ell_1}{a\mi1}\x{\ell_1}{a}\x{\ell_1}{a\pl1}\x{\ell_1}{\ell_2}\x{\ell_2}{b\mi1}\x{\ell_2}{b}\x{\ell_2}{b\pl1}}.\hspace{-0pt}\vspace{-7.5pt}\label{double_composite_cut_int}}
And so, decorating the on-shell diagram (\ref{double_composite_cut_figure}) with the integrand (\ref{double_composite_cut_int}) will guarantee that we match field theory on this particular, `doubly-composite' octa-cut. 

However, the integrand (\ref{double_composite_cut_int}) has other physical cuts: e.g.\ the singly-composite,\\[-12pt]
\vspace{-5pt}\eq{\hspace{-15pt}\raisebox{-34.65pt}{\includegraphics[scale=1]{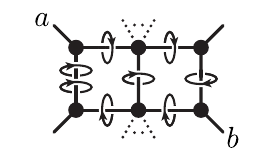}}\hspace{-15pt}.\vspace{-5pt}\label{double_composite_cut_int_other_cut}}
Conveniently, once we have matched field theory on the cut (\ref{double_composite_cut_figure}) using the integrand (\ref{double_composite_cut_int}), we automatically contribute to every octa-cut of the form  (\ref{double_composite_cut_int_other_cut}),
\vspace{-4pt}\eq{\hspace{-15pt}\phantom{.}\raisebox{-34.65pt}{\includegraphics[scale=1]{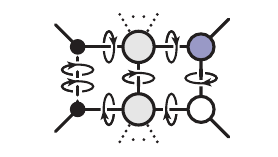}}\;\mathrm{and}\;\raisebox{-34.65pt}{\includegraphics[scale=1]{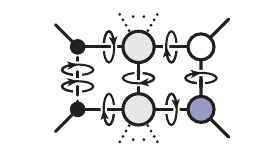}}\hspace{-15pt}.\vspace{-7.5pt}\label{first_skipped_two_loop_octacuts}}
(Although these octa-cuts are not guaranteed to match field theory until we also include terms such as (\ref{second_term_group}), which also have cuts of the form (\ref{double_composite_cut_int_other_cut}), it is clear that the cuts (\ref{first_skipped_two_loop_octacuts}) do not represent {independent} on-shell data.)

The other important way in which generalized unitarity at two-loop differs from one-loop is that the full integrand cannot be fixed by matching its co-dimension eight residues alone. Perhaps the easiest way to see this is through an explicit example. As mentioned in the introduction (see equation (\ref{elliptic_double_box_eg})), the full two-loop, 10-particle N$^3$MHV amplitude becomes extremely simple for the following component, \cite{CaronHuot:2012ab}:\\[-12pt]
\vspace{-2.5pt}\eq{\hspace{-25pt}\int\!\!\!\big(\!d\eta_1^1d\eta_2^1d\eta_3^1\hspace{-1pt}\big)\!\big(\!d\eta_4^2d\eta_5^2d\eta_6^2\hspace{-1pt}\big)\!\big(\!d\eta_6^3d\eta_7^3d\eta_8^3\hspace{-1pt}\big)\!\big(\!d\eta_9^4d\eta_{10}^4d\eta_1^4\hspace{-1pt}\big)\mathcal{A}_{10}^{(3),2}\!=\hspace{-5pt}\raisebox{-47.5pt}{\includegraphics[scale=1]{elliptic_double_box_example}}\hspace{-5pt}.\vspace{-2.5pt}\label{particular_double_box_helicity_component}}
(The notation used here is explained in \mbox{appendix \ref{overview_of_momentum_twistors_section}}.) This particular amplitude integrand has no co-dimension eight residues at all! To see this, consider the co-dimension seven residue which cuts all seven propagators:
\vspace{-2.5pt}\eq{\hspace{-20pt}\raisebox{-47.5pt}{\includegraphics[scale=1]{elliptic_double_box_example}}\raisebox{-2.75pt}{\scalebox{2}{$\Rightarrow$}}\raisebox{-47.5pt}{\includegraphics[scale=1]{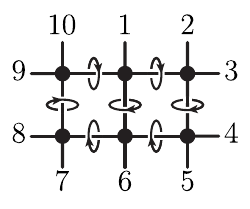}}\raisebox{-1.2pt}{\scalebox{1.4}{$\propto$}}\int\!\!\frac{dz}{\sqrt{Q(z)}}\,.\vspace{-7.5pt}\label{elliptic_heptacut_detail}}
Here, we have used $z$ to denote the remaining degree of freedom from the original eight $\{\ell_1,\ell_2\}$ on the solution to the hepta-cut equations, and $Q(z)$ is an irreducible quartic polynomial in $z$ (arising via the Jacobian for the hepta-cut). The important thing to notice is that the so-called `elliptic' differential form $dz/\sqrt{Q(z)}$ has no poles, and thus no isolated points in $z$-space where we can take a final residue. The existence of even a single component amplitude without co-dimension eight residues immediately demonstrates that two-loop integrands cannot be determined in general by their leading singularities (co-dimension $4l$ cuts) alone. 

But generalized unitarity allows us to give meaning to on-shell functions corresponding to sub-leading singularities just as easily as leading ones. For example, the following (non-isolated) on-shell function is perfectly well defined via (\ref{general_on_shell_function_definition}):
\vspace{-5pt}\eq{\fwboxR{0pt}{f_7(z)\equiv}\raisebox{-47.5pt}{\includegraphics[scale=1]{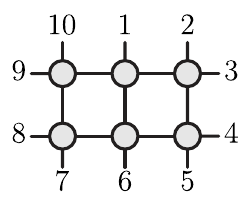}}.\vspace{-5pt}\label{massive_double_box_on_shell_diagram}}
From the discussion at the beginning of \mbox{section \ref{amplitude_cuts_section}}, it is easy to see that the on-shell function $f_7(z)$, (\ref{massive_double_box_on_shell_diagram}), involves an integral over one unfixed, internal degree of freedom---suggestively denoted `$z$'. This remaining degree of freedom clearly corresponds to the remaining loop-momentum parameter of the hepta-cut (\ref{elliptic_heptacut_detail}). (Indeed, for the component in (\ref{particular_double_box_helicity_component}), the on-shell function (\ref{massive_double_box_on_shell_diagram}) takes precisely the form (\ref{elliptic_heptacut_detail}).) 

Because we can easily compute and evaluate non-isolated on-shell functions such as (\ref{massive_double_box_on_shell_diagram}), even without any preferred points at which to match field theory, we can choose instead to match field theory at any arbitrarily chosen point $z^*$ along the hepta-cut. That is, if we choose an arbitrary reference point $z^*$ at which to to evaluate $f_7(z)$, we can match field theory {\it at this point} in loop-momentum space by decorating the on-shell function $f_7(z^*)$ with a scalar double-box integral, normalized to {\it evaluate to} the identity at the point $z^*$ along its hepta-cut. 

(We should mention here that there are other ways of eliminating the extra degree of freedom $z$ from the on-shell data (\ref{massive_double_box_on_shell_diagram}). For example, $z$ could be eliminated by integrating $f_7(z)$ over some contour, see e.g.\ \cite{Sogaard:2014jla}; however, this would both spoil our ability to match field theory at the integrand-level, and also introduce elliptic or logarithmic coefficients into the integral expansion.)

Although choosing to match field theory at arbitrarily chosen points $z^*$ along each double-box hepta-cut of the form (\ref{massive_double_box_on_shell_diagram}) would lead us to a correct representation of two-loop amplitudes, it turns out to be sufficient (and quite advantageous) to match field theory at co-dimension two points along the following hexa-cuts instead:
\vspace{-20pt}\eq{\hspace{-200pt}\hspace{-20pt}\raisebox{-57pt}{\includegraphics[scale=1]{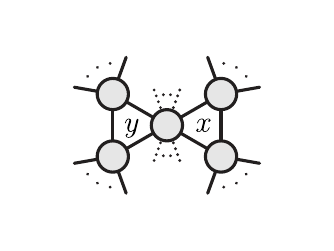}}\hspace{-20pt}.\hspace{-200pt}\vspace{-25pt}}
This is because the hexa-cut integration measure is always {rational}, while (as we already saw in (\ref{elliptic_heptacut_detail})), the hepta-cut measure is generally algebraic. This will allow us to considerably simplify our formulae (and analysis) below. And because the hexa-cut of any two-loop amplitude will be a rational differential form, {\it it} is fixed by its poles; so matching the hexa-cut will in fact ensure that we have correctly represented field theory on all its hepta-cuts.

\newpage
\subsection{Matching a Minimal (and Complete) Collection of Two-Loop Cuts}\label{on_shell_data_at_two_loops_subsection}
As we have seen above, matching field theory on certain cuts will guarantee that we match field theory on many others via residue theorems. We can therefore reconstruct the full two-loop amplitude integrand by ensuring that every cut is either matched explicitly or via a residue theorem. This can be done constructively, for example, by starting with a list of the most highly composite residues and successively adding further cuts not connected to those already on the list via residue theorems. A minimal list of independent two-loop cuts from which all other cuts will be fixed by residue theorems was given in (\ref{two_loop_on_shell_data_list_intro}) of the introduction. In this section, we describe how this list was constructed, and clarify the meaning of each of these cuts. 

We can match all of the most highly composite octa-cuts (those involving the fewest propagators), by decorating each of the on-shell diagrams (\ref{double_composite_cut_figure}) according to:\\[-12pt]
\vspace{-0pt}\eq{\fwboxL{0pt}{\hspace{-95pt}(1.a)}\phantom{.}\left\{\hspace{-20pt}\raisebox{-34.65pt}{\includegraphics[scale=1]{double_composite_octa_cut}}\hspace{-20pt}\raisebox{-2.pt}{\scalebox{1.75}{$\times$}}\hspace{-20pt}\raisebox{-34.65pt}{\includegraphics[scale=1]{first_double_box_int_picture}}\hspace{-15pt}\right\}.\vspace{-5pt}\label{first_term_group}}
These terms will automatically contribute to many other cuts (that we will ultimately match via residues theorems). One class of cuts that are completely independent, would be the single-composite double-boxes that involve two massive corners on the non-composite part of the contour:
\vspace{-5pt}\eq{\phantom{\hspace{-5pt}.}\raisebox{-47.25pt}{\includegraphics[scale=1]{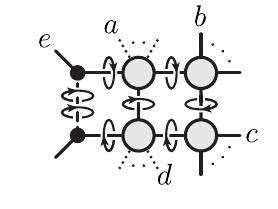}}\hspace{-5pt}.\vspace{-15pt}\label{term_two_figure_cut}}
(There are two octa-cuts with this topology because there are two solutions to the quad-cut equations for the non-composite box. The figure above represents the (parity-even) contour enclosing both cuts. As an on-shell function, (\ref{term_two_figure_cut}) corresponds to the sum of one-loop leading singularities from the box on the right: $\sum_{i=1}^2 f^i_{a,b,c,d}$.)

To be clear, figure (\ref{term_two_figure_cut}) represents {only} those double-box cuts involving massive corners $\{\ldots,\{b,\ldots,C\},\{c,\ldots,D\},\ldots\}$ with $b\!<\!C\!\equiv\!c\,\mi\,1$ and $c\!<\!D\!\equiv\!d\,\mi\,1$. The requirement that these corners involve at least two legs each is indicated in the figure (\ref{term_two_figure_cut}) according to the following convention: ranges of external legs involved at a corner will always be indicated by one of the following:
\vspace{5pt}\eq{\left\{\raisebox{-22.175pt}{\includegraphics[scale=1]{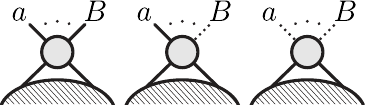}}\fwboxL{0pt}{\hspace{-60pt}\text{,}}\fwboxL{0pt}{\hspace{-120pt}\text{,}}\right\}.\label{leg_ranges_conventions_figure}\vspace{-10pt}}
The convention is that a range of legs bounded by solid-lines must include {\it at least two} legs (a `massive' corner); a range bounded by one solid- and one dashed-line involves {\it at least one} leg; and a range of legs bounded by dashed lines can be empty.

As with the double-composite leading singularities in (\ref{first_term_group}), the natural double-box integrand supporting the leading singularity (\ref{term_two_figure_cut}), 
\vspace{-2.5pt}\eq{\phantom{,\hspace{-10pt}}\raisebox{-47.25pt}{\includegraphics[scale=1]{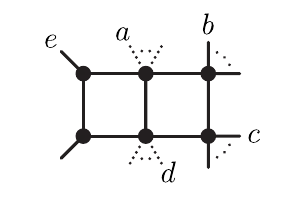}}\hspace{-10pt},\vspace{-12.5pt}}
is uniquely normalized by the criterion that it have unit residue on the cut (\ref{term_two_figure_cut}). And as with the contributions (\ref{first_term_group}), including the terms,
\vspace{-0pt}\eq{\fwboxL{0pt}{\hspace{-83.5pt}(1.b)}\phantom{.}\left\{\rule[-10pt]{0pt}{55pt}\right.\hspace{-20pt}\raisebox{-47.25pt}{\includegraphics[scale=1]{two_loop_term_2_cut}}\hspace{-10pt}\raisebox{-2.pt}{\scalebox{1.75}{$\times$}}\hspace{-20pt}\raisebox{-47.25pt}{\includegraphics[scale=1]{two_loop_term_2_int}}\hspace{-15pt}\left.\rule[-10pt]{0pt}{55pt}\right\},\vspace{-10pt}\label{second_term_group}}
will automatically ensure that many other physical cuts match via residue theorems.

There is one final class of composite leading singularities of amplitudes associated with soft-collinear divergences (see \mbox{section \ref{one_loop_IR_divergences_subsection}}): 
\vspace{-2.5pt}\eq{\phantom{\hspace{0pt}.}\raisebox{-47.25pt}{\includegraphics[scale=1]{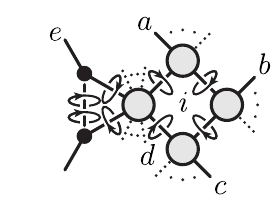}}\hspace{-0pt}.\vspace{-5pt}\label{term_three_figure_cut}}
As before, there are two octa-cuts with the topology above. It is not hard to see that these correspond to the one-loop on-shell functions $f^i_{a,b,c,d}$ of equation (\ref{one_loop_quad_cut_figure}); and matching this cut completely fixes the normalization of the penta-box integrand engineered to match this cut:
\vspace{-0pt}\eq{\fwboxL{0pt}{\hspace{-76pt}(1.c)}\phantom{.}\left\{\rule[-10pt]{0pt}{55pt}\right.\hspace{-25pt}\raisebox{-47.25pt}{\includegraphics[scale=1]{two_loop_term_3_cut}}\hspace{-0pt}\raisebox{-2.pt}{\scalebox{1.75}{$\times$}}\hspace{-20pt}\raisebox{-47.25pt}{\includegraphics[scale=1]{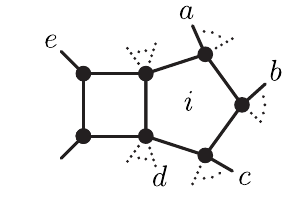}}\hspace{-5pt}\left.\rule[-10pt]{0pt}{55pt}\right\}.\vspace{-5pt}\label{third_term_group}}

Explicitly matching the cuts (\ref{double_composite_cut_figure}), (\ref{term_two_figure_cut}), and (\ref{term_three_figure_cut}) by decorating them with individually tailored integrands according to (\ref{first_term_group}), (\ref{second_term_group}), and (\ref{third_term_group}) will ensure that we match a large number of other physical cuts. In particular, residue theorems will ensure that we match field theory on cuts such as (\ref{first_skipped_two_loop_octacuts}), or any octa-cut involving a parity-even box with a single massless corner---any cut associated with an infrared divergence of the amplitude. But there remain three classes of linearly-independent on-shell data which are not not connected to those already described. 

The first class of on-shell data not related (via residue theorems) to the three terms already described are the double-boxes involving four massive corners (those involving at least two external legs): $\{a,\ldots,B\},\{b,\ldots,C\},\{d,\ldots,E\},$ and $\{e,\ldots,F\}$,\\[-12pt]
\vspace{-2.5pt}\eq{\raisebox{-47.25pt}{\includegraphics[scale=1]{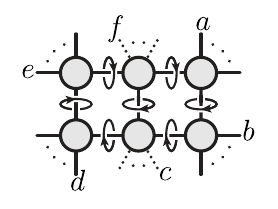}}.\vspace{-5pt}\label{massive_double_box_hepta_cut_diagram}}
When the ranges of legs $\{f,\ldots,A\}$ and $\{c,\ldots,D\}$ are both empty,\footnote{There is also an octa-cut defined when only one of the two ranges is empty; but the residue of this cut in field theory is always zero due to the non-existence of a corresponding on-shell diagram.} then there exists a composite, co-dimension eight residue corresponding to,
\vspace{2.5pt}\eq{\phantom{;}\raisebox{-35.25pt}{\includegraphics[scale=1]{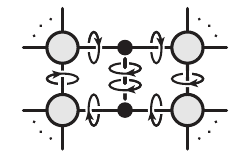}};\label{composite_ladder_double_box}\vspace{2.5pt}}
but the general hepta-cut (\ref{massive_double_box_hepta_cut_diagram}) will not have any co-dimension eight residues. Nevertheless, the integrand would be fixed by matching any point along this hepta-cut. 

Although we could include the hepta-cuts (\ref{massive_double_box_hepta_cut_diagram}) (evaluated at some arbitrary point) among our list of on-shell data, it turns out to be (algebraically) much simpler to choose instead to match field theory somewhere along its co-dimension six cut:
\vspace{-2.5pt}\eq{\phantom{.}\raisebox{-47.25pt}{\includegraphics[scale=1]{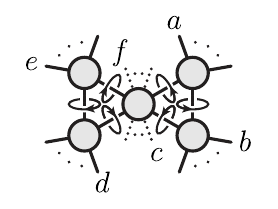}}\hspace{-0pt}.\label{double_triangle_hexa_cut}\vspace{-5pt}}
We can match (this point along) the loop-amplitude's hexa-cut by decorating the on-shell function (\ref{double_triangle_hexa_cut}) (evaluated at some point $(x^*\hspace{-1pt},y^*\hspace{-1pt})$) with a double-box integrand,
\vspace{-0pt}\eq{\fwboxL{0pt}{\hspace{-82.5pt}(2.a)}\phantom{.}\left\{\rule[-10pt]{0pt}{55pt}\right.\hspace{-20pt}\raisebox{-47.25pt}{\includegraphics[scale=1]{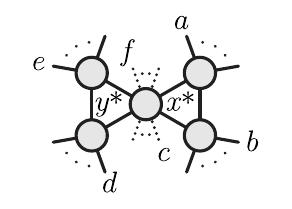}}\hspace{-15pt}\raisebox{-2.pt}{\scalebox{1.75}{$\times$}}\hspace{-15pt}\raisebox{-47.25pt}{\includegraphics[scale=1]{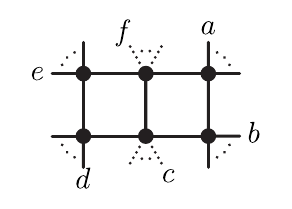}}\hspace{-20pt}\left.\rule[-10pt]{0pt}{55pt}\right\},\vspace{-5pt}\label{fourth_term_group}}
where the double-box integrand is tailored to {\it evaluate to} 1 at the specific reference point $(x^*\hspace{-1pt},y^*\hspace{-1pt})$ along its hexa-cut (\ref{double_triangle_hexa_cut}). We give a concrete formula for the normalization of the double-box integrand which meets this criterion in \mbox{appendix \ref{momentum_twistor_reps_of_integrands}}.

To be clear, the point along the hexa-cut at which we choose to match field theory could be any point along the hepta-cut (\ref{massive_double_box_hepta_cut_diagram}), or even at the location of a (composite) octa-cut such as (\ref{composite_ladder_double_box}) (when such cuts exist). But it turns out to be more convenient to systematically choose the point $(x^*\hspace{-1pt},y^*\hspace{-1pt})$ to be given by equation (\ref{choice_of_spurious_point_for_double_triangle}) as described in \mbox{appendix \ref{two_loop_on_shell_functions_section}}. There are at least two reasons for preferring the choice we make for these points. First, our choice will ensure that the double-box integrand will be dual conformally invariant. Secondly, it will ensure that both the on-shell functions and their associated integrands will be rational term-by-term (this makes the representation both more elegant and easier to evaluate numerically). 

There are two more classes of on-shell data not related to those already matched through residue theorems: {\it massive} penta-boxes, and kissing-boxes. Because penta-box cuts involve fewer external propagators than kissing-boxes, it is natural to match them first. As already mentioned above, any penta-box residue involving one or two massless corners on the side of the box have already been matched---either directly, or through residue theorems. Therefore, among the first set of independent penta-box cuts that remain to be fixed are those of the form:
\vspace{2.5pt}\eq{\hspace{-40pt}\raisebox{-47.25pt}{\includegraphics[scale=1]{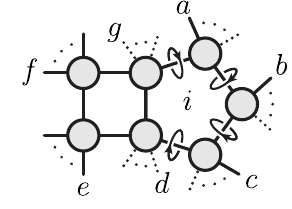}}\hspace{-5pt}\raisebox{-2.pt}{\scalebox{1.75}{$\equiv$}}\hspace{-12pt}\raisebox{-47.25pt}{\includegraphics[scale=1]{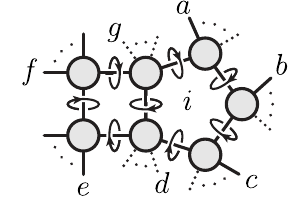}}\hspace{-5pt}\raisebox{-2.pt}{\scalebox{1.75}{$\cup$}}\hspace{-15pt}\raisebox{-47.25pt}{\includegraphics[scale=1]{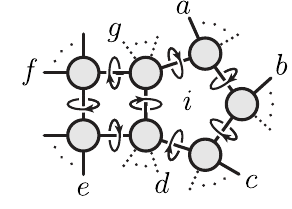}}\!\!\!.\label{massive_pentabox_cut}\vspace{-2.5pt}}
Here, observe that these cuts necessarily involve massive corners on the side of the box, $\{e,\ldots,F\}$ and $\{f,\ldots,G\}$; and the contour above encloses two distinct octa-cuts---the parity-even sum of octa-cuts involving a fixed solution to the cut on the side of the pentagon. (The parity-odd combination of penta-box cuts will be matched through residue theorems once we have included the kissing-boxes below.)

We can match field theory on all cuts of the form (\ref{massive_pentabox_cut}) by decorating each of these on-shell functions by integrands as follows:
\vspace{2.5pt}\eq{\fwboxL{0pt}{\hspace{-65.5pt}(2.b)}\phantom{.}\left\{\rule[-10pt]{0pt}{55pt}\right.\hspace{-15pt}\raisebox{-47.25pt}{\includegraphics[scale=1]{two_loop_term_5_cut_v0}}\hspace{-0pt}\raisebox{-2.pt}{\scalebox{1.75}{$\times$}}\hspace{-10pt}\raisebox{-47.25pt}{\includegraphics[scale=1]{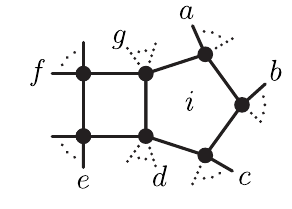}}\hspace{-10pt}\left.\rule[-10pt]{0pt}{55pt}\right\}.\vspace{-7.5pt}\label{fifth_term_group}}
Here, there are two criteria that determine the form of the integrands appearing above. The first and most obvious criterion is that these integrands have unit residue on the corresponding penta-box contour (\ref{massive_pentabox_cut}) and vanish on the other penta-box contour of this type. But it is easy to see that this alone cannot uniquely fix the form of the integrand: adding double-box integrands of the form in (\ref{fourth_term_group}) will not affect its residue on the penta-box cut. The final criterion that will fix the integrand is that it vanishes at the point  $(x^*\hspace{-1pt},y^*\hspace{-1pt})$ along any of its hexa-cuts of the form (\ref{double_triangle_hexa_cut}). (Notice that penta-boxes can support more than one hexa-cut of the form (\ref{double_triangle_hexa_cut}).) This is explained in more detail in \mbox{appendix \ref{momentum_twistor_reps_of_integrands}}.

The final class of independent on-shell data required to fully determine any two-loop amplitude integrands are the so-called `kissing-boxes' octa-cuts. Not surprisingly, we will decorate each of these with an integrand tailored to match the corresponding cut (and none others with this topology):
\vspace{2.5pt}\eq{\fwboxL{0pt}{\hspace{-57.5pt}(2.c)}\phantom{.}\left\{\rule[-10pt]{0pt}{55pt}\right.\hspace{-5pt}\raisebox{-47.25pt}{\includegraphics[scale=1]{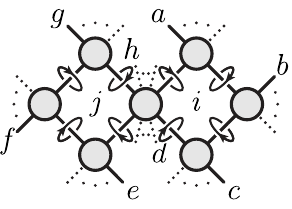}}\hspace{-0pt}\raisebox{-2.pt}{\scalebox{1.75}{$\times$}}\hspace{-5pt}\raisebox{-47.25pt}{\includegraphics[scale=1]{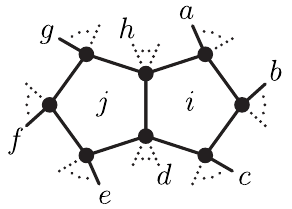}}\hspace{-10pt}\left.\rule[-10pt]{0pt}{55pt}\right\}.\vspace{-5pt}\label{sixth_term_group}}
As with the penta-box integrals, the criterion that they have unit residue on a single kissing-boxes octa-cut (and vanish on all the others) is not sufficient to fix the form of the double-pentagon integrands. To completely determine the form of the integrands needed to decorate each kissing-boxes on-shell diagram, we must also require that they do not contribute to any of the other cuts already matched---that they vanish on all penta-box contours of the form (\ref{massive_pentabox_cut}), and that the they vanish at the chosen reference point $(x^*\hspace{-1pt},y^*\hspace{-1pt})$ of any hexa-cut of the form (\ref{double_triangle_hexa_cut}). Explicit forms of the double-pentagon integrals which satisfy these criteria are given in \mbox{appendix \ref{momentum_twistor_reps_of_integrands}}.

This completes our list of on-shell data required to match field theory everywhere as a function of the loop momenta. Notice that each class of on-shell data in (\ref{two_loop_on_shell_data_list_intro}) is decorated with an integrand that is either {\it manifestly divergent}, or {\it manifestly finite}. 

\newpage
\subsection{\mbox{Local Integrand-Level Representations of All Two-Loop  Amplitudes}}\label{local_two_loop_amplitude_formula_section}\vspace{0pt}

Putting everything together from the discussion in the previous subsection, we find:\\[-12pt]
\vspace{0pt}\eq{\mathcal{A}_{n}^{(k),2}\equiv\mathcal{A}_{n,\mathrm{div}}^{(k),2}+\mathcal{A}_{n,\mathrm{fin}}^{(k),2},\vspace{-10pt}\label{two_loop_integrand_expression}}
with
\vspace{-5pt}\eq{\hspace{-540pt}\begin{array}{@{}l@{}c@{}c@{}r@{}}\displaystyle\mathcal{A}_{n,\mathrm{div}}^{(k),2}
\equiv\;&\hspace{-10pt}&\fwbox{0pt}{\,\,\,\,\,\,\left\{\rule[-10pt]{0pt}{48pt}\right.}\fwboxL{55pt}{\!\!\!\!\begin{array}{@{}c@{}}\phantom{a,b,c,d,e,f,g}\\\raisebox{-3.75pt}{\scalebox{2.25}{$\displaystyle\sum$}}\\[-2.5pt]a,\!b\end{array}}\hspace{-27pt}\raisebox{-47.25pt}{\includegraphics[scale=1]{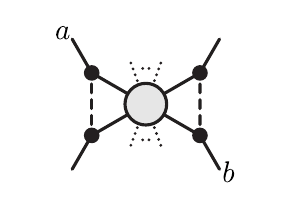}}\hspace{-20pt}\fwbox{0pt}{\times}\hspace{-14pt}\raisebox{-47.25pt}{\includegraphics[scale=1]{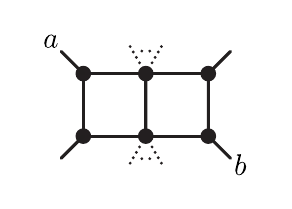}}\hspace{20pt}&\fwboxL{10pt}{\text{$(1.a)$}}\\[-10pt]%
&+\hspace{-10pt}&\fwboxL{55pt}{\!\!\!\!\begin{array}{@{}c@{}}\phantom{a,b,c,d,e,f,g}\\\raisebox{-3.75pt}{\scalebox{2.25}{$\displaystyle\sum$}}\\[-2.5pt]a,\!b,\!c,\!d,\!e\end{array}}\hspace{-27pt}\raisebox{-47.25pt}{\includegraphics[scale=1]{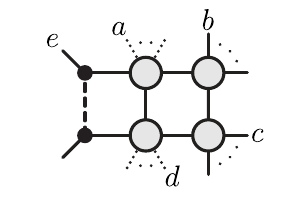}}\hspace{-5pt}\fwbox{0pt}{\times}\hspace{-5pt}\hspace{-14pt}\raisebox{-47.25pt}{\includegraphics[scale=1]{two_loop_term_2_int}}\hspace{10pt}&\fwboxL{10pt}{\text{$(1.b)$}}\\[-12pt]%
&+\hspace{-10pt}&\fwboxL{55pt}{\!\!\!\!\begin{array}{@{}c@{}}\phantom{a,b,c,d,e,f,g}\\\raisebox{-3.75pt}{\scalebox{2.25}{$\displaystyle\sum$}}\\[-2.5pt]a,\!b,\!c,\!d,\!e\end{array}}\hspace{-27pt}\hspace{-5pt}\raisebox{-47.25pt}{\includegraphics[scale=1]{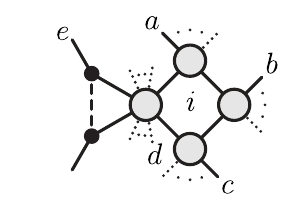}}\hspace{-0pt}\fwbox{0pt}{\times}\hspace{-5pt}\hspace{-14pt}\raisebox{-47.25pt}{\includegraphics[scale=1]{two_loop_term_3_int}}\hspace{10pt}\fwbox{0pt}{\hspace{-20pt}\left.\rule[-10pt]{0pt}{48pt}\right\}\!\!,}&\fwboxL{10pt}{\text{$(1.c)$}}\\[-0pt]%
\end{array}
\hspace{-500pt}\vspace{-10pt}\label{two_loop_integrand_formula_part_1}}
and
\vspace{-10pt}\eq{\hspace{-545pt}\begin{array}{@{}l@{}c@{}c@{}r@{}}\displaystyle \mathcal{A}_{n_{\mathrm{fin}}}^{(k),2}
\equiv\hspace{2.5pt}&\hspace{-10pt}&\fwbox{0pt}{\,\,\,\,\left\{\rule[-10pt]{0pt}{48pt}\right.\,\,\,\,}\fwboxL{55pt}{\!\!\!\!\begin{array}{@{}c@{}}\phantom{a,b,c,d,e,f,g}\\\raisebox{-3.75pt}{\scalebox{2.25}{$\displaystyle\sum$}}\\[-2.5pt]a,\!b,\!c,\!d,\!e,\!f\end{array}}\hspace{-8pt}\raisebox{-47.25pt}{\includegraphics[scale=1]{two_loop_term_4}}\fwbox{0pt}{\times}\hspace{-14pt}\raisebox{-47.25pt}{\includegraphics[scale=1]{two_loop_term_4_int}}\hspace{5pt}&\fwboxL{10pt}{\text{$(2.a)$}}\\[-10pt]%
&+\hspace{-10pt}&\fwboxL{55pt}{\!\!\!\!\begin{array}{@{}c@{}}\phantom{a,b,c,d,e,f,g}\\\raisebox{-3.75pt}{\scalebox{2.25}{$\displaystyle\sum$}}\\[-2.5pt]a,\!b,\!c,\!d,\!e,\!f,\!g\end{array}}\hspace{-8pt}\raisebox{-47.25pt}{\includegraphics[scale=1]{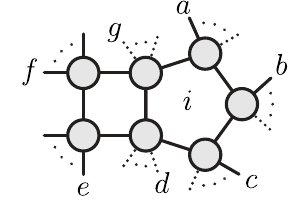}}\fwbox{0pt}{\times}\hspace{-14pt}\raisebox{-47.25pt}{\includegraphics[scale=1]{two_loop_term_5_int}}\hspace{5pt}&\fwboxL{10pt}{\text{$(2.b)$}}\\[-5pt]%
&+\hspace{-10pt}&\fwboxL{55pt}{\!\!\!\!\begin{array}{@{}c@{}}\phantom{a,b,c,d,e,f,g}\\\raisebox{-3.75pt}{\scalebox{2.25}{$\displaystyle\sum$}}\\[-3.5pt]\phantom{,}a,\!b,\!c,\!d,\\[-5pt]e,\!f,\!g,\!h\end{array}}\hspace{-8pt}\raisebox{-47.25pt}{\includegraphics[scale=1]{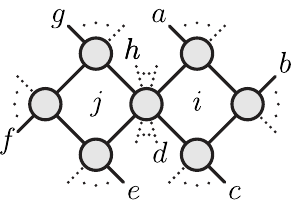}}\fwbox{0pt}{\times}\hspace{-4pt}\raisebox{-47.25pt}{\includegraphics[scale=1]{two_loop_term_6_int}}\hspace{7.5pt}\fwbox{0pt}{\hspace{-15.5pt}\left.\rule[-10pt]{0pt}{48pt}\right\}\!\!.}&\fwboxL{10pt}{\text{$(2.c)$}}\\[-6pt]%
\end{array}
\hspace{-500pt}\vspace{-20pt}\label{two_loop_integrand_formula_part_2}}

\newpage
\subsection{Making More Manifest the Finiteness of Infrared-Safe Observables}\label{two_loop_ratio_function_finiteness_section}\vspace{-8pt}
We would like to show that the representation of two-loop amplitudes given above, (\ref{two_loop_integrand_expression}), makes more (but not completely) manifest the finiteness of infrared-safe observables such as the ratio function. Recall from \mbox{section \ref{one_loop_IR_divergences_subsection}} that the ratio function can be defined order-by-order according to (\ref{general_ratio_definition}); to two-loop order, we have:
\vspace{5pt}\eq{\mathcal{R}_n^{(k),2}\equiv\mathcal{A}_{n}^{(k),2}-\mathcal{A}_{n}^{(k),1}\mathcal{A}_{n}^{(0),1}-\mathcal{A}_{n}^{(k),0}\Big(\mathcal{A}_{n}^{(0),2}-\mathcal{A}_{n}^{(0),1}\mathcal{A}_{n}^{(0),1}\Big).\vspace{0pt}\label{two_loop_ratio_function_general}}
Here, the multiplication of amplitudes could be performed at the integrand-level by using different labels for the loop momenta (and different reference points $X$); but it is not immediately obvious how the divergences of the two-loop amplitude integrands in equation (\ref{two_loop_integrand_formula_part_1}) are expected to cancel against those of (\ref{two_loop_ratio_function_general})---at least if we use the $X$-dependent representations of one-loop amplitudes of equation (\ref{local_one_loop_amplitude_expression}). 

This will be made clear in two steps. First, we will define an operation that merges $X$-dependent one-loop integrands to produce an $X$-{independent} two-loop integrand,\\[-12pt]
\vspace{0pt}\eq{\mathcal{I}_L(\ell_1,X)\merge\mathcal{I}_{R}(\ell_2,X)\mapsto\mathcal{I}_{L\otimes R}(\ell_1,\ell_2),\vspace{0pt}\label{merge_operation_sketch}} 
and show that merging is equal to multiplication for amplitudes represented by (\ref{local_one_loop_amplitude_expression}),\\[-12pt]
\vspace{0pt}\eq{\big(\mathcal{A}_n^{(k_L),1}\big)\!\!\times\!\!\big(\mathcal{A}_{n}^{(k_R),1}\big)=\mathcal{A}_{n}^{(k_L),1}\merge\mathcal{A}_{n}^{(k_R),1},\vspace{0pt}\label{amplitude_multiplication_equals_merger}}
allowing us to rewrite the products of amplitudes in (\ref{two_loop_ratio_function_general}) in terms of $X$-independent integrands according to, 
\vspace{5pt}\eq{\mathcal{R}_n^{(k),2}=\mathcal{A}_{n}^{(k),2}-\mathcal{A}_{n}^{(k),1}\merge\mathcal{A}_{n}^{(0),1}-\mathcal{A}_{n}^{(k),0}\Big(\mathcal{A}_{n}^{(0),2}-\mathcal{A}_{n}^{(0),1}\merge\mathcal{A}_{n}^{(0),1}\Big).\vspace{0pt}\label{two_loop_ratio_function_general_merger}}
Secondly, we will show that the divergent contributions to two-loop amplitudes in (\ref{two_loop_integrand_formula_part_1}) can be re-written in the very suggestive form,
\vspace{0pt}\eq{\mathcal{A}_{n,\mathrm{div}}^{(k),2}=\mathcal{A}_{n}^{(k),0}\Big(\mathcal{I}_{\mathrm{div}}\merge\mathcal{I}_{\mathrm{div}}\Big)+\Big(\mathcal{A}_{n,\mathrm{fin}}^{(k),1}\merge\mathcal{I}_{\mathrm{div}}\Big).\vspace{0pt}\label{two_loop_amplitude_divergences_as_mergers}}
Using this, and expanding each one-loop amplitude in (\ref{two_loop_ratio_function_general_merger}) according to (\ref{one_loop_divergent_finite_split}), it is easy to see that all divergent contributions cancel, resulting in:\\[-8pt]
\eqs{\mathcal{R}_n^{(k),2}\!&=\mathcal{A}_{n,\mathrm{fin}}^{(k),2}-\mathcal{A}_{n,\mathrm{fin}}^{(k),1}\merge\mathcal{A}_{n,\mathrm{fin}}^{(0),1}-\mathcal{A}_{n}^{(k),0}\Big(\mathcal{A}_{n,\mathrm{fin}}^{(0),2}-\mathcal{A}_{n,\mathrm{fin}}^{(0),1}\merge\mathcal{A}_{n,\mathrm{fin}}^{(0),1}\Big),\\&=\mathcal{A}_{n,\mathrm{fin}}^{(k),2}-\mathcal{R}_{n}^{(k),1}\merge\mathcal{A}_{n,\mathrm{fin}}^{(0),1}-\mathcal{A}_{n}^{(k),0}\mathcal{A}_{n,\mathrm{fin}}^{(0),2}.}

Let us now describe the crucial {\it merge} operation mentioned above in (\ref{merge_operation_sketch}). The merger of two $X$-dependent integrands is defined according to:\\[-8pt]
\vspace{0pt}\eqs{\mathcal{I}_L(\ell_1,X)\merge\mathcal{I}_{R}(\ell_2,X)&\equiv\left(\mathcal{I}'_{L}(\ell_1)\frac{\x{Y_L(\ell_1)}{X}}{\x{\ell_1}{X}}\right)\merge\left(\frac{\x{X}{Y_R(\ell_2)}}{\x{X}{\ell_2}}\mathcal{I}'_{R}(\ell_2)\right),\\&\equiv\mathcal{I}'_L(\ell_1)\frac{\x{Y_L(\ell_1)}{Y_R(\ell_2)}}{\x{\ell_1}{\ell_2}}\mathcal{I}'_{R}(\ell_2).\vspace{0pt}\label{merge_operation_detail}} 
We will not dwell on the many ways this operation can be motivated; but notice that merging always preserves the physical quad-cuts of the integrands being merged.

\newpage
We must now show that the (term-by-term) merger of two full amplitudes represented according to (\ref{local_one_loop_amplitude_expression}) is equivalent to direct multiplication (see equation (\ref{amplitude_multiplication_equals_merger})). When expressed in $X$-dependent terms, we consider the $q^{\mathrm{th}}$ term to be of the form,\\[-10pt]
\eq{\mathcal{I}_q(\ell,X)\equiv\mathcal{I}'_q(\ell)\frac{\x{Y_q(\ell)}{X}}{\x{\ell}{X}}\equiv\x{\mathcal{Y}_q(\ell)}{X}/\x{\ell}{X},}
where $\mathcal{Y}_q(\ell)$ includes the denominator factors of $\mathcal{I}'_{q}(\ell)$. Then the representation (\ref{local_one_loop_amplitude_expression}) becomes,
\vspace{5pt}\eq{\mathcal{A}_{n}^{(k),1}=\sum_{q}\x{\mathcal{Y}_q(\ell)}{X}/\x{\ell}{X}\equiv\x{\mathcal{Y}}{X}/\x{\ell}{X}.}
By the $X$-independence of the representation (\ref{local_one_loop_amplitude_expression}) proven in \mbox{section \ref{boundaries_of_diagrams_and_residue_theorems_section}}, we know that the object $\mathcal{Y}\!\equiv\!\sum_q\mathcal{Y}_q(\ell)$, must factorize to be proportional to the full amplitude:\\[-12pt]
\eq{\mathcal{Y}\equiv(\ell)\times\mathcal{A}_{n}^{(k),1}.}

Thus, using the $\mathcal{Y}$'s for two different amplitudes, $\mathcal{A}_{n}^{(k_L),1}\!(\ell_1)$ and $\mathcal{A}_{n}^{(k_R),1}\!(\ell_2)$,
\vspace{5pt}\eq{\x{\mathcal{Y}^L}{\mathcal{Y}^R}\equiv\sum_{q_L,q_R}\x{\mathcal{Y}^L_{q_L}}{\mathcal{Y}^R_{q_R}}=\x{\ell_1}{\ell_2}\big(\mathcal{A}_{n}^{(k_L),1}\big)\!\!\times\!\!\big(\mathcal{A}_{n}^{(k_R),1}\big),}
from which we may conclude, as desired, that:
\vspace{5pt}\eq{\hspace{-30pt}\big(\mathcal{A}_{n}^{(k_L),1}\big)\!\!\times\!\!\big(\mathcal{A}_{n}^{(k_R),1}\big)=\sum_{q_L,q_R}\frac{\x{\mathcal{Y}^L_{q_L}}{\mathcal{Y}^R_{q_R}}}{\x{\ell_1}{\ell_2}}=\sum_{q_L,q_R}\big(\mathcal{I}^L_{q_L}\merge\mathcal{I}_{q_R}^R\big)\equiv\mathcal{A}_{n}^{(k_L),1}\merge\mathcal{A}_{n}^{(k_R),1}.}

The second step of our argument is to recognize that the infrared-divergent contributions to two-loop amplitudes, (\ref{two_loop_integrand_formula_part_1}), are naturally organized according to (\ref{two_loop_amplitude_divergences_as_mergers}). This is actually quite straight-forward. Consider the first divergent contributions---those labelled $(1.a)$ in (\ref{two_loop_integrand_formula_part_1}):
\vspace{5pt}\eq{\phantom{.}\left\{\hspace{-20pt}\raisebox{-34.65pt}{\includegraphics[scale=1]{double_composite_octa_cut}}\hspace{-20pt}\raisebox{-2.pt}{\scalebox{1.75}{$\times$}}\hspace{-20pt}\raisebox{-34.65pt}{\includegraphics[scale=1]{first_double_box_int_picture}}\hspace{-15pt}\right\}.\vspace{-0pt}\label{first_term_group_second_appearance}}
All of these octa-cut coefficients are simply equal to the tree-amplitude, $\mathcal{A}_{n}^{(k),0}$, and the integrands that we attach to these cuts are easily seen to be of the form,
\eq{\hspace{-15pt}\raisebox{-34.65pt}{\includegraphics[scale=1]{first_double_box_int_picture}}\hspace{-25pt}\equiv\frac{\x{a\mi1}{a\pl1}\x{a}{b}\x{b\mi1}{b\pl1}}{\x{\ell_1}{a\mi1}\x{\ell_1}{a}\x{\ell_1}{a\pl1}\x{\ell_1}{\ell_2}\x{\ell_2}{b\mi1}\x{\ell_2}{b}\x{\ell_2}{b\pl1}}\equiv\mathcal{I}^a\merge\mathcal{I}^b.\nonumber} 
(Here, the one-loop divergent triangle integrands $\mathcal{I}^a$ were defined in equation (\ref{divergent_triangle_integrand_definition}).) Because any potentially non-planar terms vanish---$\mathcal{I}^{a}\merge\mathcal{I}^{a}\!\!=\!\mathcal{I}^{a}\merge\mathcal{I}^{a+1}\!\!=\!0$---in the combination $\mathcal{I}_{\mathrm{div}}\merge\mathcal{I}_{\mathrm{div}}\!\equiv\!\!\big(\sum_{a}\mathcal{I}^a\big)\merge\big(\sum_{b}\mathcal{I}^b\big)$, the terms $(1.a)$ of (\ref{two_loop_integrand_formula_part_1}) combine to:\\[-0pt]
\vspace{-20pt}\eq{\hspace{-40pt}\fwboxL{55pt}{\!\!\!\!\begin{array}{@{}c@{}}\phantom{a,b,c,d,e,f,g}\\\raisebox{-3.75pt}{\scalebox{2.25}{$\displaystyle\sum$}}\\[-2.5pt]a,\!b\end{array}}\hspace{-27pt}\raisebox{-47.25pt}{\includegraphics[scale=1]{two_loop_term_1}}\hspace{-20pt}\fwbox{0pt}{\times}\hspace{-14pt}\raisebox{-47.25pt}{\includegraphics[scale=1]{two_loop_term_1_int}}\hspace{-20pt}=\,\,\mathcal{A}_{n}^{(k),0}\Big(\mathcal{I}_{\mathrm{div}}\merge\mathcal{I}_{\mathrm{div}}\Big).\vspace{-10pt}}

The last two types of divergent terms in (\ref{two_loop_integrand_formula_part_1}) combine into the second term of equation (\ref{two_loop_amplitude_divergences_as_mergers}). This follows similarly from the explicit form of the integrands of these two terms, and noticing that all the potentially non-planar terms generated in the expansion of $\mathcal{A}_{n,\mathrm{fin}}^{(k),1}\merge\mathcal{I}_{\mathrm{div}}$ vanish. We will save the reader the reproduction of this exercise here, but simply note that terms $(1.b)$ and $(1.c)$ of equation (\ref{two_loop_integrand_formula_part_1}) combine to take the form $\mathcal{A}_{n,\mathrm{fin}}^{(k),1}\merge\mathcal{I}_{\mathrm{div}}$. 

Therefore, the local representation of two-loop amplitudes (\ref{two_loop_integrand_expression}), takes the form:\\[-8pt]
\eq{\hspace{-30pt}\mathcal{A}_{n}^{(k),2}\equiv\mathcal{A}_{n,\mathrm{div}}^{(k),2}+\mathcal{A}_{n,\mathrm{fin}}^{(k),2}\;\;\mathrm{with}\;\;\mathcal{A}_{n,\mathrm{div}}^{(k),2}\equiv\mathcal{A}_{n}^{(k),0}\Big(\mathcal{I}_{\mathrm{div}}\merge\mathcal{I}_{\mathrm{div}}\Big)+\Big(\mathcal{A}_{n,\mathrm{fin}}^{(k),1}\merge\mathcal{I}_{\mathrm{div}}\Big).}
From this, it is a simple exercise of expansion to see that the ratio function becomes expressed in terms of contributions:
\vspace{5pt}\eq{\boxed{\mathcal{R}_n^{(k),2}=\mathcal{A}_{n,\mathrm{fin}}^{(k),2}-\mathcal{R}_{n}^{(k),1}\merge\mathcal{A}_{n,\mathrm{fin}}^{(0),1}-\mathcal{A}_{n}^{(k),0}\mathcal{A}_{n,\mathrm{fin}}^{(0),2}.}\label{manifestly_finite_two_loop_ratio}\vspace{2.5pt}}
We should make it clear that the integrands generated by expanding the terms in $\mathcal{R}_{n}^{(k),1}\merge\mathcal{A}_{n,\mathrm{fin}}^{(0),1}$ are simple two-loop integrands---either double-pentagons or penta-boxes; the only novelty is that not all of these will correspond to planar integrands.

We should clarify an important point---not observed in the original version of this work. {\it The merger of finite one-loop integrands is not necessarily finite.} Thus, while the representation described here renders the ratio function suggestively close to finite, the individual terms are not guaranteed to be (nor are they always) finite. This will be further elaborated in future work. 

Before we conclude this section, let us briefly speculate about how this observables may extend to higher loop-orders. On general grounds, we expect that the $l$-loop integrand can always be expressed in the form,
\vspace{0pt}\eq{\mathcal{A}_n^{(k),l}=\mathcal{A}_{n,\mathrm{div}}^{(k),l}+\mathcal{A}_{n,\mathrm{fin}}^{(k),l},\quad\mathrm{with}\quad\mathcal{A}_{n,\mathrm{div}}^{(k),l}\equiv\sum_{q=1}^{l}\mathcal{A}_{n,\mathrm{fin}}^{(k),l-q}\Big(\mathcal{I}_{\mathrm{div}}\Big)^{q};\vspace{-4pt}\label{all_loop_divergence_exponentiation}}
and using this, it is easy to show that the $l$-loop ratio function, (\ref{general_ratio_definition}), becomes,
\vspace{0pt}\eq{\mathcal{R}_n^{(k),l}=\mathcal{A}_{n,\mathrm{fin}}^{(k),l}-\sum_{q=1}^{l}\mathcal{R}_{n}^{(k),l-q}\mathcal{A}_{n,\mathrm{fin}}^{(0),q}.\vspace{-4.5pt}\label{all_loop_ratio_function_finiteness}}
(This form makes use of the fact that in momentum-twistor variables $\mathcal{A}_{n}^{(0),0}\!\!=\!1$, which justifies the (not uncommon) notational simplicity, $\mathcal{A}_{n}^{(k),0}\!\hspace{-1pt}\equiv\mathcal{R}_{n}^{(k),0}$.)

Although quite suggestive, it remains to be clarified how to interpret the products of $X$-dependent factors appearing in (\ref{all_loop_divergence_exponentiation}) and (\ref{all_loop_ratio_function_finiteness}) at the integrand-level in a way that eliminates any dependence on $X$. At two-loops, the merge-operation defined in equation (\ref{merge_operation_detail}) allowed us to see that the divergences in the amplitude manifestly cancel against the products of terms appearing in the ratio function. We do not yet have a generalization of the merger that ensures this will work to all loop-orders. 

\vspace{-20pt}\section{Conclusions and Future Directions}\label{conclusions_section}\vspace{-14pt}
In this paper, we have explicitly constructed a closed-form local integrand-level representation of all two-loop amplitudes in planar, maximally supersymmetric ($\mathcal{N}\!=\!4$) Yang-Mills theory (SYM). This representation was found not through the ordinary implementation of generalized unitarity, but rather by extending the approach described for reconstructing one-loop amplitude integrands in \mbox{ref.\ \cite{Bourjaily:2013mma}}. This representation explicitly matches a small number of specific cuts of the amplitude (sufficient to reproduce all other cuts via residue theorems) by attaching to each on-shell function an integrand individually-tailored to match the corresponding cut.

Importantly, the representation of two-loop integrands we have derived, (\ref{two_loop_integrand_expression}), separates contributions that are manifestly infrared divergent, (\ref{two_loop_integrand_formula_part_1}), from contributions that are manifestly infrared finite, (\ref{two_loop_integrand_formula_part_2}). And the infrared divergent contributions of amplitudes were organized in a way that makes manifest the exponentiation of infrared divergences of amplitudes. 

There has recently been considerable interest in the computation of two-loop ratio functions, with much progress being made without the use of integral representations (see e.g.\ \cite{Dixon:2011pw, Dixon:2013eka, Dixon:2014voa, Dixon:2014xca, Dixon:2014iba} and \cite{Basso:2013vsa,Basso:2013aha,Basso:2014koa,Basso:2014nra,Basso:2014hfa}). And there has been similar progress toward understanding finite parts of MHV amplitudes without integration (see e.g.\ \cite{Golden:2013lha,Golden:2013xva,Golden:2014pua,Golden:2014xqa}). But there has also been considerable progress toward evaluating the loop integrals analytically or numerically, where the manifestly-finite integrals appearing in our representation have already played a considerable role (see e.g.\ \cite{Dixon:2011nj,Henn:2012ia,Henn:2013pwa,Henn:2014qga}). It would be very interesting to further develop these methods to construct analytic representations of all the integrands needed for two-loop ratio functions, for example. 

The enhancement of generalized unitarity to the {integrand}-level improves the traditional toolbox in several important ways. In addition to generating compact, closed-form representations of all amplitudes (without any need for a basis of integrands or the computationally challenging linear algebra needed to find coefficients), integrand-level representations preserve all of the symmetries of the theory, and make {\it more} (although not completely) manifest the finiteness of infrared-safe observables such as the ratio function (at least to two-loop order). And while having access to the all-loop recursion relations for planar SYM was important in verifying the correctness of the representations we described here, we strongly expect that the integrand-level approach will prove powerful for representing amplitudes in general field theories---even those for which the recursion relations have yet to be verified.

\vspace{\fill}
\acknowledgments We are very grateful to Simon Caron-Huot for important insights, discussions, and contributions, and to Nima Arkani-Hamed, Zvi Bern, Freddy Cachazo, and Marcus Spradlin for especially fruitful conversations. This work was supported through the hospitality of the Aspen Center for Physics (supported by NSF grant \#1066293), the Hong Kong University of Science and Technology's Institute for Advanced Study, and the California Institute of Technology. This work has been supported in part by the Harvard Society of Fellows, a grant from the Harvard Milton Fund, and a MOBILEX research grant from the Danish Council for Independent Research (JLB); and also by the David and Ellen Lee Postdoctoral Scholarship and by the Department of Energy under grant DE-SC0011632 (JT).

\newpage\appendix
\section{Momentum-Twistor Representations of Loop Amplitudes}\label{momentum_twistor_formulae_and_review_appendix}
\subsection{Kinematics, Notation, Momentum-Twistor Space, and Conventions}\label{overview_of_momentum_twistors_section}
Momentum-twistors are points in the twistor-space of dual-momentum coordinate space. Dual-momentum coordinates trivialize momentum conservation by describing the $n$ external momenta $\{p_a\}$ in terms of a closed (hence, momentum-conserving) polygon of points $\{x_a\}$ according to $p_a\!\equiv\!x_{a+1}\mi\,x_a$ (with $x_{n+1}\!\simeq\!x_1$ understood). Notice that the difference between any two of these points, $x_b$ and $x_a$ for example, represents a sum of consecutive momenta, $x_b\,\mi\,x_a\!=\!p_a\pl\,p_{a+1}\pl\ldots\pl\,p_{b-1}$, so that:
\vspace{5pt}\eq{\x{a}{b}=\x{b}{a}\equiv(x_b\,\mi\,x_a)^2=(p_a+p_{a+1}+\ldots+p_{b-1})^2.\vspace{0.pt}}

While dual-momentum coordinates make momentum conservation manifest, the on-shell condition (that $p_a^2\!=\!\x{a}{a\pl1}\!=\!0$  for all $a$) remains a non-trivial constraint on the $x_a$'s. Partly in order to trivialize these constraints, Andrew Hodges introduced {\it momentum-twistors} in \mbox{ref.\ \cite{Hodges:2009hk}}. Momentum-twistors $z_a$ are points in the twistor-space $(\mathbb{P}^3)$ of dual-momentum space---often specified as four-vectors using homogeneous coordinates. As with ordinary twistors, points in $x$-space are mapped to lines in twistor-space (and vice versa); and two points in $x$-space are null-separated iff their corresponding lines in twistor-space intersect. Two lines in twistor-space intersect iff they are linearly dependent, a condition that can be  can be tested by the determinant (the `$4$-bracket'):
\vspace{2.5pt}\eq{\ab{a\,b\,c\,d}\equiv\det\!\left(z_a,z_b,z_c,z_d\right).\label{defn_of_four_bracket}\vspace{0pt}} 
As such, any ordered list of $n$ momentum-twistors $\{z_a\}$ can be used to define a polygon whose pairwise-intersecting edges define null-separated points in $x$-space. Specifically, we may associate each line $(a\,\mi\,1\,a)\!\equiv\!\mathrm{span}\{z_{a-1},z_a\}$ in twistor-space with the point $x_a$ in dual-momentum space, and thereby ensure they satisfy $\x{a}{a\pl1}\!=\!0$.

Notice that this correspondence allows us to take an {\it unconstrained} list of $n$ points $\{z_a\}$ in momentum-twistor space and define a set of pairwise null-separated points $\{x_a\}$ in dual-momentum space, which in turn encode a manifestly momentum-conserving collection $\{p_a\}$ of null (on-shell) external momenta. This connection between momenta $p_{a}$, dual-momentum coordinates $x_a$, and momentum twistors $z_a$ (with our conventions) can be illustrated as follows:\\[-12pt]
\vspace{25.5pt}\eq{\hspace{-500pt}\begin{array}{@{}c@{}}\\[-40pt]\raisebox{5.25pt}{\includegraphics[scale=1.2]{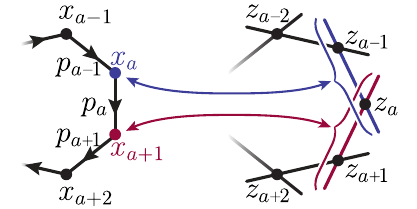}}\\[-50pt]~\end{array}\hspace{-500pt}\vspace{-20pt}}

Because loop amplitudes are integrals over points $\ell$ in dual-momentum space, they correspond to integrals over {\it lines} $(\ell)\!\equiv\!(\ell_A\,\ell_B)$ in momentum-twistor space. The precise correspondence between the more familiar loop-integration measure and the corresponding measure in momentum-twistor space is,\footnote{In this measure,  `$1/\mathrm{vol}((GL(2))$' is an instruction to mod-out by the $GL(2)$-redundancy of describing a {\it line} $(\ell_A\,\ell_B)\!\equiv\!\mathrm{span}\{\ell_A,\ell_B\}$ in terms of two points in (de-projectivized) twistor-space.}
\vspace{5pt}\eq{d^4\ell\Leftrightarrow\frac{d^4\ell_A\,d^4\ell_B}{\mathrm{vol}(GL(2))}/\ab{\ell_A\,\ell_B\,I^{\infty}}^4,}
where $I^{\infty}$ is the line in momentum-twistor space corresponding to the point `at infinity' in $x$-space. This line breaks dual-conformal invariance, and allows us to split the components of each twistor $z_a$ into those within the complement of $I^{\infty}$, denoted $\lambda_a$, and those within the span of $I^{\infty}$, denoted $\mu_a$. Thus, each momentum-twistor can be viewed as a pair of two-dimensional vectors, $z_a\!\equiv\!(\lambda_a\,\mu_a)$. 

Letting $\ab{a\,b}\!\equiv\!\ab{a\,b\,I^{\infty}}$, the precise connection between momentum-twistor $4$-brackets and ordinary kinematical invariants is the following:
\vspace{5pt}\eq{(x_b\mi\,x_a)^2=\frac{\ab{a\,\mi1\,a\,\,b\,\mi1\,b}}{\ab{a\,\mi1\,a}\ab{b\,\mi1\,b}}\equiv\frac{\ab{Aa\,Bb}}{\ab{Aa}\ab{Bb}}.}

Let $Z\!\equiv\!\big(z_1\,z_2\cdots z_n\big)$ be the $(4\times n)$-matrix whose columns are the (homogeneous coordinates for the) momentum-twistors which encode the external kinematical data. From these twistors, it is straight-forward to generate more familiar kinematical data, such as spinor-helicity variables \cite{vanderWaerden:1929} for each of the (null) momenta $p_a$,\\[-12pt]
\vspace{2.5pt}\eq{p_a^{\alpha\dot\alpha}\equiv\left(\begin{array}{@{}c@{$\;\;$}c@{}}\fwboxL{12pt}{p^0_a}\fwbox{10pt}{\,\pl\,}\fwboxR{16pt}{p^3_a}&\fwboxL{12pt}{p^1_a}\fwbox{10pt}{\,\mi\,}\fwboxR{16pt}{ip^2_a}\\\fwboxL{12pt}{p^1_a}\fwbox{10pt}{\,\pl\,}\fwboxR{16pt}{ip^2_a}&\fwboxL{12pt}{p^0_a}\fwbox{10pt}{\,\mi\,}\fwboxR{16pt}{p^3_a}\end{array}\right)\equiv\lambda_a^{\alpha}\widetilde\lambda_a^{\dot{\alpha}}.\vspace{-2.5pt}}
Upon using the line at infinity to split each $z_a$ into its components $z_a\!\equiv\!\big(\lambda_a\,\mu_a\big)$, we may define each particle's $\widetilde{\lambda}$ according to:
\vspace{2.5pt}\eq{\widetilde{\lambda}_a\equiv\mu_bQ^{b}_{a}\quad\mathrm{with}\quad Q^{b}_a\equiv\frac{\delta^b_{a-1}\ab{a\,a\pl1}\pl\,\delta^b_a\ab{a\pl1\,a\mi1}\pl\,\delta^b_{a+1}\ab{a\mi1\,a}}{\ab{a\mi1\,a}\ab{a\,a\pl1}},}
and where $\delta^{a}_b$ is the Kronecker $\delta$ symbol.

Conversely, given any momentum-conserving, massless four-momenta written in terms of spinor-helicity variables $p_a\!\equiv\!\lambda_a\widetilde{\lambda}_a$, we can define momentum-twistors $z_a$ by joining each $\lambda_a$ with $\mu_a$ constructed according to:
\vspace{2.5pt}\eq{\mu_a\equiv\widetilde{Q}_a^b\widetilde\lambda_b\quad\mathrm{where}\quad\widetilde{Q}_a^b\equiv\left\{\begin{array}{@{}c@{$\;\;\;\;\;$}r@{}}\ab{b\,a}&\text{if }1\!<\!b\!<\!a\\0&\text{otherwise}\end{array}\right.\,.}

Supermomentum-twistors are constructed by associating with each twistor $z_a$, a collection of $\mathcal{N}(\!=\!4)$ anti-commuting variables $\eta_a$ that are related to the ordinary supersymmetry variables in momentum-space, $\widetilde{\eta}$, in the same way that $\widetilde{\lambda}$ and $\mu$ are related: $\eta_a\!\equiv\!\widetilde{Q}_a^b\widetilde{\eta}_b$, and $\widetilde{\eta}_a\!\equiv\!\eta_bQ^{b}_a$.

Let us conclude this (rapid) summary of momentum-twistor kinematics by introducing the principle ingredients required to express on-shell functions (and hence amplitudes) and some of the most useful notational simplifications. A momentum-twistor superfunction of fundamental importance is the so-called `5-bracket' (also sometimes referred to as an `$R$-invariant'):\\[-12pt]
\vspace{2.5pt}\eq{\hspace{-25pt}\big[a\,b\,c\,d\,e\big]\!\equiv\!\frac{\delta^{1\times4}\big(\eta_a\ab{b\,c\,d\,e}\pl\,\eta_b\ab{c\,d\,e\,a}\pl\,\eta_c\ab{d\,e\,a\,b}\pl\,\eta_{d}\ab{e\,a\,b\,c}\pl\,\eta_e\ab{a\,b\,c\,d}\big)}{\ab{a\,b\,c\,d}\ab{b\,c\,d\,e}\ab{c\,d\,e\,a}\ab{d\,e\,a\,b}\ab{e\,a\,b\,c}}.\label{r_invariant_formula}\vspace{5pt}} 
As reviewed in \mbox{appendix \ref{two_loop_bcfw_appendix}}, all tree-amplitudes can be directly represented as sums of products of $5$-brackets. And as we will see in the following subsection, all on-shell diagrams can also be written as products of $5$-brackets, together with pre-factors involving (cross ratios of) $4$-brackets.

Often, we are interested in functions involving points in momentum-twistor space defined geometrically in terms of the external momentum-twistors. For example, it will be useful to refer to points such as ``the point where the line $(a\,b)$ intersects the plane $(c\,d\,e)$''---denoted `$(a\,b)\tcap(c\,d\,e)$'. Concretely, this point corresponds to $\mathrm{span}\{z_a,z_b\}\tcap\,\mathrm{span}\{z_c,z_d,z_e\}$, and can be concretely represented as follows:\\[-12pt]
\vspace{5pt}\eq{\hspace{-30pt}(a\,b)\tcap(c\,d\,e)\equiv z_a\ab{b\,c\,d\,e}\pl\,z_b\ab{c\,d\,e\,a}=-\big(z_c\ab{d\,e\,a\,b}\pl\,z_d\ab{e\,a\,b\,c}\pl\,z_e\ab{a\,b\,c\,d}\big).\vspace{5pt}\label{line_plane_cap_defn}}
(This formula follows trivially from the four-dimensional instance of {\it Cramer's rule}.) A similar, geometrically-defined object which proves useful is denoted `$(a\,b\,c)\tcap(d\,e\,f)$', by which we mean the rank-two subspace defined as $\mathrm{span}\{z_a,z_b,z_c\}\tcap\,\mathrm{span}\{z_d,z_e,z_f\}$:\\[-12pt]
\vspace{5pt}\eq{(a\,b\,c)\tcap(d\,e\,f)\equiv(a\,b)\ab{c\,d\,e\,f}\pl\,(b\,c)\ab{a\,d\,e\,f}\pl\,(c\,a)\ab{b\,d\,e\,f}.\vspace{5pt}\label{plane_plane_cap_defn}}

There is one final aspect of momentum-twistor variables that dramatically simplifies the complexity of formulae for on-shell functions. This is the fact that the Jacobian arising from the change of variables from momentum-space to momentum-twistor space is the full Parke-Taylor, MHV tree-level superamplitude \cite{Mason:2009qx,ArkaniHamed:2009vw}. Thus, for any on-shell function $f$, we have that:
\vspace{2.5pt}\eq{f(\lambda,\widetilde{\lambda},\widetilde{\eta})=f(Z,\eta)\frac{\delta^{2\times4}\big(\lambda\!\cdot\!\widetilde{\eta}\big)\delta^{2\times2}\big(\lambda\!\cdot\!\widetilde{\lambda}\big)}{\ab{1\,2}\ab{2\,3}\ab{3\,4}\cdots\ab{n\,1}}.\vspace{2.5pt}}
In particular, this means that the MHV tree-amplitude, when expressed in terms of momentum-twistors, is simply the identity! In the following section, we will see that the fairly trivial observation that tree-amplitudes can always be thought of as $\mathcal{A}_n^{(k)}(Z,\eta)\!\mapsto\!\mathcal{A}_n^{(k)}(Z,\eta)\!\times\!\mathcal{A}_{n}^{(0)}(Z,\eta)$ with the momentum-conserving $\delta$-functions associated with the MHV-amplitude factors will allow us to write any on-shell function in momentum-twistor variables as the product of a universal function---corresponding to the diagram where all the vertex amplitudes are replaced by MHV-amplitudes---with the actual corner amplitudes simply evaluated on the cut.

\newpage
\subsection{Explicit Momentum-Twistor Representations of On-Shell Functions}\label{explicit_formulae_for_on_shell_diagrams_appendix}
\subsubsection*{Review of One-Loop On-Shell Functions}
As described above, because MHV ($k\!=\!0$) amplitudes are the identity in momentum-twistor variables, we can consider any vertex (tree-)amplitude appearing in an on-shell diagram to include an MHV-amplitude factor. Therefore, when computing an on-shell function according to (\ref{general_on_shell_function_definition}), all of the phase-space localization of the internal particles can be viewed as arising from a diagram involving only MHV-amplitudes at each vertex,\footnote{Vertex amplitudes involving only three particles can also be $\bar{\text{MHV}}$ ($k\!=\!\mi1$).} and simply use the particular internal loop-momenta to evaluate the N$^k$MHV tree-amplitudes appearing at the vertices of the diagram.

Consider for example the box-type on-shell diagrams relevant to one-loop amplitudes. Using darker (blue) vertices to denote N$^{(k\leq0)}$MHV amplitudes,\footnote{For a leading singularity of an N$^k$MHV amplitude, the $k$-charges of the four corner amplitudes must satisfy $k\!\!=\!\!k_a\pl\, k_b\pl\, k_c\pl\, k_d\pl \,2$.} we have:\\[-12pt]
\eq{\hspace{-235pt}\raisebox{-42.25pt}{\includegraphics[scale=1]{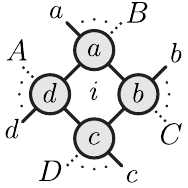}}\hspace{-6pt}=\hspace{-6pt}\raisebox{-42.25pt}{\includegraphics[scale=1]{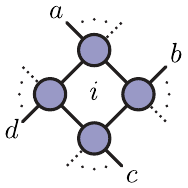}}\hspace{-7.5pt}\times\hspace{-2.5pt}\left(\!\begin{array}{@{}c@{}}\phantom{\times}\fwboxL{18pt}{\mathcal{A}_a\hspace{-1pt}\big(}\fwboxL{14pt}{Q^i_a},a,\ldots,\fwboxL{14pt}{Q^i_b}\big)\times\\\fwboxL{18pt}{\mathcal{A}_d\hspace{-1pt}\big(}\fwboxL{14pt}{Q^i_d},d,\ldots,\fwboxL{14pt}{Q^i_a}\big)\!\times\!\fwboxL{18pt}{\mathcal{A}_b\hspace{-1pt}\big(}\fwboxL{14pt}{Q^i_b},b,\ldots,\fwboxL{14pt}{Q^i_c}\big)\\
\times\fwboxL{18pt}{\mathcal{A}_c\hspace{-1pt}\big(}\fwboxL{14pt}{Q^i_c},c,\ldots,\fwboxL{14pt}{Q^i_d}\big)\phantom{\times}\end{array}\!\right)\!\!.\hspace{-200pt}\label{general_box_formula}\vspace{-2.5pt}}
Here, `$Q^i_{\bullet}$' encodes the internal momenta as follows. In momentum-twistor space, the cut conditions correspond to finding a line $(\ell)$ which intersects four given lines $\{(Aa),(Bb),(Cc),(Dd)\}$ (where `$A$' denotes $z_{a-1}$, for example) as illustrated below:\\[-12pt]
\eq{\raisebox{-72.25pt}{\includegraphics[scale=1]{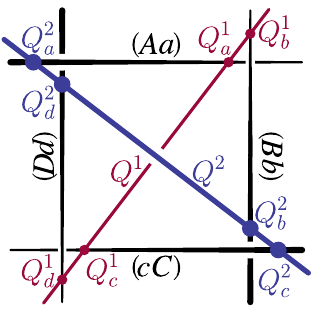}}\label{one_loop_quad_cut_geometry}}
For the $i^{\mathrm{th}}$ solution, the momentum flowing into to the top vertex ($a$) of the on-shell diagram (\ref{general_box_formula}), for example, would be represented by the point in momentum-twistor space where the `quad-cut' line $Q^i$ intersects the line $(Aa)$---denoted `$Q^i_a$' (see the figure above). We provide explicit formulae for these marked points along the quad-cuts for each solution in \mbox{Table \ref{quad_cuts_table}} below. These expressions smoothly degenerate for all boundary cases when one or more of the corner-amplitudes are massless.

\newpage 
\begin{table}[t]\caption{Explicit solutions $\ell^*\!\!\in\!\{{\color{cut1}Q^1},{\color{cut2}Q^2}\}$ to the quad-cut equations for four generic lines. Here, ${\color{cut1}Q^1}\!\equiv\!({\color{cut1}Q^1_a\,Q^1_c})\!\simeq\!({\color{cut1}Q^1_b\,Q^1_d})$ and ${\color{cut2}Q^2}\!\equiv\!({\color{cut2}Q^2_a\,Q^2_c})\!\simeq\!({\color{cut2}Q^2_b\,Q^2_d})$, and $\Delta$ is defined as in \mbox{(%
A.34%
)}. \label{quad_cuts_table}}{\small\vspace{-7pt}\eq{\hspace{-500pt}\begin{array}{|@{$\;\;\;\,\,$}l@{}l@{}l@{}c@{}l@{$\,\,\;\;\;$}|}\hline&&&&\\[-12pt]%
\text{{\normalsize${\color{cut1}Q^1_a}$}}&&&\equiv&\text{{\normalsize$z_a\pl z_A$}}\displaystyle\frac{\ab{a\,Bb\,(cC)\!\newcap\!(Dd\,A)}\pl\,\ab{A\,Bb\,(cC)\!\newcap\!(Dd\,a)}\pl\,\ab{aA\,cC}\ab{Bb\,Dd}\Delta}{2\ab{Bb\,(cC)\!\newcap\!(Dd\,A)\,A}}\\[10pt]&&&&\\[-12pt]
\text{{\normalsize${\color{cut1}Q^1_b}$}}&&&\equiv&\text{{\normalsize$z_B\pl z_b$}} \displaystyle\frac{\ab{B\,aA\,(Dd)\!\newcap\!(cC\,b)}\pl\,\ab{b\,aA\,(Dd)\!\newcap\!(cC\,B)}\pl\,\ab{aA\,cC}\ab{Bb\,Dd}\Delta}{2\ab{aA\,(Dd)\!\newcap\!(cC\,b)\,b}}\\[10pt]&&&&\\[-10pt]
\multicolumn{5}{|c|}{\text{{\normalsize${\color{cut1}Q^1_c}$}}\equiv(cC)\!\newcap\!(Dd{\color{cut1}\,Q^1_a})\quad\mathrm{and}\quad\text{{\normalsize${\color{cut1}Q^1_d}$}}\equiv(Dd)\!\newcap\!(cC{\color{cut1}\,Q^1_b})}\\
&&&&\\[-10pt]\hline\hline&&&&\\[-10pt]
\text{{\normalsize${\color{cut2}Q^2_a}$}}&&&\equiv&\text{{\normalsize$z_A\pl z_a$}}\displaystyle\frac{\ab{A\,dD\,(Cc)\!\newcap\!(bB\,a)}\pl\,\ab{a\,dD\,(Cc)\!\newcap\!(bB\,A)}\pl\,\ab{Aa\,Cc}\ab{bB\,dD}\Delta}{2\ab{dD\,(Cc)\!\newcap\!(bB\,a)\,a}}\\[10pt]&&&&\\[-12pt]
\text{{\normalsize${\color{cut2}Q^2_b}$}}&&&\equiv&\text{{\normalsize$z_b\pl z_B$}}\displaystyle\frac{\ab{b\,Cc\,(dD)\!\newcap\!(Aa\,B)}\pl\,\ab{B\,Cc\,(dD)\!\newcap\!(Aa\,b)}\pl\,\ab{Aa\,Cc}\ab{bB\,dD}\Delta}{2\ab{Cc\,(dD)\!\newcap\!(Aa\,B)\,B}}\\[10pt]&&&&\\[-10pt]
\multicolumn{5}{|c|}{\text{{\normalsize${\color{cut2}Q^2_c}$}}\equiv(Cc)\!\newcap\!(bB{\color{cut2}\,Q^2_a})\quad\mathrm{and}\quad\text{{\normalsize${\color{cut2}Q^2_d}$}}\equiv(dD)\!\newcap\!(Aa{\color{cut2}\,Q^2_b})}\\[-10pt]
&&&&\\[0pt]\hline
\end{array}\nonumber\hspace{-500pt}\vspace{-10pt}}}
\vspace{-10pt}
\end{table}

Given explicit formulae for the quad-cut solutions $Q^i$, it is not difficult to write the general expression for the one-loop box involving all MHV amplitude corners:
\vspace{5pt}\eq{\hspace{-230pt}\raisebox{-42.25pt}{\includegraphics[scale=1]{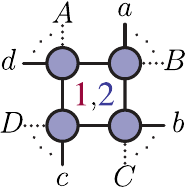}}=\left\{\begin{array}{@{}l@{}}\displaystyle\hspace{1pt}\big[{\color{cut1}Q^1_d}\,A\,a\,B\,b\big]\big[{\color{cut1}Q^1_b}\,C\,c\,D\,d\big]\!\left(1-\frac{\ab{{\color{cut1}Q^1_b}\,d\,A\,a}\ab{{\color{cut1}Q^1_d}\,b\,C\,c}}{\ab{{\color{cut1}Q^1_b}\,d\,C\,c}\ab{{\color{cut1}Q^1_d}\,b\,A\,a}}\right)^{\!\!-1}\\\displaystyle\hspace{1pt}\big[{\color{cut2}Q^2_a}\,B\,b\,C\,c\big]\big[{\color{cut2}Q^2_c}\,D\,d\,A\,a\big]\!\left(1-\frac{\ab{{\color{cut2}Q^2_a}\,c\,D\,d}\ab{{\color{cut2}Q^2_c}\,a\,B\,b}}{\ab{{\color{cut2}Q^2_a}\,c\,B\,b}\ab{{\color{cut2}Q^2_c}\,a\,D\,d}}\right)^{\!\!-1}\end{array}.\right.\hspace{-200pt}\label{mhv_box_formulae}}
Notice that we have provided different formulae for the two quad-cut solutions so that each {\it separately} degenerates smoothly in the limit where one or more of the corner amplitudes become massless.\footnote{For diagrams involving three-particle $k\!=\!\mi1$ amplitudes, some of the $5$-brackets in (\ref{mhv_box_formulae}) will degenerate---meaning that not all its arguments are distinct; in all such cases, the resulting on-shell functions is correctly obtained by simply setting any such degenerate $5$-bracket to 1.} For readers interested in more explicit detail, each degeneration was tabulated separately for both cases in \mbox{Table 3} of \mbox{ref.\ \cite{Bourjaily:2013mma}}.

Because it is easy to write momentum-twistor formulae for all tree amplitudes using BCFW recursion (see \mbox{appendix \ref{two_loop_bcfw_appendix}}), the general expression in equation (\ref{general_box_formula}) provides a closed-form expression for all one-loop box-type leading singularities as functions of external momentum-twistors.

\subsubsection*{Explicit Representations of All Two-Loop On-Shell Functions}\label{two_loop_on_shell_functions_section}
As we saw in the case of one-loop leading singularities above, all diagrams follow straightforwardly from the case when all the amplitudes are MHV ($k\!=\!0$). (As before, when three-particle amplitudes are involved, we should also include the possibility that some are $\bar{\text{MHV}}$ ($k\!=\!\mi1$); but as with one-loop, it turns out that all such on-shell functions can be found as smooth degenerations of the case where all amplitudes involve at least four particles.) And so, given these core objects, all other on-shell functions are obtained by simply multiplying these all-MHV expressions by the relevant corner amplitudes, evaluated using the momentum-twistors that encode the internal, cut momenta.

Consider for example the so-called `kissing-boxes' at two-loops. These are leading singularities with the following topology:
\vspace{-10pt}{\normalsize\eq{\hspace{-340pt}\raisebox{-57.25pt}{\includegraphics[scale=1]{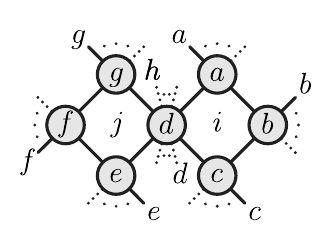}}\hspace{-10pt}=\hspace{-10pt}\raisebox{-57.25pt}{\includegraphics[scale=1]{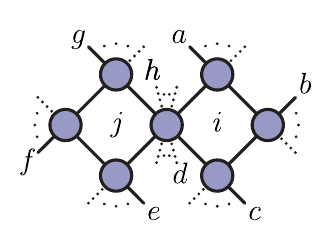}}\hspace{-15pt}\times\hspace{-3.5pt}\Bigg(\!\!\prod_v\mathcal{A}_v(\cdots)\!\!\Bigg).\hspace{-300pt}\label{general_kissing_boxes_on_shell_function_formula}\vspace{-12.5pt}}}
\hspace{-4pt}\noindent%
Notice that all of the relevant ``octa-cuts'' needed for the corner amplitudes are described in terms of a pair of quad-cuts $\{Q^i,Q^j\}$. Thus, the only non-trivial ingredient in (\ref{general_kissing_boxes_on_shell_function_formula}) is the diagram involving only MHV amplitudes; this turns out to correspond to the following product of one-loop box functions:
\vspace{-10pt}\eq{\hspace{-244.5pt}\raisebox{-57.25pt}{\includegraphics[scale=1]{kissingboxes_mhv_on_shell_function}}\hspace{-10pt}=\hspace{-5pt}\raisebox{-57.25pt}{\includegraphics[scale=1]{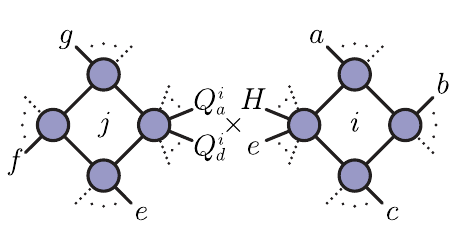}}.\hspace{-200pt}\vspace{-15pt}}
(Recall the convention for how the possible ranges of external legs are denoted in these figures, as summarized in (\ref{leg_ranges_conventions_figure}). The fact that some ranges of legs can be empty explains the somewhat unusual structure of legs appearing in the boxes above. Specifically, because the range $\{h,\ldots,A\}$ may be empty (indicated by the dotted lines in the figure), $H(\geq\!g)$ denotes the leg immediately preceding $h(\leq\!a)$.)

There is only one other class of non-composite, two-loop leading singularities (co-dimension eight residues involving eight distinct propagators): the `penta-boxes',\\[-12pt]
\vspace{-0pt}{\normalsize\eq{\hspace{-340pt}\raisebox{-47.pt}{\includegraphics[scale=1]{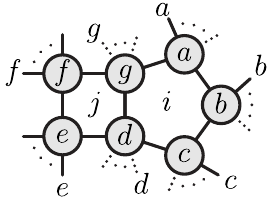}}\hspace{-5pt}=\hspace{-5pt}\raisebox{-47.pt}{\includegraphics[scale=1]{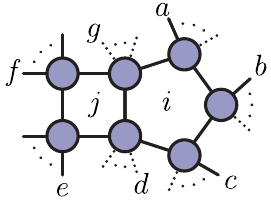}}\hspace{-5pt}\times\hspace{-3.5pt}\Bigg(\!\!\prod_v\mathcal{A}_v(\cdots)\!\!\Bigg).\hspace{-300pt}\vspace{-5pt}\label{general_pentabox_on_shell_function_formula}}}
\hspace{-4pt}%
As before (although prehaps slightly less trivially), the relevant octa-cuts (again expressed as a pair of quad-cuts $\{Q^i,Q^j\}$) follow directly from one-loop expressions described above. And again, the only non-trivial ingredient is the skeleton on-shell diagram involving only MHV amplitudes at its vertices---which is determined as the product of one-loop on-shell functions (where `$G$' is the immediate predecessor of $g$),\\[-12pt]
\vspace{-0.5pt}\eq{\hspace{-234pt}\raisebox{-47pt}{\includegraphics[scale=1]{pentabox_mhv_on_shell_function}}\hspace{-5pt}=\hspace{-0pt}\raisebox{-47pt}{\includegraphics[scale=1]{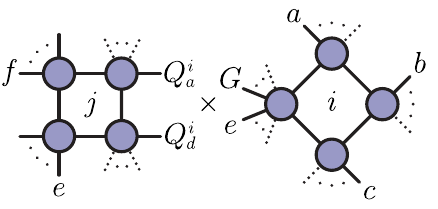}}.\hspace{-200pt}\vspace{-5.5pt}\label{pentabox_mhv_on_shell_function_formula}}

Before moving on, we should briefly remind the reader that the penta-box on-shell functions appearing in our list of independent on-shell data for two-loop amplitudes, (\ref{fifth_term_group}), were not individual penta-boxes but {\it sums} of penta-boxes,
\vspace{-6.5pt}\eq{\;\hspace{-200pt}\raisebox{-47.5pt}{\includegraphics[scale=1]{pentabox}}\hspace{0pt}\equiv\hspace{0pt}\raisebox{-3pt}{\scalebox{2}{$\displaystyle\sum_{\substack{\\[-5.5pt]\scalebox{0.55}{$j$}}}$}}\raisebox{-47.5pt}{\includegraphics[scale=1]{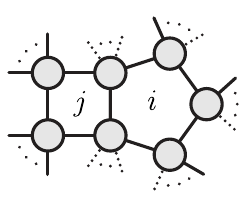}}\label{pentabox_data_summation_formula_appendix}\;.
\hspace{-200pt}\vspace{-7.5pt}}

The final class of on-shell functions needed to represent all two-loop amplitude integrands are the so-called double-triangles. Being especially explicit about the arguments of each corner amplitude, they are given by:
\vspace{-12.5pt}\begin{equation}\begin{split}\hspace{-57pt}\raisebox{-47.25pt}{\includegraphics[scale=1]{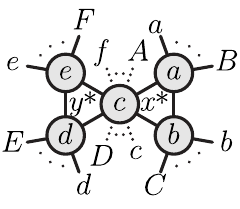}}\hspace{0pt}=\hspace{-105pt}\\[-30pt]%
\hspace{-327.25pt}%
&\hspace{-171pt}\raisebox{-47.25pt}{\includegraphics[scale=1]{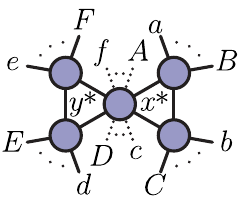}}\hspace{-7.5pt}\times\hspace{-5pt}\left(\begin{array}{@{}c@{}}\fwboxL{18pt}{\mathcal{A}_e\hspace{-1pt}\big(}\fwboxL{30pt}{T_{e}(y^*\hspace{-1pt})},e,\ldots,F,\fwboxL{30pt}{T_{f}(y^*\hspace{-1pt})}\hspace{-0pt}\big)\!\times\!\fwboxL{18pt}{\mathcal{A}_a\hspace{-1pt}\big(}\fwboxL{30pt}{T_{a}(x^*\hspace{-1pt})},a,\ldots,B,\fwboxL{30pt}{T_{b}(x^*\hspace{-1pt})}\big)\\\times\fwboxL{18pt}{\mathcal{A}_c\hspace{-1pt}\big(}\fwboxL{30pt}{T_{f}(y^*\hspace{-1pt})},\ldots,\fwboxL{35pt}{T_{a}(x^*\hspace{-1pt}),}\fwboxL{30pt}{T_c(x^*\hspace{-1pt})},\ldots,\fwboxL{30pt}{T_{d}(y^*\hspace{-1pt})}\big)\times\\%
\fwboxL{18pt}{\mathcal{A}_d\hspace{-1pt}\big(}\fwboxL{30pt}{T_{d}(y^*\hspace{-1pt})},d,\ldots,E,\fwboxL{30pt}{T_{e}(y^*\hspace{-1pt})}\hspace{-0.75pt}\big)\!\times\!\fwboxL{18pt}{\mathcal{A}_b\hspace{-1pt}\big(}\fwboxL{30pt}{T_{b}(x^*\hspace{-1pt})},b,\ldots,C,\fwboxL{30pt}{T_{c}(x^*\hspace{-1pt})}\big)\end{array}\right)\!\!,\hspace{-300pt}\\[-10pt]\end{split}\label{general_double_triangle_formula}\vspace{-20pt}\end{equation}
where $\{T(x^*\hspace{-1pt}),T(y^*\hspace{-1pt})\}$ represent the triple-cuts of the two triangles (each of which generally depends on one parameter), evaluated at an arbitrary reference point $(x,y)\!\mapsto\!(x^*\hspace{-1pt},y^*\hspace{-1pt})$. Concretely, we may parameterize these triple-cuts as follows:\\[-12pt]
\vspace{7.5pt}\eq{\left\{\begin{array}{@{}l@{}l@{}c@{}l@{}l@{}}\\[-20pt]T_a(x^*\hspace{-1pt})\,&\equiv (a+x^* A),&\hspace{20pt}&T_d(y^*\hspace{-1pt})\,&\equiv (d+y^*D)\\
T_b(x^*\hspace{-1pt})\,&\equiv(\hspace{-1pt}Bb)\tcap\big(Cc\,\,T_a(x^*\hspace{-1pt})\big),&&T_e(y^*\hspace{-1pt})\,&\equiv(\hspace{-1pt}Ee)\tcap\big(Ff\,\,T_d(y^*\hspace{-1pt})\big)\\
T_c(x^*\hspace{-1pt})\,&\equiv(\hspace{-1pt}Cc)\tcap\big(Bb\,\,T_a(x^*\hspace{-1pt})\big),&&T_f(y^*\hspace{-1pt})\,&\equiv(\hspace{-1pt}Ff)\tcap\big(Ee\,\,T_d(y^*\hspace{-1pt})\big)\\[-0pt]
\end{array}\right\}\!.\label{triple_cut_formulae_for_double_triangle}\vspace{-20pt}}
(Notice, for example, that $\{T_a(x^*\hspace{-1pt}),T_b(x^*\hspace{-1pt}),T_c(x^*\hspace{-1pt})\}$ represent points where the triple-cut line intersects the lines $\{(Aa),(Bb),(Cc)\}$, respectively.) While any sufficiently generic choice for the values of $(x^*\hspace{-1pt},y^*\hspace{-1pt})$ at which to evaluate the diagram would suffice, a particularly convenient choice for our purposes is to always take,
\vspace{5pt}\eq{\phantom{.}x^*\hspace{-1pt}\equiv\frac{\ab{(Cc\,e)\,a}}{\ab{A\,(Cc\,e)}}\quad\mathrm{and}\quad y^*\hspace{-1pt}\equiv\frac{\ab{(Ff\,b)\,d}}{\ab{D\,(Ff\,b)}}.\label{choice_of_spurious_point_for_double_triangle}}
This choice is motivated by the fact that it systematically ensures that no other physical propagators are cut, and that the double-box integrand normalized to match field theory at this point is dual-conformally invariant (as seen in the next subsection).

The final ingredient needed by (\ref{general_double_triangle_formula}) for on-shell functions of this type is the double-triangle diagram involving only MHV amplitudes. To compute this, we observe that one MHV corner of each triangle can be expanded via BCFW as an on-shell diagram, allowing us to identify the double-triangle function as a double BCFW-shift of a kissing-boxes diagram in the following way:
\vspace{0pt}\eq{\hspace{-0pt}\begin{array}{@{}c@{}c@{}c@{}}\raisebox{-47pt}{\includegraphics[scale=1]{general_mhv_double_triangle_ls_2}}&\hspace{5pt}\equiv\hspace{10pt}&\hspace{-5pt}\raisebox{-77.75pt}{\includegraphics[scale=1]{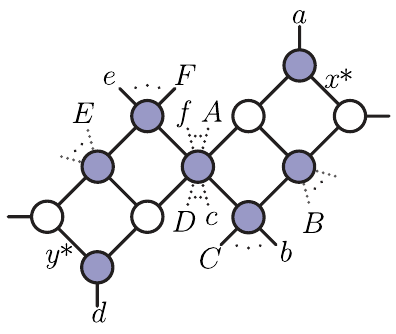}}\\&\hspace{5pt}\equiv\hspace{10pt}&\fwbox{0pt}{\big[(a+x^*\!A)\,B\,b\,C\,c\big]\big[(d+y^*\!D)\,E\,e\,F\,f\big].}\\[-27pt]~\end{array}\hspace{-0pt}\vspace{15pt}\label{double_triangle_mhv_on_shell_function}}
This follows from the fact that attaching a BCFW bridge between legs $(a\,a\pl1)$ corresponds to shifting twistor $z_a$ by some parameter $x$ in the direction of $z_{A}\!\equiv\!z_{a-1}$ (\mbox{see e.g.\ ref.\ \cite{ArkaniHamed:2010kv}}). And so, the formula above follows from applying these shifts to the on-shell function,
\vspace{-5pt}\eq{\raisebox{-57.25pt}{\includegraphics[scale=1]{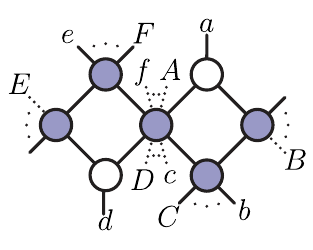}}\hspace{-5pt}=\big[a\,B\,b\,C\,c\big]\!\big[d\,E\,e\,F\,f\big],\vspace{-5pt}}
which is a particular instance of the general formula for kissing-boxes, (\ref{general_kissing_boxes_on_shell_function_formula}).

\newpage
\subsection{Explicit Momentum-Twistor Representations of Loop Integrands}\label{momentum_twistor_reps_of_integrands}
As described in the body of this work, the integrands required in our constructions are uniquely determined by a small number of simple criteria: that they have residues (or evaluate to be) of unit magnitude on a particular physical cut, that they vanish on all independent physical cuts (or reference points), etc. As such, the precise form of the integrands needed by our representations could be found using any preferred choice of kinematical variables. But for the sake of concreteness and elucidation, in this section we provide explicit solutions to these constraints in terms of numerators constructed using momentum-twistor variables. \\[-18pt]

\subsubsection*{One-Loop Amplitude Integrand Ingredients} 
The one-loop integrands used to match every quad-cut individually were uniquely fixed by the constraints described in \mbox{section \ref{one_loop_integrand_review}}. Parameterizing the integrands,
\vspace{4.5pt}\eq{\hspace{-20pt}\fwbox{430pt}{\fwbox{190pt}{\displaystyle\mathcal{I}^i_{a,b,c,d}\equiv \frac{\x{X}{Y_i(\ell)}}{\x{\ell}{a}\x{\ell}{b}\x{\ell}{c}\x{\ell}{d}\x{\ell}{X}},}\fwbox{10pt}{\mathrm{}}\fwbox{180pt}{\mathcal{I}_{\mathrm{}}^a\equiv \frac{\x{X}{Y^a}}{\x{\ell}{a\mi1}\x{\ell}{a}\x{\ell}{a\pl1}\x{\ell}{X}},}}\vspace{-0pt}}
the (unique) factors $Y_i(\ell)$ which solve all the constraints are listed in \mbox{Table \ref{chiral_box_integrands_table}}. 

\noindent\begin{table}[b]\vspace{-0pt}
\noindent\scalebox{1}{\mbox{\hspace{-0pt}\begin{minipage}[h]{\textwidth}\vspace{-70pt}
{\small\eq{\hspace{-500pt}
\begin{array}{@{}c@{}}\begin{array}{|@{}c@{}|@{$$}l@{}|}\hline\raisebox{-42.25pt}{\includegraphics[scale=1]{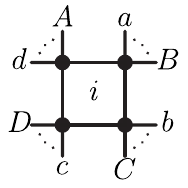}}\hspace{-0pt}&\begin{array}{@{}l@{}}\\[-22pt]\displaystyle\fwboxL{250pt}{\fwboxR{16pt}{\;Y_{\fwboxL{9pt}{\!1,\!2}}\!}\equiv\frac{1}{2}(\ell)\ab{Aa\,Cc}\ab{Bb\,Dd}\Delta\pm\frac{1}{12}\epsilon^{ijkl}\big(\ns\ns\paren{E_i}\tncap\paren{E_j\,D}\,\paren{E_k}\tncap\paren{E_l\,d}\ns\ns\big)}\\[6pt]\rule[1pt]{340pt}{0.5pt}\\[-3pt]\displaystyle\fwboxL{340pt}{\fwboxL{125pt}{\fwboxR{16pt}{\hspace{0pt}\Delta\hspace{-0pt}}\equiv\!\sqrt{(1\mi\,u\mi\,v)^2\mi\,4uv},}\fwboxL{110pt}{u\!\equiv\!\displaystyle\frac{\ab{Aa\,Bb}\ab{Cc\,Dd}}{\ab{Aa\,Cc}\ab{Bb\,Dd}}\!,}\fwboxL{105pt}{v\!\equiv\!\displaystyle\frac{\ab{Bb\,Cc}\ab{Dd\,Aa}}{\ab{Aa\,Cc}\ab{Bb\,Dd}}}}\\\fwboxL{340pt}{\;\;\{E_1,E_2,E_3,E_4\}\!\equiv\!\big\{\!\paren{Aa},\!\paren{Bb},\!\paren{Cc},\!(\ell)\!\big\}}\\[-16pt]\end{array}\\\hline\end{array}\\[-6.5pt]~\\[-6.5pt]
\begin{array}{|@{}c@{}|@{$$}l@{}|}\hline\raisebox{-42.25pt}{\includegraphics[scale=1]{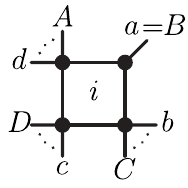}}\hspace{-0pt}&\begin{array}{@{}l@{}}\\[-18pt]\displaystyle\fwboxL{250pt}{\fwboxR{16pt}{\;Y_1}\equiv\frac{1}{2}\Big(\ns\ns\ns\paren{B\,Dd}\tncap\big(\ns Cc\,\paren{Aa\,b}\tncap(\ell)\ns\big)-\paren{B\,Cc}\tncap\big(\ns Dd\,\paren{Aa\,b}\tncap(\ell)\ns\big)\ns\ns\ns\Big)}\\[6.2pt]\rule[1pt]{340pt}{0.5pt}\\[5.2pt]\displaystyle\fwboxL{250pt}{\fwboxR{16pt}{\;Y_2}\equiv\displaystyle\frac{1}{2}\Big(\ns\ns\ns\ns\big(\ns\ns\paren{\ns Dd}\tncap\paren{Aa\,b}\,\paren{Cc}\tncap\paren{\ell\,B}\ns\big)\ns-\ns\big(\ns\paren{Cc}\tncap\paren{Aa\,b}\,\paren{\ns Dd}\tncap\paren{\ell\,B}\ns\ns\big)\ns\ns\ns\ns\Big)}\\[-1.2pt]\end{array}\\\hline\end{array}\\[-6.5pt]~\\[-6.5pt]
\begin{array}{|@{}c@{}|@{$$}l@{}|}\hline\raisebox{-42.25pt}{\includegraphics[scale=1]{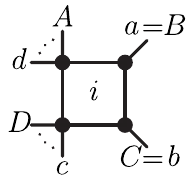}}\hspace{-0pt}&\begin{array}{@{}l@{}}\\[-18pt]\displaystyle\fwboxL{250pt}{\fwboxR{16pt}{\;Y_1}\equiv\frac{1}{2}\Big(\ns\ns\ns\paren{B\,Dd}\tncap\big(\ns Cc\,\paren{Aa}\tncap\paren{\ell\,b}\ns\big)-\paren{B\,Cc}\tncap\big(\ns Dd\,\paren{Aa}\tncap\paren{\ell\,b}\ns\big)\ns\ns\ns\Big)\hspace{200pt}}\\[6.2pt]\rule[1pt]{340pt}{0.5pt}\\[5.2pt]\displaystyle\fwboxL{250pt}{\fwboxR{16pt}{\;Y_2}\equiv\displaystyle\frac{1}{2}\Big(\ns\ns\ns\paren{Aa\,b}\tncap\big(\ns Dd\,\paren{Cc}\tncap\paren{\ell\,B}\ns\big)-\paren{Dd\,b}\tncap\big(\ns Aa\,\paren{Cc}\tncap\paren{\ell\,B}\ns\big)\ns\ns\ns\Big)\hspace{200pt}}\\[-1.2pt]\end{array}\\\hline\end{array}\\[-6.5pt]~\\[-6.5pt]\begin{array}{|@{}c@{}|@{$$}l@{}|@{}l@{}|@{}c@{}|}\hline\raisebox{-42.25pt}{\includegraphics[scale=1]{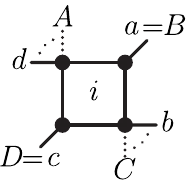}}\hspace{-0pt}&\fwboxL{140pt}{\begin{array}{@{}l@{}}\\[-18pt]\displaystyle\fwboxR{16pt}{\;Y_1}\equiv\paren{DB}\ab{\ell\,\paren{Aa\,b}\tncap\paren{Cc\,d}}\phantom{\Big(\tncap\frac{1}{2}}\\[6.2pt]\rule[1pt]{140pt}{0.5pt}\\[5.2pt]\displaystyle\fwboxR{16pt}{\;Y_2}\equiv\paren{Aa\,b}\tncap\paren{Cc\,d}\ab{\ell\,DB}\phantom{\Big(\tncap\frac{1}{2}}\\[-1.2pt]\end{array}}\hspace{-0pt}&\fwboxL{140pt}{\fwboxR{16pt}{\;Y^a}\equiv(a\mi1\,a)\ab{a\mi2\,a\mi1\,a\,a\pl1}\hspace{-20pt}}&\fwbox{60pt}{\hspace{10pt}\raisebox{-42.25pt}{\includegraphics[scale=1]{one_loop_triangle}}\hspace{-0pt}}\\\hline\end{array}\\[-2.5pt]\end{array}\hspace{-500pt}\nonumber}}\end{minipage}}}\caption{{\bf One-Loop Chiral `Box' Integrand Numerator Factors}}\label{chiral_box_integrands_table}\vspace{-40pt}\end{table}

\newpage
\subsubsection*{Two-Loop Amplitude Integrand Ingredients}
Knowing the form of the one-loop integrands required to match each particular quad-cut, it is comparatively simple to construct every two-loop integrand required to represent amplitudes according to the criteria discussed in \mbox{section \ref{two_loop_amplitude_section}}. In this subsection, we will briefly describe the form that these integrands take. 

The first term, $(1.a)$, of the local integrand representation (\ref{two_loop_integrand_formula_part_1}) is perhaps the simplest---the double-box:
\vspace{-10pt}\eq{\hspace{-335pt}\fwboxL{170pt}{\raisebox{-47.25pt}{\includegraphics[scale=1]{two_loop_term_1_int}}}\hspace{-40pt}\fwboxL{0pt}{\equiv\frac{\x{a\mi1}{a\pl1}\x{a}{b}\x{b\mi1}{b\pl1}}{\x{\ell_1}{a\mi1}\x{\ell_1}{a}\x{\ell_1}{a\pl1}\x{\ell_1}{\ell_2}\x{\ell_2}{b\mi1}\x{\ell_2}{b}\x{\ell_2}{b\pl1}}.}\hspace{10pt}\vspace{-15pt}}
This integrand's numerator is completely fixed by the criterion that it have unit residue on the corresponding, physical (doubly-composite) octa-cut. 

The second term, $(1.b)$, is fixed in the identical way---by ensuring that it has unit residue on the corresponding, single-composite octa-cut. The third term, $(1.c)$, however, is slightly more interesting:
\vspace{2.5pt}\eq{\hspace{-335pt}\fwboxL{170pt}{\raisebox{-47.25pt}{\includegraphics[scale=1]{two_loop_term_3_int}}}\hspace{-30pt}\fwboxL{0pt}{\equiv\frac{\x{Y_i(\ell_1)}{e}\x{e\pl1}{e\mi1}}{\x{\ell_1}{\!a}\x{\ell_1}{\!b}\x{\ell_1}{\!c}\x{\ell_1}{\!d}\x{\ell_1}{\ell_2}\x{\!\ell_2}{\!e\mi1}\x{\ell_2}{\!e}\x{\ell_2}{\!e\pl1}}.}\vspace{-2.5pt}\label{composite_penta_box_int_formula}\vspace{-5pt}}
Again, its numerator is determined by the criterion that it have unit reside on the corresponding {\it composite} octa-cut. Notice the role played by the one-loop numerator (given in \mbox{Table \ref{chiral_box_integrands_table}}) in the integrand that is ultimately needed for the pentabox, (\ref{composite_penta_box_int_formula}). (Recall that the one-loop numerators $Y_i(\ell)$ change form depending on which (if any) of the legs are massless---equivalently, which labels $\{a,b,c,d\}$ are consecutive.)

Perhaps the most interesting integrands needed in the expansion of two-loop integrands are those of the class $(2.a)$---the finite double-boxes associated with the double-triangle hexa-cuts (evaluated at particular reference points):
\vspace{-5pt}\eq{\hspace{-10pt}\raisebox{-47.25pt}{\includegraphics[scale=1]{two_loop_term_4_int}}\hspace{-10pt}.\vspace{-10pt}\label{finite_double_box_fig_eph}} 
As described in \mbox{section \ref{two_loop_amplitude_section}}, when both ranges of legs $\{f,\ldots,A\}$ and $\{c,\ldots,D\}$ in (\ref{finite_double_box_fig_eph}) are non-empty, such integrands do not have {any} co-dimension eight residues at all, and therefore have no preferred points in loop-momentum space where we can match field theory via a residue. Nevertheless, it turns out that we need only match field theory at any {\it arbitrary} point along its two-dimensional hexa-cut---according to the double-triangle on-shell function described in \mbox{appendix \ref{two_loop_on_shell_functions_section}}.

Except for the novelty of fixing the {\it value} of the integrand at a particular point along its hexa-cut (instead of at a place where the hexa-cut integrand supports a co-dimension two residue), it is completely straight-forward to normalize the integrand uniquely so that evaluates to the identity when its double-triangle hexa-cut is evaluated at the particular point $(x^*\hspace{-1pt},y^*\hspace{-1pt})$:
\vspace{-7.5pt}\eq{\hspace{-325pt}\fwboxL{170pt}{\raisebox{-47.25pt}{\includegraphics[scale=1]{two_loop_term_4_int}}}\hspace{-40pt}\fwboxL{0pt}{\equiv\frac{\x{T(x^*\hspace{-1pt})}{T(y^*\hspace{-1pt})}}{x^*y^*\x{\ell_1}{a}\x{\ell_1}{b}\x{\ell_1}{c}\x{\ell_1}{\ell_2}\x{\ell_2}{d}\x{\ell_2}{e}\x{\ell_2}{f}},}\vspace{-7.5pt}\label{finite_double_box_int_formula}}
where $\{T(x^*\hspace{-1pt}),T(y^*\hspace{-1pt})\}$ are the triple-cut points of the right- and left-triangles, respectively. Using the form of these cuts in momentum-twistor space given in (\ref{triple_cut_formulae_for_double_triangle}), the numerator of (\ref{finite_double_box_int_formula}) becomes,
\vspace{5pt}\eq{\hspace{-30pt}\x{T(x^*\hspace{-1pt})}{T(y^*\hspace{-1pt})}\!\equiv\!\ab{(a\pl x^*\!A)(Cc)\tcap\big(Bb(a\pl x^*\!A)\!\big)(d\pl y^*\!D)(Ff)\tcap\big(Ee\,(d\pl y^*\!D)\!\big)},\vspace{-0pt}}
using the particular choice of $(x^*\hspace{-1pt},y^*\hspace{-1pt})$ described above,
\vspace{5pt}\eq{x^*\!\equiv\ab{(Cc\,e)\,a}/\ab{A\,(Cc\,e)}\quad\mathrm{and}\quad y^*\!\equiv\ab{(Ff\,b)\,d}/\ab{D\,(Ff\,b)}.\label{spurious_triple_cut_point_formula_2}}
Notice that for this choice of $(x^*\hspace{-1pt},y^*\hspace{-1pt})$, (\ref{finite_double_box_int_formula}) becomes dual-conformally invariant.

The penultimate class, $(2.b)$, of integrands required in the representation (\ref{two_loop_integrand_formula_part_2}) are the penta-box integrands. These integrands are fixed by the requirement that they have unit residues on the co-dimension eight contour enclosing {both} cuts of box for a given pentagon quad-cut (a contour that is parity-even on the box-part), and also that the integrand vanish at the points $(x^*\hspace{-1pt},y^*\hspace{-1pt})$ for each of its four-mass double-triangle hexa-cuts. It is not difficult to construct the unique numerator which solves these constraints:
\vspace{0pt}\eq{\hspace{-315pt}\fwboxL{170pt}{\raisebox{-47.25pt}{\includegraphics[scale=1]{two_loop_term_5_int}}}\hspace{-30pt}\fwboxL{0pt}{\equiv\frac{\x{\widehat{Y_i}(\ell_1)}{f}\x{g}{e}\Delta[Q^i,e,f,g]}{\x{\ell_1}{a}\x{\ell_1}{b}\x{\ell_1}{c}\x{\ell_1}{d}\x{\ell_1}{\ell_2}\x{\ell_2}{e}\x{\ell_2}{f}\x{\ell_2}{g}},}\vspace{-7.5pt}\label{finite_pentabox_integral_formula}}
where $\Delta[a,b,c,d]$ is the familiar square-root normalizing the four-mass box integral,\\[-12pt]
\vspace{2.5pt}\eq{\hspace{-20pt}\Delta[a,b,c,d]\equiv\!\sqrt{(1\,\mi\,u\,\mi\,v)^2\,\mi\,4uv}\;\;\mathrm{with}\;\;u\equiv\frac{\x{a}{b}\x{c}{d}}{\x{a}{c}\x{b}{d}},\;v\equiv\frac{\x{b}{c}\x{d}{a}}{\x{a}{c}\x{b}{d}},\label{four_mass_delta_defintion}}
and where $\widehat{Y_i}(\ell_1)$ is the chiral numerator for one-loop, pentagon sub-integral (see \mbox{Table \ref{chiral_box_integrands_table}}), but corrected in order to explicitly vanish at the reference points $(x^*\hspace{-1pt},y^*\hspace{-1pt})$ on all its four-mass hexa-cuts:
\vspace{2.5pt}\eq{\widehat{Y_i}(\ell_1)\equiv Y_i(\ell_1)\hspace{3pt}-\hspace{-23pt}\sum_{\substack{\text{four-mass hexacuts}\\\text{not involving prop}\\\x{\ell_1}{\rho}\text{, }\rho\in\{a,b,c,d\}}}\hspace{-20pt}Y_i(T(x^*\hspace{-1pt}))\frac{\x{\ell_1}{\rho}}{\x{T(x^*\hspace{-1pt})}{\rho}}.}

The requirement that these integrands vanish somewhere along each of its four-mass hexa-cuts is very important to allowing us to uniquely specify the form given in  (\ref{finite_pentabox_integral_formula}): the requirement that the integral have unit residue on the directly relevant octa-cut is not strong enough to uniquely fix the integrand, because adding terms in the numerator proportional to any of the pentagon's propagators would not spoil this criterion. It is only because we demand that these integrals have no support on any of the lower, already matched on-shell data that we find the unique form of the integrand, (\ref{finite_pentabox_integral_formula}).

The last integrand required to represent all two-loop amplitude integrands are the double-pentagons associated with the kissing-boxes octa-cuts. These integrands are again fixed by the criteria that they have unit residue on the corresponding octa-cut and that they do not affect any of the already-fixed on-shell data. In particular, this means that they must vanish on all combinations of penta-box residues that are parity-even on the side of the box, and that the integrands vanish at all the points $(x^*\hspace{-1pt},y^*\hspace{-1pt})$ of its four-mass hexa-cuts. The unique solution to these constraints can be written as follows:
\vspace{2.5pt}\eq{\hspace{-42.5pt}\raisebox{-47.25pt}{\includegraphics[scale=1]{two_loop_term_6_int}}\hspace{-12.5pt}\equiv\!\frac{N}{\x{\ell_1}{\!a}\x{\ell_1}{\!b}\x{\ell_1}{\!c}\x{\ell_1}{\!d}\x{\ell_1}{\!\ell_2}\x{\ell_2}{\!e}\x{\ell_2}{\!f}\x{\ell_2}{\!g}\x{\ell_2}{\!h}}\!,\vspace{-2.5pt}}
where the numerator $N$ is given by
\vspace{2.5pt}\eq{\hspace{-30pt}N=\x{Y_i(\ell_1)}{Y_j(\ell_2)}\hspace{4pt}-\hspace{-44pt}\sum_{\substack{\text{four-mass hexacuts not}\\\text{involving props: }\x{\ell_1}{\rho},\x{\ell_2}{\lambda},\\\text{with }\rho\in\{a,b,c,d\},\,\lambda\in\{e,f,g,h\}}}\hspace{-40pt}\x{Y_i(T(x^*\hspace{-1pt}))}{Y_j(T(y^*\hspace{-1pt}))}\frac{\x{\ell_1}{\rho}\x{\ell_2}{\lambda}}{\x{T(x^*\hspace{-1pt})}{\rho}\x{T(y^*\hspace{-1pt})}{\lambda}},\hspace{-0pt}}
where $Y_i(\ell_1)$ and $Y_j(\ell_2)$ are the one-loop numerators which match the chiral boxes listed in \mbox{Table \ref{chiral_box_integrands_table}}. Because the one-loop numerators $Y_i(\ell)$ were fixed by imposing the constraint that the resulting integrands vanish on all parity-even four-mass contours involving the propagator $\x{\ell}{X}$, the numerator above will not contribute to any of the contours which are fixed by the penta-box terms. 

\newpage
\section{\mbox{Explicit BCFW Representations of Two-Loop Amplitudes}}\label{two_loop_bcfw_appendix}
As described in ref.\ \cite{ArkaniHamed:2010kv} (see also \cite{ArkaniHamed:2012nw}), all $l$-loop integrands for scattering amplitudes in planar SYM can be found by the BCFW recursion relations. In terms of on-shell diagrams, the recursion relations correspond to:
\vspace{-0.3cm}\eq{\hspace{-230pt}\raisebox{-50pt}{\includegraphics[scale=1]{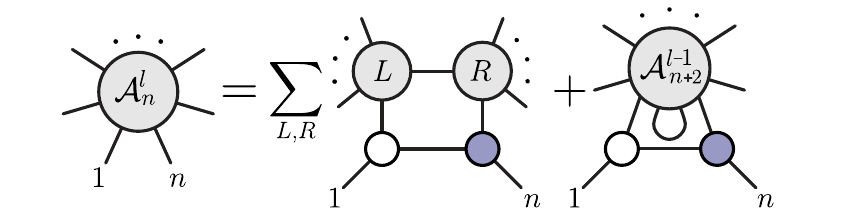}}\hspace{-30pt}.\hspace{-200pt}\vspace{-0.35cm}\label{bcfw_all_loop_recursion_figure}}
Letting $\mathcal{A}_{n}^{(k),l}$ denote the $l$-loop, $n$-particle, N$^{k}$MHV amplitude integrand, and being explicit about the ranges of the terms involved, we write the two contributions above---the so-called `bridge terms' and `forward-limits'---using the shorthand:
\vspace{5pt}\eq{\mathcal{A}_n^{(k),l}\;=\hspace{-.05cm}\sum_{\text{{\scriptsize$\hspace{0pt}\begin{array}{@{}c@{}c@{}c@{$\!$}c@{}c@{}c@{}l}\\[-14pt]n&=&n_L&+&n_R&-&2\\k&=&k_L&+&k_R&+&1\\l&=&l_L&+&l_R\\[-4pt]\end{array}\hspace{-0pt}$}}}\!\!\!\mathcal{A}_{n_L}^{(k_L),l_L}\!\!\bigotimes_{\mathrm{BCFW}}\!\!\mathcal{A}_{n_R}^{(k_R),l_R}+\mathrm{FL}\Big(\mathcal{A}_{n+2}^{(k+1),l-1}\Big).\vspace{-.0cm}\label{bcfw_all_loop_recursion}}

In momentum-twistor variables, the BCFW bridge corresponds to a shift $z_n\!\to\!z_n\pl\,\alpha\, z_{n-1}$; for $n_R\!>\!3$, the so-called `bridge terms' are found to be,
\vspace{7.5pt}\eq{\hspace{-.5cm}\mathcal{A}_{n_L}^{(k_L),l_L}\!\!\bigotimes_{\mathrm{BCFW}}\!\!\mathcal{A}_{n_R>3}^{(k_R),l_R}\equiv\mathcal{A}_{n_L}^{(k_L),l_L}(1,\hspace{-1pt}\ldots\hspace{-1pt},A,\ahat)[1\,A\,a\,\,n\mi1\,n]\mathcal{A}_{n_R}^{(k_R),l_R}(\ahat,a,\hspace{-1pt}\ldots\hspace{-1pt},n\mi1,\hat{n}),\vspace{0pt}\nonumber}
where $\ahat\equiv(aA)\newcap(n\mi1\,n\,1)$ and $\hat{n}\equiv(n\,n\mi1)\newcap(A\,a\,1)$; when $n_R\!=\!3$ and $n_L\!=\!n\mi1$, the bridge simply results in,
\vspace{7.5pt}\eq{\mathcal{A}_{n-1}^{(k),l_L}\!\!\bigotimes_{\mathrm{BCFW}}\!\!\mathcal{A}_{3}^{(-1),0}\equiv\mathcal{A}_{n-1}^{(k),l_L}(1,\ldots,n\mi1).\vspace{-0pt}\nonumber}
And so the bridge terms in (\ref{bcfw_all_loop_recursion}) are fairly straight-forward momentum-twistor space---the operations involved being the same regardless of the loop-levels of the amplitudes being bridged. (At tree-level, there are no `forward-limit' contributions, so only the bridge terms are needed; thus, the discussion so far suffices to represent all tree-level ($l\!=\!0$) amplitudes.)

More interesting are the `forward-limit' terms, $\mathrm{FL}\big(\mathcal{A}_{n+2}^{(k+1),l-1}\big)$. It is easy to see that (\ref{bcfw_all_loop_recursion}) gives rise to $l$ levels of nested forward-limits. As described in \mbox{ref.\ \cite{ArkaniHamed:2012nw}}, it is generally difficult to determine which terms of the lower-loop amplitude remain non-vanishing in the forward-limit (even the number of terms which contribute becomes scheme-dependent beyond one loop). Nevertheless, once we have chosen how to recurse each lower-loop amplitude, it is possible to identify all the terms that remain non-vanishing. The recursion scheme we will use {\it always} takes the legs identified in the forward limit as the `bridge' legs ($\{1,n\}$ in the figure (\ref{bcfw_all_loop_recursion_figure}) ) for further recursion. 

Expressed in terms of ``kermit'' functions corresponding to the (nested) forward-limits of $5$-brackets, an explicit solution to the recursion relations for any amplitude through two-loop-order is given by:\\[-10pt]
\eq{\hspace{-200pt}\boxed{\hspace{45pt}\hspace{-5pt}\begin{array}{l}
\hspace{-43pt}\mathcal{A}_n^{(k),l\leq2}(1,\hspace{-1pt}\ldots\hspace{-1pt},n\mi1,n)
\\[15pt]\\[-30pt]\hspace{-37pt}=\phantom{+}\hspace{-4pt}\mathcal{A}_{n-1}^{(k),l}(1,\hspace{-1pt}\ldots\hspace{-1pt},n\mi1)
\\[15pt]\\[-25pt]\hspace{-30pt}+\hspace{-19pt}\displaystyle\sum_{\text{{\scriptsize$\hspace{17.5pt}\begin{array}{@{}c@{}c@{}c@{$\!$}c@{}c@{}c@{}l@{$\,\,$}l}\\[-14pt]n&=&n_1&+&n_2&-&2&\fwboxL{0pt}{(n_2\geq4)}\\k&=&k_1&+&k_2&+&1\\l&=&l_1&+&l_2\\[-8pt]~\\[-4pt]\end{array}\hspace{-17.5pt}$}}}\!\!\!\hspace{-0.375cm}\underset{\;\;\;\;\;\;\;\text{{\normalsize$\hspace{41.75pt}\begin{array}{c}\\[-23pt]\text{{\footnotesize ~ }}\\\\[-18pt]\hspace{15pt}\text{{\footnotesize$\ahat\equiv(aA)\newcap(n\mi1\,n\,1),\;\hat{n}\equiv(n\,n\mi1)\newcap(A\,a\,1)$}}\end{array}\hspace{-41.75pt}$}}}{\mathcal{A}_{n_1}^{(k_1),l_1}(1,\hspace{-1pt}\ldots\hspace{-1pt},A,\ahat)\mathcal{A}_{n_2}^{(k_2),l_2}(\ahat,a,\hspace{-1pt}\ldots\hspace{-1pt},n\mi1,\hat{n})\big[1\,A\,a\,\,n\mi1\,n\big]\,\,}\\[15pt]\\[-10pt]\hspace{-30pt}+\hspace{-19pt}\displaystyle\sum_{\text{{\scriptsize$\hspace{17.5pt}\begin{array}{@{}c@{}c@{}c@{$\!$}c@{}c@{}c@{}l@{$\,\,$}l}\\[-14pt]n&=&n_1&+&n_2&-&4&\fwboxL{0pt}{(n_1,n_2\!\geq\!4)}\\k&=&k_1&+&k_2&\phantom{+}&\phantom{1}\\l&=&l_1&+&l_2&+&1\\[-8pt]~\\[-4pt]\end{array}\hspace{-17.5pt}$}}}\hspace{-.575cm}\underset{\;\;\;\;\text{{\normalsize$\hspace{28pt}\begin{array}{c}\\[-20pt]\text{{\footnotesize ~ }}\\\\[-23pt]\hspace{60.5pt}\text{{\footnotesize$\ahat\equiv(aA)\newcap(\ell\,1),\;\hat{n}\equiv(n\,n\mi1)\newcap(\ell\,1),\; \hat{\ell}\equiv(\ell)\newcap(n\mi1\,n\,1)$}}\end{array}\hspace{-50.5pt}\hspace{-28pt}$}}}{\mathcal{A}_{n_1}^{(k_1),l_1}\hspace{-0.05cm}(\hat{\ell},1,\hspace{-1pt}\ldots\hspace{-1pt},A,\ahat)\mathcal{A}_{n_2}^{(k_2),l_2}(\ahat,a,\hspace{-1pt}\ldots\hspace{-1pt},n\mi1,\hat{n},\hat{\ell}\,)K_1[a,n]\,}
\\[15pt]\\[-10pt]\hspace{-30pt}+\hspace{-29pt}\displaystyle\sum_{\text{{\scriptsize$\hspace{22.5pt}\hspace{5pt}\begin{array}{@{}c@{}c@{}c@{$\!$}c@{}c@{$$}c@{$$}c@{$\!$}c@{}l@{$\,\,$}l}n&=&n_1&+&n_2&+&n_3&-&7&\fwboxL{0pt}{(n_3\!\geq\!4)}\\k&=&k_1&+&k_2&+&k_3&\phantom{+}&\phantom{1}\\l&=&l_1&+&l_2&+&l_3&+&2\\[-4pt]\end{array}\hspace{-22.5pt}$}}}
\hspace{-0.95cm}\underset{\;\;\;\;\text{{\normalsize$\begin{array}{c}\\[-20pt]\text{{\footnotesize ~ }}\\\\[-23pt]\hspace{51pt}\text{{\footnotesize$\begin{array}{l@{$$}c@{}ll@{}c@{}llc@{}l}\hat{n}&\equiv&(n\,n\mi1)\newcap(\ell_1\,1),&\hat{\ell_1}&\equiv&(\ell_1)\newcap(n\mi1\,n\,1),&\hat{\ell_2}&\equiv&(\ell_2)\newcap(\ell_1\,1),\\\hat{a}&\equiv&(aA)\newcap(\ell_2\,\hat{\ell_1})&\hspace{1pt}\hat{b}&\equiv&(bB)\newcap(\ell_2\,\hat{\ell_1}),&\hat{\hat{\ell_2}}&\equiv&(\hat{\ell_2}\hat{\ell_1})\newcap(B\,b\,\hat{a})
\end{array}$}}\end{array}$}}}{\mathcal{A}_{n_1}^{(k_1),l_1}\hspace{-0.05cm}(\hat{\ell_1},1,\hspace{-1pt}\ldots\hspace{-1pt},A,\ahat)\mathcal{A}_{n_2}^{(k_2),l_2}(\hat{a},a,\hspace{-1pt}\ldots\hspace{-1pt},B,\hat{b})\mathcal{A}_{n_3}^{(k_3),l_3}(\hat{b},b,\hspace{-1pt}\ldots\hspace{-1pt},n\mi1,\hat{n},\hat{\ell_1},\hat{\hat{\ell_2}})K_2[a,b,n]\,}
\\[15pt]\\[-10pt]\hspace{-30pt}+\hspace{-29pt}\displaystyle\sum_{\text{{\scriptsize$\hspace{22.5pt}\hspace{5pt}\begin{array}{@{}c@{}c@{}c@{$\!$}c@{}c@{}c@{$$}c@{$\!$}c@{}l@{$\,\,$}l}n&=&n_1&+&n_2&+&n_3&-&8&\fwboxL{0pt}{(n_2,n_3\!\geq\!4)}\\k&=&k_1&+&k_2&+&k_3&\phantom{+}&\phantom{1}\\l&=&l_1&+&l_2&+&l_3&+&2\\[-4pt]\end{array}\hspace{-22.5pt}$}}}
\hspace{-0.95cm}\underset{\;\;\;\;\text{{\normalsize$\begin{array}{c}\\[-20pt]\text{{\footnotesize ~ }}\\\\[-23pt]\hspace{40pt}\text{{\footnotesize$\begin{array}{l@{$$}c@{}ll@{}c@{}llc@{}l}\hat{n}&\equiv&(n\,n\mi1)\newcap(\ell_1\,1),&\hat{\ell_1}&\equiv&(\ell_1)\newcap(n\mi1\,n\,1),&\hat{\ell_2}&\equiv&(\ell_2)\newcap(\ell_1\,1),\\\hat{b}&\equiv&(bB)\newcap(\ell_2\,\hat{\ell_1})&\hspace{1pt}\hat{a}&\equiv&(aA)\newcap(\ell_2\,\hat{\ell_1}),&\hat{\hat{\ell_2}}&\equiv&(\hat{\ell_2}\hat{\ell_1})\newcap(A\,a\,\hat{\ell_1})
\end{array}$}}\end{array}$}}}{\mathcal{A}_{n_1}^{(k_1),l_1}\hspace{-0.05cm}(\hat{\ell_1},1,\hspace{-1pt}\ldots\hspace{-1pt},A,\ahat)\mathcal{A}_{n_2}^{(k_2),l_2}(\hat{a},a,\hspace{-1pt}\ldots\hspace{-1pt},B,\hat{b},\hat{\hat{\ell_2}})\mathcal{A}_{n_3}^{(k_3),l_3}(\hat{b},b,\hspace{-1pt}\ldots\hspace{-1pt},n\mi1,\hat{n},\hat{\ell_1},\hat{\ell_2})K_2[b,a,n].\,}
\end{array}\hspace{-5pt}}\hspace{-200pt}\label{full_bcfw_recursion_formua}\vspace{-0pt}}

\noindent Here, the one-loop `kermit' $K_1[a,n]$ is given by,
\vspace{-30pt}\eq{\hspace{-2cm}\begin{array}{rl}\\[-2.5pt]~\\K_1[a,n]\equiv&\displaystyle -d^4\ell\,\,\frac{\ab{\ell\,(\hspace{-1pt}1\,A\,a)\newcap(\hspace{-1pt}1\,n\mi1\,n)}^2}{\ab{\ell\,1A}\ab{\ell\,A\,a}\ab{\ell\,a\,1}\ab{\ell\,1n\,\mi1}\ab{\ell\,n\mi1\,n}\ab{\ell\,n\,1}},\\[10pt]=&\hspace{-0cm}\displaystyle -d\hspace{-1pt}\log\left(\!\frac{\ab{\ell\,1A}}{\ab{\ell\,A\,a}}\!\right)\!d\hspace{-1pt}\log\left(\!\frac{\ab{\ell\,A\,a}}{\ab{\ell\,a\,1}}\!\right)\!d\hspace{-1pt}\log\left(\!\frac{\ab{\ell\,1\,n\mi1}}{\ab{\ell\,n\mi1\,n}}\!\right)\!d\hspace{-1pt}\log\left(\!\frac{\ab{\ell\,n\mi1\,n}}{\ab{\ell\,n\,1}}\!\right);\\[-2.5pt]\end{array}\hspace{-2cm}\vspace{2pt}\label{introducing_mister_kermit_one}}
and the two-loop `kermit' $K_2[a,b,c]$ is given by,
\vspace{-30pt}\eq{\hspace{-200pt}\begin{array}{rl}\\[-2.5pt]~\\K_2[a,b,n]\!\equiv\!&\displaystyle\frac{\ab{\ell_1(\hspace{-1pt}1\hspace{1pt}\ell_2)\newcap(\hspace{-1pt}n\mi1\hspace{1pt}n\hspace{1pt}1)}^2\ab{\ell_2(\ell_1\,1)\newcap(\hspace{-1pt}Aa\,\widehat{\ell_1})}^2\ab{\widehat{\ell_1}\hspace{1pt}Bb\hspace{1pt}\widehat{a}}^3}{\ab{\ell_1\hspace{0pt}1\hspace{1pt}n\mi1}\hspace{-1pt}\ab{\ell_1\hspace{0pt}n\mi1\hspace{1pt}n}\hspace{-1pt}\ab{\ell_1\hspace{0pt}n1}\hspace{-1pt}\ab{\ell_1\hspace{1pt}\ell_2}\hspace{-1pt}\ab{\ell_2\hspace{1pt}\widehat{\ell_1}A}\hspace{-1pt}\ab{\ell_2\hspace{1pt}Aa}\hspace{-0pt}\ab{\ell_2\hspace{1pt}a\widehat{\ell_1}}\hspace{-1pt}\ab{\widehat{\ell_2}\widehat{\ell_1}\hspace{1pt}Bb}\hspace{-1pt}\ab{\widehat{\ell_2}\widehat{\ell_1}B\widehat{a}}\hspace{-1pt}\ab{\widehat{\ell_2}Bb\hspace{1pt}\widehat{a}}\hspace{-1pt}\ab{\widehat{\ell_2}\widehat{\ell_1}b\widehat{a}}},\\[10pt]\!\equiv\!&\hspace{-0cm}\displaystyle d\hspace{-1pt}\log\!\left(\rho_1\right)\cdots d\hspace{-1pt}\log\!\left(\rho_8\right),\\[-2.5pt]\end{array}\hspace{-200pt}\vspace{-0pt}\label{introducing_mister_kermit_two}\nonumber\vspace{2.5pt}}
where the $d\!\log$-coordinates $\{\rho_1,\ldots,\rho_8\}$ are given by:
\eq{\hspace{-200pt}\left\{\begin{array}{@{}l@{$\,$}c@{$\,$}l@{$,\;\;\;$}l@{$\,$}c@{$\,$}l@{$,\;\;\;$}l@{$\,$}c@{$\,$}l@{$,\;\;\;$}l@{$\,$}c@{$\,$}l@{}}\rho_1&\equiv&\displaystyle\frac{\ab{\ell_1\hspace{1pt}n\mi1\hspace{1pt}1}}{\ab{\ell_1\hspace{1pt}n\mi1\hspace{1pt}n}}&\rho_2&\equiv&\displaystyle\frac{\ab{\ell_1\hspace{1pt}n\mi1\hspace{1pt}n}}{\ab{\ell_1\hspace{1pt}n\hspace{1pt}1}}&\rho_3&\equiv&\displaystyle\frac{\ab{\ell_2\hspace{1pt}a\hspace{1pt}\widehat{\ell_1}}}{\ab{\ell_2\hspace{1pt}A\hspace{1pt}\widehat{\ell_1}}}&\rho_4&\equiv&\displaystyle\frac{\ab{\ell_2\hspace{1pt}A\hspace{1pt}\widehat{\ell_1}}}{\ab{\ell_2\hspace{1pt}A\hspace{0pt}a}}%
\\[10pt]\rho_5&\equiv&\displaystyle\frac{\ab{\widehat{\ell_1}\hspace{1pt}\widehat{\ell_2}\hspace{1pt}b\hspace{1pt}\widehat{a}}}{\ab{\widehat{\ell_1}\hspace{1pt}\widehat{a}\hspace{1pt}Bb}}&
\rho_6&\equiv&\displaystyle\frac{\ab{\widehat{\ell_1}\hspace{1pt}\widehat{\ell_2}\hspace{1pt}B\hspace{1pt}\widehat{a}}}{\ab{\widehat{\ell_1}\hspace{1pt}\widehat{a}\hspace{1pt}Bb}}&
\rho_7&\equiv&\displaystyle\frac{\ab{\widehat{\ell_1}\hspace{1pt}\widehat{\ell_2}\hspace{1pt}B\hspace{0pt}b}}{\ab{\widehat{\ell_1}\hspace{1pt}\widehat{a}\hspace{1pt}Bb}}&
\rho_8&\equiv&\displaystyle\frac{\ab{\widehat{\ell_2}\hspace{1pt}\widehat{a}\hspace{1pt}B\hspace{0pt}b}}{\ab{\widehat{\ell_1}\hspace{1pt}\widehat{a}\hspace{1pt}Bb}}\end{array}\right\},\hspace{-200pt}\nonumber\vspace{2.5pt}}
in terms of the shifted momentum-twistors, defined according to:
\vspace{2.5pt}\eq{\hspace{-200pt}\fwboxL{0pt}{~\hspace{-10pt}\raisebox{-1pt}{$\left\{\rule{0pt}{20pt}\right.$}}\begin{array}{c}\\[-20pt]\text{{\normalsize ~ }}\\\\[-28pt]\hspace{0pt}\text{{\normalsize$\begin{array}{l@{$$}c@{}ll@{}c@{}llc@{}l}\hat{n}&\equiv&(n\,n\mi1)\newcap(\ell_1\,1),&\hat{\ell_1}&\equiv&(\ell_1)\newcap(n\mi1\,n\,1),&\hat{\ell_2}&\equiv&(\ell_2)\newcap(\ell_1\,1)\\\hat{a}&\equiv&(aA)\newcap(\ell_2\,\hat{\ell_1}),&\hspace{1pt}\hat{b}&\equiv&(bB)\newcap(\ell_2\,\hat{\ell_1}),&\hat{\hat{\ell_2}}&\equiv&(\hat{\ell_2}\hat{\ell_1})\newcap(Bb\,\hat{a})\\[-4pt]
\end{array}$}}\end{array}\fwboxL{0pt}{\hspace{-5pt}\raisebox{-1pt}{$\left.\rule{0pt}{20pt}\right\}.$}}\hspace{-200pt}\vspace{-5pt}}

\newpage
\section{Implementation of Two-Loop Results in {\sc Mathematica}}\label{mathematica_appendix}
\subsection*{Obtaining and Initializing the {\sc Mathematica} Package {\tt two\uscore loop\uscore amplitudes} }\label{mathematica_setup_appendix}
In order to make the tools described in this paper most useful to researchers, we have prepared a {\sc Mathematica} package called `{\tt two\uscore loop\uscore amplitudes}' which implements our results. In addition to providing explicit, analytic, efficiently-evaluatable representations of loop-amplitude integrands, the {\tt two\uscore loop\uscore amplitudes} package also serves as a reliable reference for the many results tabulated above (as any transcription error would obstruct numerical consistency checks).

The package and a notebook illustrating its functionality are included with the submission files for this paper on the {\tt arXiv}, which can be obtained as follows. From this work's abstract page on the {\tt arXiv}, look for the ``download'' options (in the upper-right corner of the page), follow the link to ``other formats'' (below the option for ``PDF''), and download the ``source files'' for the submission. The source will contain\footnote{On certain systems, the `source' file from the {\tt arXiv} is often saved to disk without any extension; this can be ameliorated by manually appending ``{\tt .tar.gz}'' to the name of the downloaded file.} the primary package {\tt two\uscore loop\rule[-1.05pt]{7.5pt}{.75pt}amplitudes.m}, together with a notebook {\tt two\uscore loop\rule[-1.05pt]{7.5pt}{.75pt}amplitudes\rule[-1.05pt]{7.5pt}{.75pt}demo.nb} which has detailed examples of the package's functionality.

Upon obtaining the source files, one should open and evaluate the {\sc Mathematica} notebook `{\tt two\uscore loop\rule[-1.05pt]{7.5pt}{.75pt}amplitudes\rule[-1.05pt]{7.5pt}{.75pt}demo.nb}'; in addition to walking the user through  example computations, this notebook will copy  {\tt two\uscore loop\rule[-1.05pt]{7.5pt}{.75pt}amplitudes.m} to the user's {\tt ApplicationDirectory[]}; this will make the package available to run in any future notebook via the command ``{\tt <<two\uscore loop\uscore amplitudes.m}'':\\[-15pt]

\mathematica{.8}{\raisebox{-2pt}{{\tt<<two\rule[-1.05pt]{7.5pt}{.75pt}loop\rule[-1.05pt]{7.5pt}{.75pt}amplitudes.m}}}{\raisebox{-180pt}{\includegraphics[scale=.875]{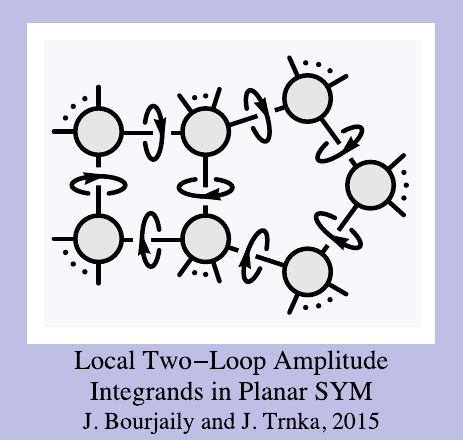}}}\\[-20pt]

\newpage
\subsection{Glossary of the Primary Functions of the {\sc Mathematica} Package}\label{mathematica_glossary_appendix}

\subsubsection*{Abstract Symbols for Objects \& Functions Related to Loop Amplitudes}
\vspace{-7.5pt}
\defn{ab}{\vardefms{abcd}}{represents a symbol for the momentum-twistor $4$-bracket involving twistors labeled by the sequence \var{abcd}. The arguments of \fun{ab}{\tt[]} can include geometrically defined  points in twistor-space including, for example, \fun{shift}{\tt[\{}\var{a},\var{b}{\tt\}},\var{$\alpha$}{\tt]} or \fun{cap}{\tt[\{}\var{a},\var{b}{\tt\}},{\tt\{}\var{c},\var{d},\var{e}{\tt\}]}, geometrically defined lines such as \fun{cap}{\tt[\{}\var{a},\var{b},\var{c}{\tt\}},{\tt\{}\var{d},\var{e},\var{f}{\tt\}]}, or even differences of lines (or points) represented by \fun{dif}{\tt[}\var{a},\var{b}{\tt]}.}

\defntb{boxes}{\vardef{i}}{\vardef{legList},\vardef{kLists}}{a symbol representing a collection of one-loop, box-type leading singularities indexed by the $R$-charges of the corner amplitudes $\{\var{k_a},\ldots,\var{k_d}\}\!\!\in$\var{kLists}, with the topology specified by the external legs \var{legList}, and involving the $Q^{\var{i}}$ solution to the quad-cut equations for the internal momentum. When the function \fun{toAnalytic}{\tt[]} is applied to an expression involving \mbox{\fun{boxes}{\tt[}\var{i}{\tt]\![}$\{\!\{\var{a},\ldots\},\ldots,\{\var{d},\ldots\}\}${\tt]}}, for example, it will be replaced by the on-shell function(s) $f^{\var{i}}_{\var{a},\ldots,\var{d}}$ times its corresponding chiral one-loop integrand, $\mathcal{I}^{\var{i}}_{\var{a},\ldots,\var{d}}$.}

\defn{cap}{{\tt \{}\vardefms{a}{\tt\}},{\tt\{}\vardefms{b}{\tt\}}}{when appearing as an argument of \fun{R}{\tt[\vardefms{abcde}]} or \fun{ab}{\tt[\vardefms{abcd}]}, for example, \mbox{\fun{cap}{\tt[\var{a}]\![\var{b}]}} represents the geometrically-defined object `(\var{a})$\tcap$(\var{b})' in momentum-twistor space. Such geometrically-defined points or lines can be expanded concretely using equations (\ref{line_plane_cap_defn}) or (\ref{plane_plane_cap_defn}) of \mbox{appendix \ref{momentum_twistor_formulae_and_review_appendix}}, respectively.}

\defn{dif}{\vardef{x},\vardef{y}}{represents the difference between two twistor arguments when appearing as an argument of \fun{ab}{\tt[$\cdots$]}\hspace{-1pt}. For example, for three lines in momentum-twistor space $\{${\tt\var{x},\var{y},\var{z}}$\}$, \mbox{\fun{ab}{\tt[\fun{dif}[\var{x},\var{y}],\var{z}]}}$\,\equiv\,$\mbox{\fun{ab}{\tt[\var{x},\var{z}]}}$-$\mbox{\fun{ab}{\tt[\var{y},\var{z}]}}. It behaves similarly when $\{$\var{x},\var{y}$\}$ represent points in momentum-twistor space.}

\defn{R}{\vardefms{abcde}}{represents the $5$-bracket superfunction (also known as the `$R$-invariant') involving twistors given by the sequence \var{abcde}. See equation (\ref{r_invariant_formula}). }

\defn{shift}{{\tt\{}\vardef{zA},\vardef{zB}{\tt\}},\vardef{$\alpha$}}{represents the point in momentum-twistor space (\var{zA}$+$\var{$\alpha$}$\,$\var{zB}), where $\{$\var{zA},\var{zB}$\}$ are twistors, and \var{$\alpha$} is a scalar.}

\defn{treeAmp}{\vardef{n},\vardef{k}}{abstractly represents the \var{n}-particle, N$^{\text{\var{k}}}$MHV tree-amplitude, $\mathcal{A}_{\var{n}}^{(\text{{\tt \var{k}}}),0}$, for the purposes of expanding loop amplitude integrands---e.g.\ as seen in the output of the function \fun{localLoopIntegrand}{\tt[\var{n},\var{k},\var{ell}]}.}

\newpage
\subsubsection*{Analytic \& Symbolic Representations of Scattering Amplitude Integrands}
\vspace{-7.5pt}
\defn{localLoopIntegrand}{\vardef{n},\vardef{k},\vardefo{ell}2}{returns a symbolic representation of the local integrand representation of the \var{ell}-loop, \var{n}-particle, N$^{\text{\var{k}}}$MHV amplitude integrand, $\mathcal{A}_{\text{\var{n}}}^{(\text{\var{k}}),\text{\var{ell}}}\!,$ as derived in this work. For \var{ell}$\,=\!0$, the output is simply the symbolic `\fun{treeAmp}{\tt[}\var{n},\var{k}{\tt]}' (see above); for \var{ell}$\,=\!1$, the function returns the representation described in \mbox{section \ref{one_loop_integrand_review}} (as derived in ref.\ \cite{Bourjaily:2013mma}), written in terms of \fun{boxes}{\tt[]\![]}'s, \fun{scalarTriangle}{\tt[]}'s and \fun{treeAmp}{\tt[]}'s; and for \var{ell}$\,=\!2$ (the default value), it returns the representation described in \mbox{section \ref{two_loop_amplitude_section}}, written in terms of (abstract symbols representing) the six types of on-shell diagrams used to encode the result.}

\defn{rAmp}{\vardef{n},\vardef{k},\vardefo{ell}0}{returns the particular BCFW representation of the \var{ell}-loop, \var{n}-particle, N$^{\text{\var{k}}}$MHV amplitude integrand, $\mathcal{A}_{\text{\var{n}}}^{(\text{\var{k}}),\text{\var{ell}}}\!,$ expressed in terms of momentum-twistor variables (with $5$-brackets \fun{R}{\tt[}\vardefms{abcde}{\tt]}, $4$-brackets \fun{ab}{\tt[$\cdots$]}, and so-called `kermit' functions), recursed according to the scheme corresponding to equation (\ref{full_bcfw_recursion_formua}) given in \mbox{appendix \ref{two_loop_bcfw_appendix}}. For $2$-loop amplitudes (\var{ell}$\,=\!\!2$), the output of \fun{rAmp}{\tt[]} is {\it not} symmetrized with respect to the loop-momentum variables.}

\subsubsection*{Explicit Expressions for Loop Amplitudes \& Integrand Ingredients}
\vspace{-7.5pt}
\defntb{chiralIntegrand}{\vardefms{ij}}{\vardef{legList}}{for any one- or two-loop on-shell diagram decorated by a {\it finite} loop integrand---one-loop boxes, double-triangles, pentaboxes, and kissing-boxes involving the external legs indicated by \var{legList} (ordered according to the figures throughout this work), \fun{chiralIntegrand} returns the corresponding loop integrand expression in terms of $4$-brackets \fun{ab}{\tt[$\cdots$]} ({\it without} symmetrization of the loop-momentum variables).
\\[-22pt]

\ind The use of \fun{chiralIntegrand} can be illustrated by the following examples:\\[5pt]
\mathematica{0.945}{
nice[chiralIntegrand[1][\{\!\{2\}\!,\!\{3,4\}\!,\!\{5\}\!,\!\{6,1\}\!\}]]\hspace{14cm}
nice[chiralIntegrand[1,1][\{\!\{1\}\!,\!\{2\}\!,\!\{3\}\!,\!\{\}\!,\!\{4\}\!,\!\{5\}\!,\!\{6\}\!,\!\{\}\!\}]]\hspace{0cm}}{
\vspace{-7.5pt}\scalebox{0.95}{$\displaystyle\frac{\ab{(\ell_1)\,(123)\tncap(456)}\ab{(X)\,25}}{\ab{(\ell_1)12}\ab{(\ell_1)23}\ab{(\ell_1)45}\ab{(\ell_1)56}}$}\rule[-10pt]{0pt}{20pt}\hspace{14cm}$~$
\rule[-10pt]{0pt}{20pt}\scalebox{0.95}{$\rule[-10pt]{0pt}{35pt}\displaystyle\frac{\ab{(\ell_1)\,(612)\tncap(234)}\ab{(\ell_2)\,(345)\tncap(561)}\ab{1346}}{\ab{(\ell_1)61}\ab{(\ell_1)12}\ab{(\ell_1)23}\ab{(\ell_1)34}\ab{(\ell_1)(\ell_2)}\ab{(\ell_2)34}\ab{(\ell_2)45}\ab{(\ell_2)56}\ab{(\ell_2)61}}$}}
}

\defntb{fromRform}{\vardef{n}}{\vardef{expression}}{converts any momentum-twistor $5$-brackets (encoded by the symbols \fun{R}{\tt[}\vardefms{abcde}{\tt]}) in \var{expression} into superfunctions of the momentum-twistors of the form $f\!\times\!\delta^{k\times4}\big(C\!\cdot\!\eta\big)$ encoded by lists $\{f,C\}$, where $f$ is an ordinary function (of momentum-twistors) and $C$ is a $(k\times$\var{n}$)$-matrix of ordinary functions.}

\newpage
\defn{localPoles}{\vardef{n},\vardefo{ell}0}{returns the product of all physical poles that can appear in an \var{n}-point, \var{ell}-loop amplitude integrand. That is, the product of all {\it local} poles involving the external momenta (four-brackets of the form \fun{ab}{\tt[}$a\mi1,a,b\mi1,b${\tt]}), all local propagators \fun{ab}{\tt[}$(\ell_i),a\mi1,a${\tt]}, and (if \var{ell}$\,\geq\!2$) all internal propagators \mbox{\fun{ab}{\tt[}$(\ell_i),(\ell_j)${\tt]}}. This is useful for verifying that expressions are free of spurious poles: if all momentum-twistor components are integers, then multiplying any amplitude by \fun{localPoles}{\tt[]} should always evaluate to an integer.}

\defn{quadCuts}{\vardef{legList}}{for a box whose corners are given by the legs specified by \var{legList}, \fun{quadCuts} returns $\{\{{\color{cut1}Q^1_a},\ldots,{\color{cut1}Q^1_d}\},\{{\color{cut2}Q^2_a},\ldots,{\color{cut2}Q^2_d}\}\}$, specifying the points along the lines $(Aa),\ldots,(Dd)$ which lie along the quad-cuts (see \mbox{Table \ref{quad_cuts_table}}).}

\defntb{supercomponent}{\vardefms{component}}{\vardef{superFunction}}{in the {\tt two\uscore loop\rule[-1.05pt]{7.5pt}{.75pt}amplitudes} package, a \var{superFunction} must be represented by a pair $\{f,C\}$: an {\it ordinary} function $f(Z)$ of momentum-twistors times a {\it fermionic} $\delta$-function of the form,\\[-12pt]
\vspace{-0pt}\eq{\delta^{k\times4}\big(C\!\cdot\!{\eta}\big)\equiv\prod_{I=1}^4\left\{\bigoplus_{a_1<\!\cdots<a_k}\!\!(a_1\!\cdots a_k)\,{\eta}_{a_1}^I\!\!\cdots{\eta}_{a_k}^{I}\right\},\vspace{-0pt}}
with $C\!\equiv\!\big(c_1,\ldots,c_n\big)$ an $(n\times k)$-matrix of functions, $(a_1\cdots a_k)\!\equiv\!\mathrm{det}\hspace{-1pt}(c_{a_1},\ldots,c_{a_k})$, and where $\eta\!\equiv\big(\eta_1,\ldots,\eta_n\big)$ denotes the momentum-twistor fermionic (anti-commuting) variables which label each state. To be clear, we consider each particle to be a Grassmann coherent state (see ref.\ \cite{ArkaniHamed:2008gz}) expressed in the form,\\[-10pt]
\vspace{-2.5pt}\eq{\left|a \right> \equiv \left|a\right>_{\{\}}\!\pl\, \eta_a^I  \left|a\right>_{\{I\}}\!\pl\, \frac{1}{2!}  \eta_a^I  \eta_a^J \left|a \right>_{\{I,J\}}\!\pl\,\frac{1}{3!}  \eta_a^I  \eta_a^J  \eta_a^K \left|a\right>_{\{I,J,K\}}\!\pl\,\eta_a^1  \eta_a^2  \eta_a^3  \eta_a^4 \left|a\right>_{\{1,2,3,4\}}\!.\nonumber\vspace{-0cm}}
Thus, if we let \var{$r_a$} denote the $R$-charge of the $a^{\mathrm{th}}$ particle according to,
\vspace{2.5pt}\eq{\begin{array}{|l|l|l@{\hspace{1cm}}|l|}
\hline \text{field }&\text{helicity}&R\text{-charge} (\text{\var{$r_a$}})&\text{short-hand for \var{$r_a$}}\\\hline
|a\rangle_{\{\}}&\;\;+1&{\tt \{\}}& {\tt p}\\
|a\rangle_{\{I\}}&\;\;+\frac{1}{2}&{\tt \{I\}}&{\tt p/2}(\Leftrightarrow{\tt \{4\}})\\
|a\rangle_{\{I,J\}}&\;\;\phantom{+}0&{\tt \{I,J\}}&\text{---}\\
|a\rangle_{\{I,J,K\}}&\;\;-\frac{1}{2}&{\tt \{I,J,K\}}&{\tt m/2}(\Leftrightarrow{\tt \{1,2,3\}})\\
|a\rangle_{\{1,2,3,4\}}&\;\;-1&{\tt \{1,2,3,4\}}&{\tt m}\\\hline
\end{array}\nonumber\vspace{0.5pt}}
then \fun{superComponent}[\var{$r_1$},\ldots,\var{$r_n$}][\var{superFunction}] returns the {\it component} function of \var{superFunction} (an ordinary function) proportional to,
$\prod_{a=1}^{n}\!\!\left(\prod_{I\in r_a}\!\!{\eta}_a^{I}\right)$,
\eq{\text{\fun{superComponent}[\var{$r_1$},\ldots,\var{$r_n$}][\var{superFunction}]}=\int\!\!\prod_{a=1}^{n}\prod_{I\in r_a}d{\eta}_a^{I}\big(\text{\var{superFunction}}\big);\nonumber}
this is the component of \var{superFunction} involving states $\big\{|1\rangle_{\text{\var{$r_1$}}},\ldots,|n\rangle_{\text{\var{$r_n$}}}\big\}$. (To work with momentum-space components, the package of \mbox{ref.\ \cite{Bourjaily:2010wh}} may be used. 
}

\defn{symmetrize}{\vardef{loopIntegrand}}{given an argument \var{loopIntegrand} expressed in terms of $4$-brackets, {\tt ab[$\cdots$]}, \fun{symmetrize}{\tt[]} returns its symmetrization with respect to the loop-momentum variables---that is, it adds to \var{loopIntegrand} the same expression, but with references to loop-momentum variables exchanged.
}

\defn{toAnalytic}{\vardef{symbolicExpression}
}{replaces all the symbolic representations of terms occurring in \var{symbolicExpression} (as generated, for example, by the function \fun{localLoopIntegrand}), with superfunctions and (symmetrized) integrands---expressed in terms of $5$-brackets \fun{R}{\tt[\vardefms{abcde}]} and $4$-brackets \fun{ab}{\tt[\vardefms{abcd}]}.
}

\defn{toFullAnalytic}{\vardef{symbolicExpression}
}{is the same as \fun{toAnalytic}, but converts every superfunction expressed in terms of (products of) $5$-brackets \fun{R}{\tt[\vardefms{abcde}]} into a list $\{f,C\}$, where $f$ is an ordinary functions (of momentum-twistors) and $C$ is a $(k\times$\var{n}$)$-matrix of ordinary functions. That is, calling \fun{toFullAnalytic}{\tt[\var{exprn}]} is the same as calling \fun{fromRform}{\tt[]\![\fun{toAnalytic}{\tt[\var{exprn}]}]}.}

\defntb{octaCut}{\vardef{i},\vardef{j}}{\vardef{legList}}{for any two-loop, non-composite leading singularity---either kissing-boxes or a pentabox (indicated by whether \var{legList} has 8 or 7 entries, respectively)---\fun{octaCut} returns the momentum-twistors which encode the internal loop-momenta on the particular octacut solution labeled by $\{\text{\var{i}},\text{\var{j}}\}$.
For kissing-boxes labelled by \var{legList}$\,\,\equiv\!\!\big\{\!\{a,\ldots\},\ldots,\{h,\ldots\}\!\big\}$, it would return $\{Q^{\text{\var{i}}}_a,\ldots,Q^{\text{\var{j}}}_h\}$, while for a pentabox labelled by \var{legList}$\,\,\equiv\!\!\big\{\!\{a,\ldots\},\ldots,\{g,\ldots\}\!\big\}$, it would return $\{Q^{\text{\var{i}}}_a,\ldots,Q^{\text{\var{j}}}_g,Q^{\text{\var{j}}}_{\ell^*}\}$---where $Q^{\text{\var{j}}}_{\ell^*}$ refers to the internal line. }

\defntb{onShellFunction}{\vardefms{ij}}{\vardef{legList},\vardef{kList}}{for any on-shell function corresponding to a non-composite, one- or two-loop leading-singularity or a double-triangle on-shell function (evaluated at the point $(x^*\hspace{-1pt},y^*)$ as described in \mbox{appendix \ref{two_loop_on_shell_functions_section}}) whose topology of external legs is specified by \var{legList}, \fun{onShellFunction} will return the combination of on-shell functions involving the cut specified by \var{ij} (which can be a single integer $1$ or $2$ for one-loop, or a sequence of two integers form $\{1,2\}$ for two-loops, or 0 for a double-triangle) and whose corner amplitudes have $R$-charges specified by the list \var{kList}---e.g.,\vspace{-10pt}

\mathematica{0.945}{
nice/@\!\big\{\!onShellFunction\![1]\!\![\{\!\{1\}\!,\!\{2\!,\!3\}\!,\!\{4\!,\!5\}\!,\!\{6\!,\!7\!,\!8\}\!\}\!,\!\{\!-\!1\!,\!0\!,\!0\!,\!1\!\}\!]\!,\hspace{14cm}
onShellFunction\![0]\!\![\{\!\{1\!,2\}\!,\!\{3\!,\!4\}\!,\!\{\}\!,\!\{5\!,\!6\}\!,\!\{7\!,\!8\}\!,\!\{\}\!\}\!,\!\{\!0\!,\!0\!,\!0\!,\!0\!,\!0\!\}\!]\!,\hspace{14cm}
onShellFunction\![1\!,\!2]\!\![\!\{\!\{\!1\!\}\!,\!\{\!2\!\}\!,\!\{\!3\!\}\!,\!\{\}\!,\!\{\!4\!,\!5\!,\!6\!\}\!,\!\{\!7\!,\!8\!\}\!,\!\{\}\!\}\!,\!\{\!-\!1\!,\!0\!,\!-\!1\!,\!-\!1\!,\!1\!,\!0\!,\!0\!\}\!]\!,\hspace{14cm}
onShellFunction\![1\!,\!1]\!\![\!\{\!\{\!1\!\}\!,\!\{\!2\!\}\!,\!\{\!3\!\}\!,\!\{\}\!,\!\{\!4\!\}\!,\!\{\!5\!\}\!,\!\{\!6\!,\!7\!,\!8\!\}\!,\!\{\}\!\}\!,\!\{\!-\!1\!,\!0\!,\!-\!1\!,\!0\!,\!-\!1\!,\!0\!,\!1\!\}\!]\!\big\}\hspace{0cm}}{
\vspace{-9.5pt}\scalebox{1}{$\displaystyle \big\{R[1,3,4,5,6]R[(56)\tcap(134),6,7,8,1],$}\rule[-0pt]{0pt}{0pt}\hspace{14cm}$~$
\scalebox{1}{$\displaystyle R[Q[1,8],2,3,4,Q[5,4]]R[Q[5,4],6,7,8,Q[1,8]],$}\rule[-0pt]{0pt}{0pt}\hspace{14cm}$~$
\scalebox{1}{$\displaystyle R[1,3,6,7,8]R[3,4,5,6,(67)\tcap(813)],$}\rule[-0pt]{0pt}{0pt}\hspace{14cm}$~$
\scalebox{1}{$\displaystyle R[4,5,6,8,1]R[(56)\tcap(481),6,7,8,(81)\tcap(456)]\big\}$}\rule[-0pt]{0pt}{0pt}\hspace{0cm}$~$\vspace{-0pt}}
The list of examples above illustrate how to specify (in order): a one-loop box, a double-triangle, a penta-box, and kissing-boxes leading singularities. In the above example, we should point out that `\fun{$Q$}{\tt\![\var{a},\var{b}]}' is how \fun{shift}{\tt[}$\{$\var{a},\var{b}$\},$\var{$\alpha$}{\tt]} is formatted by \fun{nice}{\tt[]}.}

\subsubsection*{Kinematical Specification, Reference Data, \& Numerical Evaluation}

All evaluation routines refer to a set of momentum-twistors stored as the global variable {\tt Zs}---a list of four-vectors, the last four of which are understood as denoting the reference loop-momenta $(\ell_1,\ell_2)$. The variable {\tt Zs} can be re-defined by the user at will, but problems may arise if the number of twistors in the list {\tt Zs} is not $(n\pl4)$.

\vspace{-2.5pt}

\defn{evaluate}{\vardef{expression}}{uses the kinematical data specified by the global variable {\tt Zs} to evaluate all $4$-brackets \fun{ab}{\tt[$\cdots$]} as determinants (see equation (\ref{defn_of_four_bracket})). If \var{expression} involves superfunctions expressed in terms of $5$-brackets \var{R}{\tt[\vardefms{abcde}]}, then it converts all of these to the form generated by \fun{fromRform} prior to evaluation; also, if the output of \fun{rAmp} for a 2-loop amplitude is detected, then \fun{evaluate} will call \mbox{\fun{symEvaluate}{\tt[\var{expression}]}} in order to directly symmetrize the loop-momentum variables.
}

\defn{referenceKinematics}{\vardef{n}}{defines the global variable {\tt Zs}---which specifies the kinematical data to be used for evaluation---to correspond to a {\it very} convenient point in the space of external kinematics and internal loop-momentum. In particular, it chooses the \var{n} external momentum-twistors to be,
\vspace{5pt}{\tt \eq{\left(\begin{array}{@{}c@{}@{}c@{$\{$}c@{$,$}c@{$,$}c@{$,$}c@{$\}$}}\text{{\tt Zs[\![1]\!]}}&\equiv&1&1&1&1\\\text{Zs[\![2]\!]}&\equiv&1&2&3&4\\[-4pt]
&\multicolumn{1}{c}{$$}&\multicolumn{1}{c}{\vdots\hspace{0.5pt}~}&\multicolumn{1}{c}{\vdots\hspace{0.5pt}~}&\multicolumn{1}{c}{\vdots\hspace{0.5pt}~}&\multicolumn{1}{c}{\vdots\hspace{0.5pt}~}\\\text{Zs[\![\var{n}]\!]}&\equiv&\binom{\text{\var{n}}-1}{\text{\var{n}}-1}&\binom{\text{\var{n}}}{\text{\var{n}}-1}&\binom{\text{\var{n}}+1}{\text{\var{n}}-1}&\binom{\text{\var{n}}+2}{\text{\var{n}}-1}\end{array}\right),\vspace{5pt}}}
\hspace{-5pt}and similarly for the four twistors which specify the lines $(\ell_1,\ell_2)$ for the internal loop-momenta (the last four-entries of {\tt Zs}). These twistors are convenient for many reasons---all $4$-brackets involving ordered sets of twistors are (small) positive integers, and these values avoid hitting any accidental, spurious poles.}

\defn{randomPositiveKinematics}{\vardef{n}}{defines the global variable {\tt Zs} (a list of \mbox{(\var{n}\,\pl\,4)} \mbox{$4$-tuples} (the last four of which specify the point $(\ell_1,\ell_2)$ in loop-momentum space for evaluation of the integrand)) to be a {\it randomly chosen} positive matrix: {\tt Zs}$\in\!\!G_+(4,\text{\var{n}}\pl\,4)$. (A {\it positive} matrix is one for which all maximal minors involving ordered lists of columns are positive.) Although this kinematical data is generated at random using integers, the twistors are then rescaled in order to reduce the magnitudes of $4$-brackets---reducing the size of the integers appearing in the ratios generated by the evaluation of amplitudes.}

\newpage
\defnNA{showTwistors}{}{returns a formatted table illustrating the kinematical data currently defined by the global variable {\tt Zs} (where $\ell_1\!\equiv\!(A,B)$ and $\ell_2\!\equiv\!(C,D)$); this is the set of twistors used for evaluation, for example, by the function \fun{evaluate}\hspace{-1pt}{\tt[]}.\\[-25pt]

\mathematica{0.9}{referenceKinematics[10]\hspace{14cm}$~$
showTwistors}{{\tt \vspace{-10pt}\eq{\fwboxL{11.1cm}{\begin{array}{|@{$\,\,$}c@{$\,\,$}c@{$\,\,$}c@{$\,\,$}c@{$\,\,$}c@{$\,\,$}c@{$\,\,$}c@{$\,\,$}c@{$\,\,$}c@{$\,\,$}c@{$\,\,$}|}\hline \text{Z}_{\text{1}}&\text{Z}_{\text{2}}&\text{Z}_{\text{3}}&\text{Z}_{\text{4}}&\text{Z}_{\text{5}}&\text{Z}_{\text{6}}&\text{Z}_{\text{7}}&\text{Z}_{\text{8}}&\text{Z}_{\text{9}}&\text{Z}_{\text{10}}\\\hline
\text{1}&\text{1}&\text{1}&\text{1}&\text{1}&\text{1}&\text{1}&\text{1}&\text{1}&\text{1}\\
\text{1}&\text{2}&\text{3}&\text{4}&\text{5}&\text{6}&\text{7}&\text{8}&\text{9}&\text{10}\\\text{1}&\text{3}&\text{6}&\text{10}&\text{15}&\text{21}&\text{28}&\text{36}&\text{45}&\text{55}\\\text{1}&\text{4}&\text{10}&\text{20}&\text{35}&\text{56}&\text{84}&\text{120}&\text{165}&\text{220}\\
\hline\end{array}\,\begin{array}{|@{$\,\,$}c@{$\,\,$}c@{$\,\,$}|}\hline\text{A}_{\fwboxL{0pt}{\phantom{1}}}&\text{B}\\\hline\text{2}&\text{1}\\\text{21}&\text{12}\\\text{121}&\text{78}\\\text{506}&\text{364}\\\hline
\end{array}\,\begin{array}{|@{$\,\,$}c@{$\,\,$}c@{$\,\,$}|}\hline\text{C}_{\fwboxL{0pt}{\phantom{1}}}&\text{D}\\\hline\text{2}&\text{1}\\\text{25}&\text{14}\\\text{169}&\text{105}\\\text{819}&\text{560}\\\hline
\end{array}}\label{example_twistor_kinematics_10pt}\vspace{5pt}}}}
}

\defn{symEvaluate}{\vardef{expression}}{called automatically by \fun{evaluate}{\tt[]} if \var{expression} contains any two-loop `kermit' functions (as would be generated by \fun{rAmp}{\tt[\var{n},\var{k},2]}); because the output of \fun{rAmp}{\tt[]} does {\it not} automatically symmetrize over the loop-momentum variables, \fun{symEvaluate} is called in order to do this symmetrization numerically (and quickly). To be clear, \fun{symEvaluate} generates the same output as \fun{evaluate}{\tt[}\fun{fromRform}{\tt[]\![}\var{expression}{\tt]]}, but in combination with the same result, swapping the loop-momentum-twistors representing the lines $(\ell_1,\ell_2)$. (Local integrand representations obtained using \fun{localLoopIntegrand}{\tt[]} are automatically symmetrized with respect to the loop-momenta.\\[-22pt]

\ind (To {\it prevent} the output of \fun{rAmp}{\tt[]} form being symmetrized, one can use \fun{evaluate}{\tt[\fun{explicitKermits}{\tt[\var{expression}]]}}; this replaces all `kermits' with expressions using $4$-brackets, and will prevent \fun{evaluate}{\tt[]} from calling \fun{symEvaluate}.)
}

\newpage
\subsubsection*{Miscellaneous (but Generally-Useful) Functions Defined by the Package}
\vspace{-7.5pt}
\defn{complement}{\vardef{listA},\vardefms{listsB}}{returns a list of the elements of \var{listA} not in the (sequence of one or more) lists \var{listsB}. It is essentially the same as {\sc Mathematica}'s function {\tt Complement}\hspace{-1pt}{\tt[]}, but where the output is not sorted---both saving computation time, and leaving the ordering of \var{listA} in place. }

\defnNA{memory}{}{returns the amount of memory currently being used by the notebook's kernel---simply a formatted version of {\sc Mathematica}'s function \mbox{{\tt MemoryInUse}\hspace{-1pt}{\tt[]}}.}

\defn{nice}{\vardef{expression}}{formats \var{expression} to display `nicely' by making replacements such as \fun{ab}{\tt[$\cdots$]}$\mapsto\!\ab{\cdots}$, $\alpha${\tt [1]}$\mapsto\!\alpha_1$, etc., by writing any level-zero matrices in {\tt MatrixForm}, and making other simplified, notational replacements.}

\defn{niceTime}{\vardef{timeInSeconds}}{converts a time measured in seconds \var{timeInSeconds}, to human-readable form. For example,\\[3pt]
\mathematica{0.9}{niceTime[299\,792\,458]\hspace{14cm}$~$
niceTime[3.1415926535]}{{\tt 9 years, 182 days}\hspace{14cm}$~$ {\tt 3 seconds, 141 ms}}}

\defn{random}{\vardef{objectList}}{returns a random element from (the first level of) \var{objectList}.}

\defn{timed}{\vardef{expression}}{evaluates \var{expression} and prints a message regarding the time required for evaluation.}

~\newpage
\providecommand{\href}[2]{#2}\begingroup\raggedright\endgroup

\end{document}